\documentclass[11pt]{article}
\usepackage{graphicx}
\usepackage{natbib}
\setcitestyle{authoryear,round,aysep={},yysep={,},citesep={,}}
\usepackage[margin=1in]{geometry}   
\usepackage{setspace}
\usepackage{float}
\usepackage{parskip}
\usepackage{amsmath}
\usepackage{amsfonts}
\usepackage[T1]{fontenc}
\usepackage{newtxtext,newtxmath}    
\usepackage{dsfont}
\usepackage{longtable}
\usepackage{booktabs}
\usepackage{caption}
\usepackage{tabularx}
\usepackage{xcolor}
\usepackage{listings}
\usepackage{array}
\usepackage{comment}
\usepackage{threeparttable} 
\usepackage{bbm} 
\usepackage{subcaption}
\usepackage{tikz}
\usepackage{pdflscape}
\usepackage{hyperref}
\hypersetup{pdfauthor={}, pdfcreator={}, pdfproducer={}}   
\usepackage{cleveref} 
\usetikzlibrary{shapes.geometric, arrows.meta, positioning, calc}


\usepackage{needspace}

\definecolor{codegreen}{rgb}{0,0.6,0}
\definecolor{codegray}{rgb}{0.5,0.5,0.5}
\definecolor{codepurple}{rgb}{0.58,0,0.82}
\definecolor{backcolour}{rgb}{0.95,0.95,0.92}

\lstdefinestyle{mystyle}{
    backgroundcolor=\color{backcolour},   
    commentstyle=\color{codegreen},
    keywordstyle=\color{magenta},
    numberstyle=\tiny\color{codegray},
    stringstyle=\color{codepurple},
    basicstyle=\ttfamily\footnotesize,
    breakatwhitespace=false,         
    breaklines=true,                 
    captionpos=b,                    
    keepspaces=true,                 
    numbers=left,                    
    numbersep=5pt,                  
    showspaces=false,                
    showstringspaces=false,
    showtabs=false,                  
    tabsize=2
}

\title{Beyond Exposure: AI and the Division of Labor in Public Sector Jobs and Teams}
\author{}
\date{}

 \title{AI and the Division of Labor in Public Sector Jobs and Teams}
 \author{Andrew Ledingham, Michael Hollins, Matthew Lyon, David Gillespie, \\
 Umar Yunis-Guerra, Jamie Siviter, David Duncan and Oliver P. Hauser}
 \date{August 2026}

\begin{document}

\baselineskip24pt
\maketitle 

\begin{abstract}

How artificial intelligence (AI) affects an organization depends not only on which jobs are exposed, but on how more (versus less) exposed tasks are combined within roles. Using 283,484 UK Civil Service vacancies, we extract 2.7 million tasks and estimate their automation potential with an ensemble of eight foundation models. We find that differences among tasks within roles account for 65.3 percent of total exposure variance, while profession and grade together explain only 4.0 percent, suggesting that seemingly similar roles often differ substantially in their exposure and potential for redesign. Combining roles' mean exposure with their within-role dispersion distinguishes where workforce adjustment, job redesign, targeted task automation, or limited intervention may be most relevant. Taking account of which tasks can change as AI expands into a role, we simulate redesign at the role and team level. Role redesign cuts administration from 30.0 to 24.6 percent of working time and raises assurance tasks from 8.7 to 14.5 percent. When conducted at the team level, redesign merges overlapping roles, cutting the number of distinct roles by 17.7 percent while headcount falls by only 6.4 percent; the posts removed come from junior and middle grades, so restructured teams become weighted toward senior staff. We contribute to the automation, augmentation and productivity debates and advance our understanding of job redesign in the age of AI. 

\end{abstract}\clearpage
\section{Introduction}\label{section:Introduction}

The rapid diffusion of generative artificial intelligence has renewed longstanding questions about how technological change affects jobs, workers, and organizations \citep{brynjolfsson2017can, acemoglu2022artificial}. Research has responded by constructing measures of occupational exposure to AI, generally by assessing how the capabilities of AI systems map onto occupational tasks \citep{FeltenEtAl2021, Gmyrek2025generative, EloundouEtAl2024GPTsAreGPTs}. These measures suggest that generative AI may affect a large share of cognitive and information-processing work, including occupations associated with highly educated and high-wage workers \citep{FeltenEtAl2023, felten2023occupational}, and an often implied suggestion is that exposed jobs will experience partial or full automation and lead to job replacement through AI \citep{frank2019toward}. Emerging studies document changes in employment, vacancies, and work following the introduction of generative AI, although both the size and direction of these effects vary across technologies, settings, and groups of workers \citep{GathmannGrimmWinkler2024AITaskChanges, AghionEtAl2025DifferentUsesAIShapeLaborDemand, HampoleEtAl2025AILaborMarket, HumlumVestergaard2025UnequalAdoptionChatGPT, PrytkovaPetit2025EmploymentImpactUS}. Some present evidence of displacement in particular contexts and countries \citep{BrynjolfssonChandarChen2025Canaries, klein2025generative, demirci2025ai, HosseiniMaasoumLichtinger2025SeniorityBiased, dominski2025advancing}, while others find no evidence of it \citep{chandar2025tracking, HumlumVestergaard2025}.

Exposure, however, is neither adoption nor an employment outcome. We define a task's AI exposure as its automation potential: the estimated share of its substantive content that current AI systems could perform. This technical potential does not establish whether an organization will deploy AI, whether deployment will be economically or institutionally feasible, or whether the affected work will be automated, augmented, or reorganized, and many jobs may simply not be economically attractive or technically feasible to automate in the near future \citep{svanberg2024beyond}. Legal and accountability requirements, data and infrastructure, procurement, worker capabilities, managerial choices, and organizational capacity for change all stand between technological capability and realized impact, particularly in our public sector setting (Section \ref{section:Discussion}).

Measuring exposure also raises two challenges. First, exposure is commonly measured at the occupational level, using standardized task taxonomies such as the Occupational Information Network (O*NET), the International Standard Classification of Occupations (ISCO-08), or national occupational classifications \citep{FeltenEtAl2021, Gmyrek2025generative, EloundouEtAl2024GPTsAreGPTs, HampoleEtAl2025AILaborMarket, henseke2025exposed, brynjolfsson2018can}. These resources enable comparisons across occupations and economies, but generally assign a common task bundle to all jobs within an occupation, although actual roles differ by domain, seniority, organizational setting, and the allocation of responsibilities within an employer. Contemporaneous studies examine heterogeneity in exposure across firms \citep{HampoleEtAl2025AILaborMarket} and across occupations, industries, and regions \citep{fenoaltea2024follow}, complementing our within-organization focus. Second, studies employ different definitions, scales, prompts, sources of judgment, and aggregation rules, drawing on expert or worker assessments, language models, or semantic similarity between occupational tasks and descriptions of technological capabilities or observed AI use \citep{Gmyrek2025generative, EloundouEtAl2024GPTsAreGPTs, HampoleEtAl2025AILaborMarket, PrytkovaPetit2025EmploymentImpactUS, HandaEtAl2025}. LLM-based scores may also vary substantially across models even when the underlying tasks and instructions remain fixed \citep{yin2026}, so exposure estimates may reflect the measurement instrument as well as properties of the work being measured.

Nevertheless, understanding a role's exposure through its bundle of associated tasks is well-established \citep{AcemogluRestrepo2022TasksAutomationWageInequality, FeltenEtAl2021, GmyrekBergBescond2023GenerativeAIJobs, klein2025generative}. In particular, \citet{HampoleEtAl2025AILaborMarket} emphasize two sufficient statistics of AI exposure. A higher mean exposure increases the scope for AI to substitute for the tasks performed by a role, while exposure concentration in particular tasks creates scope for within-job reallocation. We measure this concentration as the variance of exposure across a role's tasks, which is zero when all tasks are equally exposed and rises as exposure becomes concentrated in some tasks. Depending on the task mix of a role, its substitution effect may therefore be partly offset by its exposure dispersion. Because the effects of AI on labor demand vary according to how the technology is used \citep{AghionEtAl2025DifferentUsesAIShapeLaborDemand} and which workers or tasks are affected \citep{GathmannGrimmWinkler2024AITaskChanges, PrytkovaPetit2025EmploymentImpactUS}, we examine whether roles that were more exposed before the release of ChatGPT subsequently experienced different recruitment trajectories.

Considering these two dimensions jointly distinguishes four forms of organizational impact. Roles with high average exposure and little variation among their tasks, which we classify as Displacement roles, provide the strongest case for substitution, because comparatively little unexposed work remains around which to reorganize the role. In Augmentation roles, high average exposure coexists with substantial task-level variation, so redesign is more plausible because exposed and less-exposed activities remain bundled together. Concentrated roles, with lower average exposure but a small number of highly exposed tasks, may be candidates for targeted automation, while Insulated roles, with both low average exposure and limited dispersion, are likely to experience less substantial change. The labels do not predict whether any particular role will disappear, but indicate whether workforce adjustment, job redesign, targeted task automation, or limited intervention is the more relevant response.

Task heterogeneity nevertheless creates only an opportunity for reallocation. Shifting the focus from job displacement to the automation of specific tasks highlights the opportunity for productivity gains through reinvesting newfound time into other activities \citep{acemoglu2018race}. As \citet{agrawal2023we} argue, ``automation of some tasks can lead to augmentation of labor elsewhere'' (p. 155), and complementarities and redesigned processes are key to productivity gains \citep{brynjolfsson2017can}. Experimental and field studies demonstrate that generative AI can improve performance in particular tasks and workflows \citep{noy2023experimental, dell2023navigating, BrynjolfssonLiRaymond2025}, and studies in education \citep{Choo2024} and healthcare \citep{Meynhardt2025} argue that the time saved can be reallocated to other tasks. Realizing that opportunity, however, requires organizations to decide which activities should be automated, which responsibilities should remain with people, and how released capacity should be used, so AI presents an automation--augmentation paradox rather than a simple technological choice \citep{RaischKrakowski2021AutomationAugmentationParadox}. Evidence that AI improves performance within an existing workflow does not determine how a job should be redesigned once some of its tasks require less human time.

When AI can perform part of a job but not the whole role, will released capacity be redirected toward less automatable work, used to supervise AI-generated outputs, or absorbed by further routine activities? We examine this question through two exploratory redesign exercises. In role redesign, an LLM is asked to improve the task composition of job roles. In team redesign, a similar process is applied to the roles within a team, alongside an AI-driven reduction in team headcount, so that tasks can move between roles. These exercises do not predict how managers and employees will redesign jobs in practice, but test whether AI-assisted redesign, by itself, directs human effort toward work with lower automation potential and how it reshapes team structures. Our central argument is that exposure locates the organizational design problem but does not solve it. How exposed a role is depends on how tasks are bundled within a particular organization; organizations respond to that exposure through the adjustment margins available to them; and how released capacity should be used cannot be read directly from the exposure measure or delegated to the technology.

We study these questions in the context of the UK Civil Service (UKCS), a large, multi-unit organization with a shared recruitment platform, standardized grades and professions, and substantial variation in organizational functions. We analyze 283,484 vacancies advertising 748,690 roles across 83 departments and the full Civil Service grade structure between 2018 and 2026. Rather than assigning standardized occupational tasks to these roles, we extract the 2,729,344 tasks described in the vacancies, and eight foundation models estimate the automation potential of each task. We aggregate their judgments into an ensemble exposure score and reweight the vacancy sample by department, grade, and profession to represent the covered Civil Service workforce rather than recruitment flows alone.


We address three related research questions. First, how much variation in AI exposure arises between roles, and how much arises among the tasks within a role? Second, is pre-existing exposure associated with subsequent changes in recruitment demand following the release of ChatGPT? Third, when an LLM is asked to redesign exposed roles and teams without being shown their exposure scores, how does it propose that tasks and team structures should change? The exposure analysis measures the technical automation potential of existing work, the labor-demand analysis examines an observed outcome without directly observing AI adoption or establishing a causal effect, and the redesign analysis studies proposals generated by an LLM rather than changes implemented by the UKCS. We focus on changes within existing work rather than on the creation of entirely new jobs or shifts in the growth of jobs across industries \citep{autor2015untangling, acemoglu2020robots}.

We report three main findings. First, most variation in exposure occurs within roles rather than between them, and most of the remaining variation occurs between roles that share the same profession and grade. Exposure also varies more across departments and professions than across grades, suggesting that it is more closely related to the content and organizational setting of work than to hierarchical seniority alone, in line with the argument that high-skilled jobs will not be uniformly affected by AI \citep{fenoaltea2024follow}. Across the workforce, the technical opportunity, measured as salary weighted by mean exposure, totals approximately £7.5 billion, around 36 percent of the £21.1 billion salary bill. This figure indicates the scale of technically exposed work, not realized savings, and its meaning differs across the four groups. For Displacement roles, the exposure-weighted value may understate what is at stake, because if posts are not refilled as staff leave, the relevant value becomes the full salary cost of the role. For Augmentation roles, the value depends on how released time is redeployed.

Second, mean exposure predicts subsequent recruitment demand. Following \citet{HampoleEtAl2025AILaborMarket}, we estimate a two-way fixed-effects panel model for 949 department-by-profession-by-grade occupations, using exposure measured before ChatGPT's release. Mean exposure is only weakly associated with postings during the first post-release year, but becomes a statistically significant negative predictor from the second year onward. At the sample median of three monthly postings, a one-standard-deviation increase in mean exposure is associated with a 10.4 percent reduction in postings during the second post-release year, a relationship that is robust across alternative samples, pre-release windows, specifications, and task-weighting assumptions. Within-role dispersion has a positive coefficient, consistent with task reallocation partly offsetting substitution, but it is not statistically significant in the principal specifications. Because the analysis is observational, it does not establish that AI adoption caused the reduction in postings.

Third, AI-assisted redesign changes the content of individual roles only at the margin. In role redesign, applied to 10,000 roles, the model mostly replaces exposed tasks with reviewing AI output, shifting employees from direct execution toward oversight of automated processes, and proposes using the time this frees mainly for routine reporting and monitoring, leaving whole-role exposure largely unchanged. These results qualify the expectation that automation will naturally make jobs more human-centric, since without an explicit objective concerning comparative advantage \citep{acemoglu2018race, LoaizaRigobon2024}, the model does not consistently direct released capacity toward less automatable work.

Team redesign, applied to 80 teams alongside an AI-driven reduction in headcount, instead changes how work is organized. Organization design research treats the division of work into roles as a central design choice, because how tasks are grouped determines how much people must coordinate and understand one another's work \citep{PuranamRaveendranKnudsen2012, kiggundu1981task}, and automation of some tasks can lead workers to specialize in the remaining ones because they no longer face the costs of coordination \citep{gong2024automation}. The model instead merges overlapping roles in most teams, leaving fewer, larger roles, and the reduction falls on junior and middle grades rather than senior ones, consistent with expertise moderating whether automation leads to higher or lower wages for specific jobs \citep{autor2025expertise}. Much of the human work the model adds back is again oversight of AI output. Exposure can therefore inform job redesign, but it does not determine how human time should be used or how work should be divided between roles.

The paper makes three contributions. First, it contributes to the task-based AI-exposure literature \citep{FeltenEtAl2021, EloundouEtAl2024GPTsAreGPTs, GmyrekBergBescond2023GenerativeAIJobs, henseke2025exposed} by developing an organization-specific measure that preserves variation among roles conventionally treated as identical, and our eight-model ensemble, human comparison, NLP validation, and prompt-ablation analysis provide a more robust basis for using LLMs to measure automation potential. Second, we connect role-level exposure to subsequent labor demand within a large public-sector organization and translate exposure into four forms of organizational impact through the joint mean--dispersion framework. Third, we contribute to research on job redesign \citep{hackman1976motivation, bartling2012screening, gibbs2022new} and the automation--augmentation tension by examining what an LLM proposes when asked to redesign real roles, both individually and within teams. \citet{karasek1979job} observed that job redesign should consider expanding decision latitude for workers to reduce mental strain and job dissatisfaction, whereas the redesigns we study mostly fill released time with routine reporting, monitoring, and oversight of AI output. Productive augmentation and the grouping of work into roles therefore require deliberate choices about the future content and organization of work, echoing the argument that organizations need a holistic view to adapt to technological change \citep{stephany2024price}.

Section \ref{section:DatasetDesc} describes the UK Civil Service setting and vacancy data, and Section \ref{section:Methodology} presents the task-extraction, exposure-scoring, reweighting, classification, and redesign methods. Section \ref{section:Results} reports the results, and Section \ref{section:Discussion} discusses their implications and the study's limitations. The Supplementary Materials provide the prompts, worked examples, validation exercises, and robustness checks.
\section{Research setting and data}\label{section:DatasetDesc}

The UKCS is a very large employer with a headcount of 557,825 in 2026 and operates using centralized recruitment platforms. Most recruitment flows through Civil Service Jobs, the recruitment platform run by the Government Recruitment Service to advertise roles across the UK Government.\footnote{See \url{https://www.civilservicejobs.service.gov.uk/}. Exceptions include large graduate schemes such as the Fast Stream and certain professions such as security and intelligence that recruit through separate platforms.} As vacancies across many departments are published in a common format with standard grades and professions, Civil Service Jobs is an unusually rich source for analyzing workforce tasks. 

We obtained an extract of roles published between February 2018 and May 2026 held in the Government Recruitment Information Database (GRID), the internal dataset of previously advertised roles. Each vacancy records its title, department, grade, profession, job description and summary, posting and closing dates, and the number of jobs available (Appendix \ref{section:AppendixDatasetDescription}). A single vacancy may therefore advertise many roles, as in national recruitment campaigns for Border Force Officers or Prison Officers. We refer to ``vacancies'' as unique entries in the dataset and ``roles'' as the number of jobs available.

Our dataset comprises 283,484 unique vacancies (748,690 roles). Three quarters of vacancies advertise for just one role each, accounting for 28 percent of roles, whereas four percent of vacancies advertise for 10+ roles, accounting for almost half of roles. Eleven newly created departments and agencies only begin reporting after October 2019. Between October 2019 and March 2026 (our substantive sample), an average of 3,521 vacancies (9,199 roles) were posted each month (Figure \ref{fig:VacanciesOverTime}).

The dataset spans 83 departments, 31 professions (plus an ``Other'' category), and all grades in the UKCS structure. Civil Service grades reflect increasing levels of responsibility: Administrative Assistant (AA) and Administrative Officer (AO) cover routine administration and frontline services; Executive Officer (EO) covers operational decision-making; Higher Executive Officer (HEO) and Senior Executive Officer (SEO) cover policy development and project management; Grade 7 (G7) and Grade 6 (G6) cover complex policy and strategic planning; and the Senior Civil Service (SCS) - Deputy Directors, Directors and Permanent Secretaries - leads major policy areas and government strategy. The 83 departments together employ 491,710 FTE, or 94\% of the UKCS workforce.\footnote{See the \hyperlink{https://www.gov.uk/government/statistics/civil-service-statistics-2026/statistical-bulletin-civil-service-statistics-2026}{UK Civil Service Statistical Bulletin 2026.}} Every grade other than AA, G6 and SCS represents at least 10\% of roles; the low shares for AA, G6 and SCS are expected given their small share of overall headcount. Around a quarter of roles were originally assigned to the ``Other'' profession, so we passed each of these vacancies to an LLM, which reassigned the profession where the title and job description indicated a different one (Appendix \ref{sec:AppendixProfessionReassignment}). After this step, Operational Delivery accounts for 30.8\% of vacancies and 53\% of roles, representing the majority of our sample.

Our goal with this dataset is to make inferences about the UKCS. However, vacancies represent the inflow of roles, and they may not be representative of the stock of the UKCS workforce. We therefore compare our dataset's proportions against the official UKCS workforce statistics and find that our vacancies are broadly in line across department, grade, and profession (Appendices \ref{section:DeptDataDescAppendix}, \ref{section:GradeDataDescAppendix}, and \ref{section:ProfessionDataDescAppendix}). To correct for any remaining discrepancies, we use an iterative proportional fitting procedure (Appendix \ref{app:Raking}), which allows us to extrapolate from vacancy data to the current workforce. Because vacancies capture the flow of recruitment rather than the stock of roles, they are also informative in their own right, as they let us observe how hiring patterns and the tasks described in adverts change over time (Section \ref{sec:DemandForLabor}).       
\section{Methodology}\label{section:Methodology}

This section describes how we use our GRID extract from Section \ref{section:DatasetDesc} to construct a measure of AI exposure for the UKCS. First, we build a prompt ablation study to refine and improve our prompt by identifying the most important components. Second, with our prompt we describe how we extracted tasks and assigned AI exposure scores. Third, we reweight vacancies to represent the UKCS headcount and aggregate scores to the job level. Finally, we review how we aggregated these vacancy-level scores into a consolidated view across most of the UKCS, and proceeded to consider job and team redesign. A visual representation of the full analytical workflow described in this section is provided in Figure \ref{fig:full_methodology}, and a full worked example is outlined in Appendix \ref{sec:AppendixExampleWalkthrough}.

\subsection{Refining the prompt}\label{sec:PromptRefining}

Existing work using AI exposure scores has prompts that differ substantially in scale, anchoring and output format. \cite{EloundouEtAl2024GPTsAreGPTs} classify O*NET tasks categorically (E0–E3) against a binary threshold of whether an LLM could halve completion time at equivalent quality; \cite{colombo2025} use five-shot prompts across three open-weight models on a 1–5 integer scale; the ILO \citep{Gmyrek2025generative} scores continuously on 0–1 but derives its anchors from worker surveys and expert panels rather than the model alone; and \cite{karpathy2026} uses a continuous 0–10 scale with descriptive anchors and worked examples at each band. Yin et al. (2026) replicate the Eloundou rubric verbatim across three frontier models and find mean exposure diverging by a factor of 3.6 with Cohen’s $\kappa$ of 0.36 for Gemini and Claude, demonstrating that the scoring instrument is itself a source of measurement error. No existing paper, however, reports a systematic sensitivity analysis of the prompt.

We designed an initial scoring prompt around the characteristics recurring in some of these studies: a continuous 0.0 to 1.0 scale, descriptive band anchors with a worked example at each band, framing in terms of current AI systems, and a required reasoning step before the score. To establish which of these components are most influential in assigning exposure scores, we conducted an ablation study using Anthropic’s Claude Sonnet 4.6 model over eleven prompt variants, each assigning exposure scores to an identical list of 5,623 tasks extracted from a random sample of 500 job adverts. A control rerun of the unmodified prompt establishes the noise floor and each variant's divergence from baseline is reported as a multiple of it. 

Removing the numeric bands shifts the score distribution 27 times further from the baseline than simply rerunning the identical prompt does (i.e., 27 times the noise floor, with scores falling by 13\% on average and 64\% of tasks moving down), and removing the reasoning instruction or the rubric produces shifts of 12 and 9.5 times the noise floor respectively; by contrast, other design choices (such as adding a system persona, dropping an exposure-type field or the band descriptions, or using explicit percentage thresholds) leave 92--95\% of scores within 0.1 of the baseline, close to the 95\% obtained by a plain rerun. Full results for all eleven variants are reported in Appendix~\ref{sec:AppendixPromptAblation}. 

The ablation study informed the design of the final prompt rather than validating it directly. The final prompt retains the three components the ablation study identified as essential (namely numeric bands, a reasoning step and a full rubric). It differs from the baseline in defining exposure as the share of a task's substantive content that current AI systems could perform, rather than the time that could be saved, in keeping with the definition of exposure scores used throughout this paper.

\subsection{Assigning AI exposure scores}\label{sec:AssigningExposureScores}

To assign AI exposure scores to each task, we proceeded in two phases. In the first phase, we gave each job description in GRID to Claude Sonnet 4.6. Per vacancy, the LLM extracted each task and assigned every task an automation potential (or exposure) score along with a line of reasoning.\footnote{Appendix \ref{sec:AppendixExtractAndScorePrompt} contains the extract and score prompt which were informed by the most important elements of the prompt ablation study.} By construction, these scores ranged between 0 and 1 at 0.1 intervals, meaning eleven scores were possible per task.\footnote{The eleven-band discretization is enforced at parsing: values outside $[0,1]$ are rejected, and intermediate values are snapped to the nearest band. This happened in very few cases. See Table \ref{tab:coercion-audit} in Appendix \ref{sec:AppendixTaskErrorRates} for extraction statistics.} 

In the second phase, we gave each vacancy's job description and its previously extracted tasks to seven other foundation models. Each model was instructed to provide a score per task along with a justification.\footnote{Appendix \ref{sec:AppendixScorePrompt} contains the scoring prompt.} Accordingly, for each task, we produced eight exposure scores and eight justifications given by each model.

We score tasks with eight models rather than one for three reasons. First, any single model carries its own calibration quirks, such as systematically harsher or more generous scores, so relying on one model would tie every downstream result to those idiosyncrasies. Second, we do not want the analysis to depend on proprietary models that may be deprecated or changed without notice. Six of the eight models in Table~\ref{tab:FMMetadata} are open-weight (the exceptions being Claude Sonnet 4.6 and Nova 2 Lite), so the exercise can be replicated and rerun in future without access to any one vendor. Third, the models differ in provider, size, release date, license and reasoning capability, so agreement among them is more informative than agreement among near-copies of the same system. Because the models do not agree perfectly on absolute levels, our primary exposure measure is the composite (the mean of the eight scores for each task, see Appendix~\ref{app:Aggregation}), rather than a single model's score or the most common score. We return to how closely the models agree, and to what that implies for our results, in Section~\ref{sec:ExposureModelResults}.

\begin{table}[!htbp]
    \centering
    \footnotesize
    \setlength{\tabcolsep}{4pt}
    \renewcommand{\arraystretch}{1.2}
    \newcommand{\modelid}[2]{\begin{tabular}[c]{@{}l@{}}#1\\[-2pt]\texttt{\scriptsize #2}\end{tabular}}
    \begin{threeparttable}
    \caption{Foundation models used to assign AI exposure scores}
    \label{tab:FMMetadata}
    \begin{tabular}{@{}lllllc@{}}
        \toprule
        Model (AWS Bedrock ID) & Provider & Release & Parameters & License & Reasoning \\
        \midrule
        \modelid{Claude Sonnet 4.6}{anthropic.claude-sonnet-4-6} & Anthropic & Feb 2026 & & Proprietary & Off\tnote{$\dagger$} \\
        \modelid{GPT-OSS-120B}{openai.gpt-oss-120b-1:0} & OpenAI & Aug 2025 & 117B / 5.1B & Apache 2.0 & Yes \\
        \modelid{Qwen3-235B-A22B}{qwen.qwen3-235b-a22b-2507-v1:0} & Alibaba & Jul 2025 & 235B / 22B & Apache 2.0 & No \\
        \modelid{Nova 2 Lite}{amazon.nova-2-lite-v1:0} & Amazon & Dec 2025 & & Proprietary & Off\tnote{$\dagger$} \\
        \modelid{Llama 4 Maverick}{meta.llama4-maverick-17b-instruct-v1:0} & Meta & Apr 2025 & 400B / 17B & Llama 4 Community\tnote{*} & No \\
        \modelid{Nemotron 3 Super}{nvidia.nemotron-super-3-120b} & NVIDIA & Mar 2026 & 120B / 12B & NVIDIA Open Model & Off\tnote{$\dagger$} \\
        \modelid{GLM-5}{zai.glm-5} & Z.AI & Feb 2026 & 744B / 40B & MIT & Off\tnote{$\dagger$} \\
        \modelid{Ministral 3 14B}{mistral.ministral-3-14b-instruct} & Mistral & Dec 2025 & 14B (dense) & Apache 2.0 & No \\
        \bottomrule
    \end{tabular}
    \begin{tablenotes}[flushleft]
        \footnotesize
        \item \textit{Note.} Parameters are total\,/\,active for mixture-of-experts models; Ministral 3 14B is dense. Blank cells denote parameter counts not disclosed for proprietary models. Proprietary models have closed weights, and all others have open weights. Reasoning indicates whether the deployed checkpoint emits an explicit reasoning trace.
        \item[$\dagger$] Supports extended thinking or reasoning, disabled for this exercise.
        \item[*] Open weights under the Llama 4 Community License, which carries usage restrictions and is not OSI-approved.
    \end{tablenotes}
    \end{threeparttable}
\end{table}

Conceptually, we chose to define AI exposure at the task level as \textit{the automation potential of a task using current AI systems}. Therefore, the score represents the estimated fraction of the task's substantive content that could be performed by current AI systems. Two important corollaries follow from this definition. First, the score represents the technical automation potential that competent AI deployment could perform. This is distinct from currently realized automation. The gap reflects adoption limits, workflow friction, procurement barriers, validation overhead, and organizational readiness, which we discuss further in Section \ref{section:Discussion}. Second, the scores are necessarily time-bound to AI systems as of late 2025 / early 2026. While our analysis is deliberately concerned with what current AI systems can do, and all of our results should be read in that light, it is still useful to know how far the scores depend on the version of the models used. To test how much this matters, we also rescored 93,000 tasks from 10,000 vacancies with earlier versions of two models in the ensemble, Llama 3 70B and Nova Lite, and compared their scores with those of the current versions.

\subsection{Weighting and aggregating exposure scores}\label{sec:WeightingAggregating}

Two further steps convert task-level scores into estimates for the UKCS workforce. First, because vacancies in GRID are not a random sample of UKCS posts, we reweight them using iterative proportional fitting (``raking'') so that the weighted sample matches published FTE totals by department, grade and profession, yielding a weight $w_i$ for each vacancy $i$ that sums to the estimation population's total FTE (Appendix~\ref{app:Raking}). Second, we aggregate task scores up to the vacancy (job) level. A task's ensemble score $\bar x_{ij}$ is the mean of the eight model scores. Tasks are then combined using geometric decay weights $\tau_{ij} \propto \delta^{\,j-1}$ based on their order in the advert, on the assumption that earlier-listed duties occupy more of the role (we set $\delta = 0.75$ and report $\delta \in \{0.5, 0.9, 1.0\}$ as sensitivity checks). This gives each vacancy a mean exposure $\mu_i$, an exposure dispersion $\sigma_i^2$ (the decay-weighted variance of its task scores), and a measure of cross-model disagreement $D_i$ (Appendix~\ref{app:Aggregation}). We use $\mu_i$ and $\sigma_i^2$ to classify roles in Section~\ref{sec:role_impact}, with $w_i$ applied whenever results are reported for the workforce rather than the sample.

\subsection{Role impact}\label{sec:role_impact}
Having established role-level statistics to describe the task-level variation in job roles, we then move on to assess the future organizational impacts of AI. To do so, we use the calculated values of $\mu_i$ and $\sigma_{i}^{2}$ to split our population into different impact groups.

Our purpose in doing so is to move beyond a single population wide average towards something actionable. We distinguish the differing real-world consequences of AI exposure across roles (job elimination, role redesign, or negligible effect), so that categories of impact can be identified and addressed differently. To do this, we apply a single consistent rule throughout: at each stage, roles are split by taking the top quartile against the remaining three-quarters, along whichever axis is relevant at that stage.

Applying this rule twice yields four groups. The first application splits the population by mean exposure alone, separating the quarter of roles with the highest average exposure from the remaining three-quarters. The same rule is then applied a second time, independently within each of those two groups, this time to dispersion rather than mean. Among the high-mean quarter, the quarter with the least dispersion is separated from the rest; among the low-mean three-quarters, the quarter with the most dispersion is separated from the rest. Table~\ref{tab:CategoryConditions} sets out the resulting four categories formally, together with the mean and variance conditions that define each.

\begin{table}[!htb]
    \centering
    \small
    \begin{threeparttable}
    \caption{Category definitions under the quartile classification}
    \label{tab:CategoryConditions}
    \begin{tabular}{lll}
        \toprule
        \textbf{Category} & \textbf{Description} & \textbf{Condition} \\
        \midrule
        Displacement & High mean, low variance  & $\mu \ge \mu_{0.75}$ and $\sigma^2 < \sigma^2_{0.25 \mid \text{high}}$ \\
        Augmentation & High mean, high variance & $\mu \ge \mu_{0.75}$ and $\sigma^2 \ge \sigma^2_{0.25 \mid \text{high}}$ \\
        Concentrated & Low mean, high variance  & $\mu < \mu_{0.75}$ and $\sigma^2 \ge \sigma^2_{0.75 \mid \text{low}}$ \\
        Insulated    & Low mean, low variance   & $\mu < \mu_{0.75}$ and $\sigma^2 < \sigma^2_{0.75 \mid \text{low}}$ \\
        \bottomrule
    \end{tabular}
    \begin{tablenotes}[flushleft]
        \footnotesize
        \item \textit{Note.} $\mu$ = mean task exposure, $\sigma^2$ = variance of task exposure, both at the role level. $\mu_{0.75}$ is the weighted 75th percentile of $\mu$ across the whole population. $\sigma^2_{0.25 \mid \text{high}}$ and $\sigma^2_{0.75 \mid \text{low}}$ are the weighted 25th and 75th percentiles of $\sigma^2$, computed separately within the high-mean and low-mean strata respectively.
    \end{tablenotes}
    \end{threeparttable}
\end{table}

We consider that these four categories fall into three broad groups by likely organizational impact: \textit{Displacement} roles, where the case for the role persisting in its current form is weakest; \textit{Augmentation} and \textit{Concentrated} roles, where exposure coexists with within-role variation and therefore points toward productivity gains rather than role loss; and \textit{Insulated} roles, where exposure is low throughout. 

Using the UKCS Workforce Statistics\footnote{See end of Section \ref{section:DatasetDesc}.}, we join on vacancy $i$'s estimated salary.\footnote{89 percent of vacancies have a matched salary by their department $\times$ grade $\times$ profession. The remaining 11 percent match by department $\times$ grade.} For each vacancy, we measure the size of the technical opportunity as its salary $s_i$, weighted by its raking weight $w_i$ and its mean ensemble exposure score $\mu_i$, such that $y_i = s_i \, w_i \, \mu_i$. We apply the same measure to every role, whatever its category, so that a role's opportunity scales with the share of its task content that current AI systems could perform. 

\subsection{Occupational redesign}\label{sec:MethodologyRedesign}

The exposure scores and classification above show where AI could perform work, but not what happens to the work that remains or how any time AI frees might be used. To explore this, we use LLMs to redesign exposed work at two levels and examine what they propose. Role redesign holds each role fixed and shows how its tasks change. Team redesign then uses a similar process to redesign individual roles within a team setting, alongside an AI-driven reduction in overall team headcount. This allows us to assess how organizational structures change and how work moves around a team, rather than within a single role in isolation. For both, our starting point is roles or teams that combine high average exposure with substantial variation across tasks or roles. They contain both work that AI could take over and work that remains with people, which makes them natural candidates for redesign. We use an LLM rather than managers or postholders because it can apply the same instructions consistently to thousands of roles and teams, which lets us look at the broad pattern of impact across the organization rather than at a handful of individual cases. For role redesign, we draw a sample of 10,000 Augmentation roles, stratified so that it reflects the exposure profile of the quadrant as a whole. For team redesign, we use all teams identified with high exposure and within a size range.\footnote{See Appendix \ref{sec:AppendixTeamMethodology} for full details of team selection.} In both, the model identifies work that AI could take over, reorganizes the work that remains, and fills any time this frees with new work that still needs a person. We then compare each role or team before and after the full redesign. Both exercises are exploratory. They describe what the model proposes, not changes the Civil Service has made or is likely to make.

\subsubsection{Role redesign}
For each sampled role, the model is given the job description and task list. First, it removes any tasks it judges AI can now perform and replaces each with a new task that would make the role more productive. Second, if these changes free up time, it proposes additional work to fill that time. We estimate the time freed from where each new task sits in the role's task list, using the decay weight parameter introduced in Section \ref{sec:WeightingAggregating}. In a variant, we require that the additional work be something AI could not already do well. To describe how the mix of work changes, we classify every task, before and after redesign, into nine broad categories of work, with a tenth for tasks that fit none of them. This lets us compare each role's task mix before and after redesign and identify the kinds of work that redesigned roles come to focus on. Using the scores for the new tasks, we also recalculate each role's mean and variance of exposure and assign it to a quadrant, holding the thresholds fixed at their baseline values, as calculated in Table~\ref{tab:CategoryConditions}.

\subsubsection{Team redesign}
Team redesign applies the same logic as role redesign to a whole team, so that changes to one role can be absorbed by the roles around it. Each team is described by its roles, the number of posts held in each role, each role's grade, and each role's tasks, with weekly hours derived from the order of tasks in the advert. Each team is assumed to lose exactly one post. First, the model chooses the role that loses the post. It favors the role whose work looks most amenable to AI. For that role only, it also identifies the tasks that cannot be automated away because they still need a person, and these are protected. The rest of the work of the departing post leaves with it. Second, the model reviews all other tasks in the team and marks those AI could automate. This parallels the first step of role redesign, except that automated tasks are removed without being replaced. Third, the remaining and protected tasks are regrouped into a new set of roles, each with its own title, grade and number of posts, where the posts sum to the original total minus one and no role's workload exceeds the working hours of its postholders. The model may leave a role unchanged, merge roles or move tasks between them. Finally, the model fills spare capacity in the new roles with work that still requires a person, either as new tasks or as additional time on an existing task, up to the capacity available. This step brings together the replacement tasks and additional work of role redesign, but adds them to the newly defined roles rather than to the original ones. Our comparison focuses on how team structure changes, including the number of roles and how work is consolidated between them, and on the grades from which the model removes posts.

In both exercises, the model never sees our exposure scores, and a separate model scores each new task using the same rubric as the main analysis. Full methodological details and prompts are reported in Appendices~\ref{sec:AppendixJobRedesign} and~\ref{sec:AppendixTeamRestructure}.

\clearpage
\newgeometry{margin=1cm}   
\begin{landscape}
\thispagestyle{empty}      

\begin{center}
    \includegraphics[width=\linewidth]{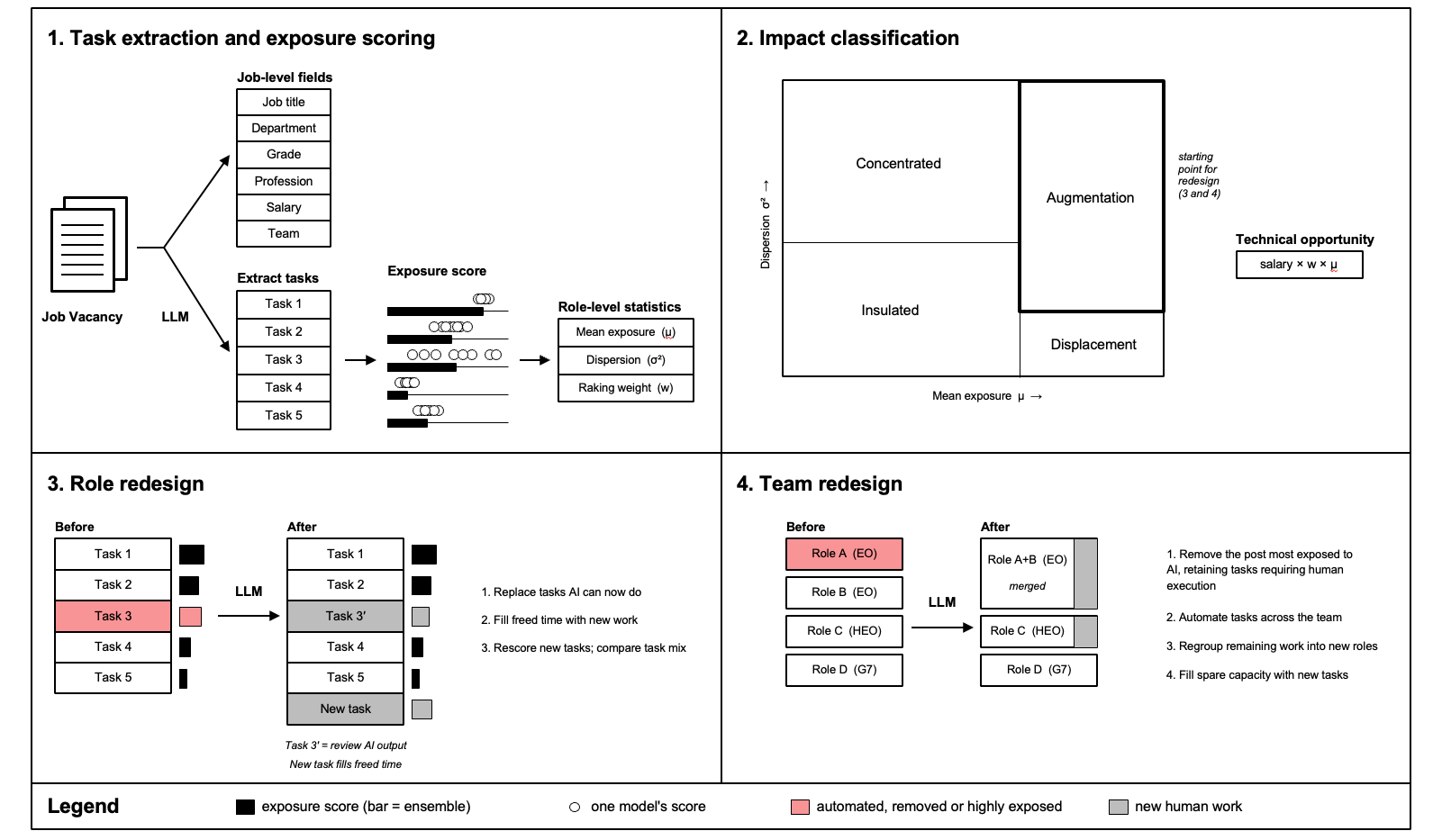}
    \captionof{figure}{Visual representation of the analytical workflow}
    \label{fig:full_methodology}
    \caption*{\footnotesize{This visual shows the workflow for one single job description; we repeat this for all 283,484 in our dataset. Step 1 uses an LLM to extract each unique task from the job description. When coupled with metadata from the job description, this provides us with a detailed role profile for each job. Step 2 categorizes each role based on the mean and variation of the task exposure scores, to understand the differing impacts on job roles. Step 3 assesses how the task mix is likely to change through role-level redesign. Step 4 replicates the redesign process, but at a team level to understand how organizational structures may change.}}
\end{center}

\vfill
\centering \thepage       
\end{landscape}
\restoregeometry
\baselineskip24pt

\section{Results}\label{section:Results}

This section presents the results in five stages. First, we describe the extracted task corpus and show how our vacancy-level approach captures heterogeneity among otherwise similar roles. Second, we validate the exposure scores by examining agreement across models and comparing the ensemble scores with human judgments and an independently constructed NLP measure. Third, we aggregate task exposure across the workforce, decompose its variation within and between roles, and classify roles according to their mean exposure and within-role dispersion. Fourth, we test whether pre-release exposure predicts subsequent recruitment demand. Finally, we examine how an LLM redesigns roles where the task mix provides the greatest scope for augmentation.

\subsection{Tasks}\label{sec:Tasks}

\subsubsection{Extraction results}
Our final dataset comprises 2,729,344 tasks from 283,484 vacancies. The mean number of tasks per vacancy is 9.6 and the median is 8. The task extraction process does result in some extreme, albeit rare, values for the number of tasks. The minimum number of tasks identified in a vacancy is three\footnote{Every vacancy in our estimation sample has at least three extracted tasks due to a data-quality filter we apply so that role-level dispersion is never estimated from fewer than three observations.} and the maximum is 72. For the majority of vacancies, the number of tasks extracted is within the expected range: 91\% of vacancies have 5 or more tasks extracted and 89\% have fewer than 16 tasks. The outliers reflect the varying data quality in the job description field, but it does also partly capture the genuine breadth of vacancies in the UKCS.\footnote{We also compared LLM task extraction to human task extraction. Across 200 manually sampled job descriptions, the LLM extracted an average of 12 tasks per description, compared to 8 extracted by a human reviewer. This suggests that the model breaks down job descriptions into more granular components, potentially offering a finer view of task exposure and role complexity. We present two examples of this comparison in Appendix \ref{sec:AppendixExtraction}.} Finally, the number of tasks extracted is consistent across all years, and Civil Service job description writing practices have not clearly changed since the release of ChatGPT, allaying any potential concerns of endogeneity from AI being used to write job descriptions.\footnote{We observe a minor increase in the mean number of tasks extracted over the time period, from 9.67 in 2019 to 10.20 in 2026.} Given this, we are satisfied that the job descriptions in this dataset are of sufficient quality to proceed with further analysis. 

\subsubsection{Heterogeneity across similar roles}\label{sec:ONETIscoComparison}
This bottom-up task extraction differs from the standard approach in the literature which relies on top-down task lists assigned to occupations such as O*NET \citep{FeltenEtAl2023, EloundouEtAl2024GPTsAreGPTs} and the International Standard Classification of Occupations (ISCO-08; \citealp{Gmyrek2025generative}). Such approaches rest on standardized classifications that assign one task list per job title, so every worker sharing a job title receives the same exposure score, effectively imposing a within-profession homogeneity. 

In Table \ref{tab:onet-isco-summary}, we present the value of our granular, role-level approach by comparing, for three professions (Economists, HR Managers, and Lawyers), the spread of automation-potential scores across individually-scored GRID roles against the single estimate each framework would assign to the profession as a whole. For example, ISCO-08 lists 12 tasks under ``Economist'', which it also applies to ``Econometrician'', ``Economic Advisor'', ``Economic Analyst'' and ``Labor Economist''. Similarly, O*NET lists 13 tasks for economists, ranging from analyzing data to providing economic advice and reviewing documents. Any exposure score assignment against these lists of tasks forces all economists to be treated equally. By contrast, our GRID sample contains 2,818 distinct economist vacancies, comprising 24,048 unique tasks. Therefore, when exploring how economists are exposed to AI, we can use a distribution of exposure scores that reflects how each vacancy's work varies, incorporating important dimensions such as seniority and domain (see Appendix \ref{sec:AppendixRoleHeterogeneity} for a visual representation of these results).

\begin{table}[htbp]
  \centering
  \small
  \begin{threeparttable}
  \caption{Task exposure for three occupations in GRID, O*NET, and ISCO-08}
  \label{tab:onet-isco-summary}
  \begin{tabular}{lrrcrcrc}
  \toprule
   & \multicolumn{3}{c}{GRID} & \multicolumn{2}{c}{O*NET} & \multicolumn{2}{c}{ISCO-08} \\
  \cmidrule(lr){2-4} \cmidrule(lr){5-6} \cmidrule(lr){7-8}
  Job title & Roles & Tasks & Mean (SD) & Tasks & Mean (SD) & Tasks & Mean (SD) \\
  \midrule
  Economists  & 2,818 & 24,048 & 0.39 (0.11) & 13 & 0.53 (0.11) & 12 & 0.61 (0.07) \\
  HR Managers & 3,600 & 33,978 & 0.35 (0.10) & 23 & 0.32 (0.13) & 11 & 0.40 (0.11) \\
  Lawyers     & 3,770 & 34,700 & 0.36 (0.11) & 15 & 0.42 (0.18) & 10 & 0.38 (0.17) \\
  \bottomrule
  \end{tabular}
  \begin{tablenotes}[flushleft]
    \footnotesize
    \item \textit{Note.} GRID and O*NET task weights use geometric decay ($\delta = 0.75$), following advert order for GRID and O*NET's own importance ranking. ISCO-08 tasks are weighted uniformly ($\delta = 1.0$) because their order carries no importance signal. For GRID, the mean and SD are averages across individually scored roles. For O*NET and ISCO-08, each is a single weighted estimate for the occupation.
  \end{tablenotes}
  \end{threeparttable}
\end{table}

\subsection{Exposure scores}\label{sec:ExposureScores}

As described in Section \ref{sec:AssigningExposureScores} and Appendix \ref{app:Aggregation}, we score each task using eight foundation models, and report the average value as the ``ensemble'' exposure score.\footnote{0.28\% of tasks were scored by seven models.} The mean ensemble task automation score, i.e. the vacancy-level exposure score, is 0.39 and the median is also 0.39. The standard deviation is 0.16, a little over one-and-a-half times the size of a single increment on the underlying 0.1-spaced scoring rubric. About two-thirds of tasks fall within one standard deviation of the mean. 

These task-level exposure scores represent the average judgment of eight LLMs, but this raises the question of how reliable they are. We address this in two stages. First, we perform an internal consistency check to explore how much the models agree with each other. Second, we run two external validation exercises against independently sourced scores, comparing our LLM scores to human-labeled scores on a random subset of tasks, and to NLP-derived exposure measures following \cite{HampoleEtAl2025AILaborMarket}. 

\subsubsection{Model consistency}\label{sec:ExposureModelResults}

\cite{yin2026} demonstrate that model inconsistency can be a source of measurement error when LLMs assign AI exposure scores. In the first place, to mitigate the systematic differences, our primary specification uses the ensemble score i.e. the mean across eight models. Nevertheless, it is important to examine model disagreement closely: we report results briefly here, and in more detail in Appendix \ref{AppendixModelConsistency}. 

To characterize model disagreement precisely, we calculate a battery of inter-rater metrics to answer four complementary questions. First, do these models agree in absolute terms, correcting for chance? We report a mean pairwise quadratic-weighted Cohen's $\kappa$ of 0.693, indicating that on average, our models show ``substantial'' agreement per \cite{LandisKoch1977}. Our Krippendorff's ordinal alpha $\alpha$, measuring all eight raters simultaneously, is 0.673, just inside the ``tentative'' band in conventional thresholds \citep{Krippendorff2004}.\footnote{However, this rises to 0.719 if we exclude Sonnet 4.6, showing that this model is a meaningful contributor to model disagreement.} In addition, our intraclass correlation coefficient is $ICC(2,1)=0.687$, placing agreement in the ``moderate'' band as per \cite{KooLi2016}. Taken together, these statistics show that when considering the absolute level of AI exposure scores, agreement between models is only moderate.

Second, given this observed level of disagreement, we next ask whether models rank tasks in the same way. We calculate the pairwise Spearman's $\rho$, ranging from $0.71$ to $0.87$ (mean $0.76$), indicating fairly consistent rank ordering of tasks. Similarly, Kendall's rank coefficient $\tau_b=0.647$ on average, likewise implying a reasonably strong rank consistency. Together with the previous finding, we infer that the models agree substantially on the relative ordering of tasks, but that this agreement weakens once absolute levels are considered.

Third, we can use Equation \ref{eq:task-var}, the measure of annotator spread, to understand where model disagreement is highest at the task level, allowing us to examine whether disagreement concentrates on particular kinds of tasks. The highest-disagreement tasks are heavily concentrated in administrative and clerical duties. They are overwhelmingly in the most junior grade (82 percent in AO roles), in Operational Delivery professions, and disproportionately from the HM Courts and Tribunals Service. In these instances, the models roughly split into two camps, half scoring near zero and the other half scoring highly. Using the reasoning provided by each model, we find that these disagreements are driven by differing interpretations of the rubric, typically on the need for human/physical involvement. Examples can be found in \Cref{tab:task-example-photocopying,tab:task-example-reading-aloud,tab:task-example-divisions} in Appendix \ref{AppendixModelConsistencyModelAgreement}. 

Fourth, the exposure scores assigned reflect AI systems available in late 2025 and early 2026, so we also test how much they would change with earlier models. For a sample of 10,000 vacancies, we rescored the same tasks with older versions of two models in the ensemble, Llama 3 70B and Nova Lite. The older models judge tasks to be less exposed, with average task scores about 0.1 lower than those of their successors. The overall level of exposure therefore depends on when it is measured and is consistent with estimates rising as AI systems improve. The older models also rank roles somewhat differently. The correlation between role-level scores from the older and newer versions is 0.54 for Llama and 0.36 for Nova, compared with 0.68 between two current models from different providers. Moving to an earlier model version therefore changes which roles appear most exposed by more than switching between current providers does. Our exposure estimates should accordingly be read as the assessment of current AI systems rather than a fixed property of each role.

Therefore, models differ meaningfully on the absolute scores they assign for task AI exposure, underscoring the benefit of using an ensemble average score built from several different models. Nevertheless, models tend to agree on which tasks they judge as more or less automatable than others. Moreover, where we witness high cross-model disagreement, these disagreements are based on varied but defensible interpretations of the prompt's rubric. Together, we are satisfied that our models provide scores that are internally valid. Accordingly, the next two subsections compare our scores against alternative methods to test their external validity.

\subsubsection{Human-labeled scores}\label{sec:HumanScores}
Having examined agreement among the models, we next compare their ensemble scores with human judgments. Fifteen expert reviewers specializing in the development and evaluation of AI tools for Civil Service workflows independently scored the same 71 tasks, drawn from 10 roles, using the rubric supplied to the LLMs. The ensemble scores align closely with the human consensus, with Pearson's $r=0.894$ at task level and $r=0.913$ at role level. The ensemble also outperforms every individual model against the human mean, whose task-level correlations range from 0.676 to 0.874. These results provide evidence that the ensemble identifies broadly the same relative pattern of exposure as the reviewers. Some of this advantage is to be expected, because averaging across models cancels out errors that are specific to each one. We therefore randomly split the reviewers 500 times into a group of eight and a group of seven, and compare how closely the ensemble and the group of eight each match the average score of the group of seven. The ensemble's average correlation with the group of seven is $r = 0.87$, compared with $r = 0.89$ for the group of eight. We therefore conclude that the ensemble agrees with expert consensus almost as closely as a second panel of experts.

Agreement is weaker on absolute score levels. The ensemble is modestly more conservative than the human consensus overall (bias $=-0.039$; MAE $=0.081$), scoring below the human mean in eight of the ten roles. This difference becomes larger among tasks assigned higher exposure scores. Human reviewers also disagree substantially with one another: Krippendorff's $\alpha=0.361$, while the average correlation between an individual reviewer and the human consensus is 0.664. Exposure assessment therefore retains an important element of judgment regardless of whether it is performed by people or models. However, given the degree of disagreement among human experts, we take the ensemble score to be a useful, and potentially more consistent, basis for assigning exposure ratings than any single human rater would provide. Further comparisons between the ensemble and human reviewers at task- and role-level are reported in Appendix \ref{section:AppendixHumanReview}.

\subsubsection{NLP-derived exposure scores}\label{sec:NLPScores}
AI exposure scores can be assigned by LLMs and/or humans. An alternative approach uses techniques from natural language processing to measure the semantic overlap between occupations and technologies \citep{PrytkovaPetit2025EmploymentImpactUS}, or tasks and AI \citep{HampoleEtAl2025AILaborMarket, GathmannGrimmWinkler2024AITaskChanges}. As with the human-labeled scores, we use an NLP-derived method to compare the consistency between methods. 

We draw on the Anthropic Economic Index, which reports the observed share of real-world Claude usage matched to 680 O*NET task descriptions.\footnote{We use the O*NET work activities at a task level for the United Kingdom from the Anthropic Economic Index \href{https://www.anthropic.com/economic-index\#data-explorer}{data explorer}. This provides a list of 680 tasks that Claude users commonly perform.} We embed this reference set and every task in our own corpus using a sentence-transformer model, and score each of our tasks by its cosine similarity to the single closest of the 680 reference tasks. This produces a continuous measure of how closely a task semantically overlaps with real-world AI use, constructed entirely without reference to our LLM-based scores. The relationship between this similarity measure and our ensemble exposure score is positive and statistically significant at both task level (r = 0.284, p $<$ 0.001) and vacancy level (r = 0.386, p $<$ 0.001). In Appendix \ref{section:AppendixNLPExposure}, we plot the deciles of NLP-derived exposure against our LLM-derived exposure. The relationship is positive and monotonic, demonstrating the underlying agreement between measures. Tasks in the top decile of similarity to real-world AI-use language carry exposure scores 44\% higher, in relative terms, than those in the bottom decile (vacancy level: 31\% higher).

\subsection{AI exposure mean and variance across the workforce}\label{sec:AIExposureAggregated}

\subsubsection{Aggregate exposure and its variance decomposition}

Satisfied that our exposure scores are sufficiently consistent and accord with alternatively derived measures, we now move to consider how they aggregate into the workforce-level statistics. Weighting each occupation by its raking weight, the population-average automation exposure across Civil Service roles is $\bar{\mu} = 0.363$ (i.e. the average role's tasks are estimated to be 36.3\% automatable).

Breaking the population along the department, grade, and profession dimensions reveals that automation exposure varies markedly by what a role does but comparatively little by how senior it is: grade produces the narrowest spread (0.30–0.37, a range of just 7 percentage points), whereas department (0.29–0.45) and profession (0.31–0.48) both span over twice that range. This underlines that exposure is more a property of a role’s task content than of its position in the hierarchy. 

However, these department/grade/profession means describe only the average exposure within each group; they say nothing about how much roles differ from one another even within the same group, or how much a single role's own tasks differ in their individual automatability. To capture that, we now turn to a full variance decomposition, splitting the total dispersion in exposure across the workforce into a within-role component (heterogeneity among the tasks that make up a single posting) and a between-role component (how much roles' average exposure differs from one another).

Decomposing total variance, $V_{\text{total}} = V_{\text{within}} + V_{\text{between}} = 0.0145 + 0.0077 = 0.0221$, shows that differences among tasks within a role account for most of the dispersion in exposure (65.3\%).\footnote{To check whether within-role heterogeneity reflects how advertisements are written rather than job content, we test whether the number of extracted tasks predicts within-role variance. The relationship is weak (Pearson $r = 0.14$): task count explains about 2\% of the variance in dispersion. Heterogeneity is therefore not primarily a mechanical artifact of how many tasks an advertisement happens to list.} The remaining 34.7\% reflects differences between roles, and standard occupational categories capture little of it. The 244 profession $\times$ grade cells, the closest analogue in our data to a standard occupational classification, account for only 4.0\% of total variance, so 88.6\% of between-role variation lies among roles that share a profession and grade. Adding department (7,293 cells) raises the between-cell share to 13.4\%: organizational setting explains roughly a third of the between-role variation that profession and grade leave unexplained. Yet even among roles with the same department, profession and grade, the finest categories in the Civil Service's own job architecture, 61.4\% of between-role variation remains.

Exposure is therefore better understood as a property of how tasks are bundled into roles in particular organizational settings than as a property of occupations. Occupation-level scores, which give every role in an occupation the same exposure, would miss most of the variation between roles.

\subsubsection{A classification of workforce impact}\label{sec:workforce-impact}
\begin{figure}[!t]
  \centering
  \includegraphics[width=0.8\textwidth]{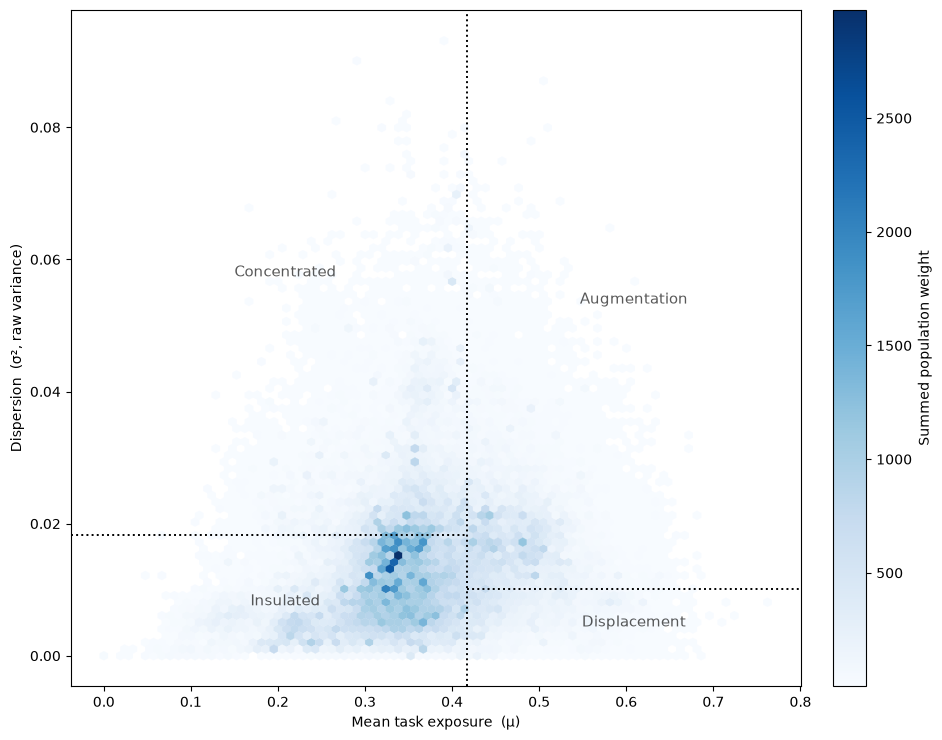}
  \caption{FTE-weighted AI-exposure plane for UK Civil Service roles}
  \label{fig:expsoure-scatter}
\end{figure}

Figure \ref{fig:expsoure-scatter} makes this joint distribution of mean and variance visible at the level of the individual role, plotting each role's decay-weighted mean exposure and dispersion, colored by its FTE representation from the raking weight. We split this distribution into four distinct categories, using the conditions outlined in Section \ref{sec:role_impact}.

\textit{Displacement} roles are the most exposed under this classification. Task-level exposure is high and uniform, with no substantial set of low-exposure tasks, so the classification offers the least basis, of the four groups, for expecting these roles to persist in their current form, although this does not mean they disappear on any particular timescale. \textit{Augmentation} and \textit{Concentrated} roles both combine meaningful exposure with task-level variation and are better read as productivity opportunities than as roles at risk, since a plausible outcome is the role persisting in reshaped form. The two groups differ in where that variation sits. \textit{Augmentation} roles are already highly exposed on average, making them the stronger candidates for near-term role redesign (Section \ref{sec:Redesign}). \textit{Concentrated} roles sit just below that mean threshold, with most tasks scoring low but a handful scoring much higher, making them candidates for automating those specific tasks and redistributing any saved time to the rest of the role. \textit{Insulated} roles show both low average exposure and little internal spread, with no substantial exposed subset of tasks to redesign around. Many roles sit just inside this quadrant. Around a sixth of the sample (40,000 roles, 81,000 FTE) falls within a narrow range of mean exposure (0.30--0.38) and dispersion (0.010--0.018), so an exact, absolute cut point would be fragile, with small changes to either threshold reclassifying a large share of the workforce. Table \ref{tab:CategoryProfile} presents summary statistics for each of the four groups.

Accordingly, this four-way split should be read as one illustrative typology rather than a uniquely correct partition of the workforce. The single quartile threshold used throughout was chosen for parsimony, so that one rule could be applied consistently across both axes, not because the joint distribution exhibits a natural break at that point. Appendix \ref{sec:RoleSensitivity} reports how the group sizes move under alternative threshold choices.

\begin{table}[!htb]
    \centering
    \small
    \newcommand{\hd}[1]{\begin{tabular}[b]{@{}c@{}}#1\end{tabular}}
    \begin{threeparttable}
    \caption{Summary statistics for role categories}
    \label{tab:CategoryProfile}
    \begin{tabular}{@{}lrccrr@{}}
        \toprule
        Category & FTE & \hd{Mean\\exposure ($\mu$)} & \hd{Mean\\variance ($\sigma^2$)} & \hd{Salary cost\\(\% of total)} & \hd{Technical\\opportunity} \\
        \midrule
        Insulated    & 276,730 & 0.319 & 0.0095 & \pounds 12.91bn (61.3\%) & \pounds 4.12bn \\
        Concentrated &  92,053 & 0.346 & 0.0272 &  \pounds 3.59bn (17.1\%) & \pounds 1.24bn \\
        Augmentation &  92,197 & 0.473 & 0.0193 &  \pounds 3.34bn (15.8\%) & \pounds 1.58bn \\
        Displacement &  30,730 & 0.484 & 0.0065 &   \pounds 1.23bn (5.8\%) & \pounds 0.60bn \\
        \midrule
        Total        & 491,710 & --- & --- & \pounds 21.08bn & \pounds 7.54bn \\
        \bottomrule
    \end{tabular}
    \begin{tablenotes}[flushleft]
        \footnotesize
        \item \textit{Note.} FTE is full-time equivalent staff in the reweighted Civil Service workforce. Mean exposure and mean variance are averages of the role-level values within each category. Salary cost is the annual salary bill for each category, with its share of the total Civil Service salary bill in parentheses. Technical opportunity is salary weighted by mean exposure, summed across the roles in each category.
    \end{tablenotes}
    \end{threeparttable}
\end{table}

\subsection{AI exposure and the demand for labor}\label{sec:DemandForLabor}

Inspired by \cite{HampoleEtAl2025AILaborMarket}, we test whether AI exposure predicts subsequent labor demand for UK Civil Service roles. We model both the mean exposure and dispersion, as defined in Equations \ref{eq:mean} and \ref{eq:variance}, jointly throughout. For each occupation $i$ (a department $\times$ profession $\times$ grade combination), we fix $\mu_i$ and $\sigma_i^2$ once, as averages over the entire pre-release period (1 October 2019 to 30 November 2022). We then estimate a two-way fixed-effects panel regression of role postings on these fixed moments, interacted with indicators for each year since ChatGPT's public release, with the entire pre-release period as the reference category:

\begin{equation}
  \operatorname{arcsinh}(n_{it}) = \alpha_i + \gamma_t +
    \sum_{k=1}^{4} \beta_{1k}\left(\mu_i \cdot D_{t}^{k}\right) +
    \sum_{k=1}^{4} \beta_{2k}\left(\sigma_i^2 \cdot D_{t}^{k}\right) +
    \varepsilon_{it},
  \label{eq:event-study}
\end{equation}

where $n_{it}$ is the number of roles advertised for occupation $i$ in month $t$, $\operatorname{arcsinh}(n)=\ln\!\left(n+\sqrt{n^2+1}\right)$, $D_t^k$ indicates the $k$-th period following ChatGPT's release, each spanning 12 months except Year 1, which spans 11 months because November 2022 (the month of release itself) falls within the pre-release reference period, and Year 4, which is truncated to 5 months by data availability (Year 1: Dec 2022–Oct 2023; Year 2: Nov 2023–Oct 2024; Year 3: Nov 2024–Oct 2025; Year 4: Nov 2025–Mar 2026), and $\alpha_i$ and $\gamma_t$ are occupation and month fixed effects.\footnote{We use the inverse hyperbolic sine rather than $\log(n_{it})$ because it is defined at $n=0$ and behaves like $\log(2n)$ for $n \gg 0$.} Standard errors are clustered by occupation. The sample comprises 949 occupations with sufficient pre-release posting volume, observed over their full monthly panel of 74,022 occupation-months.

We find that mean exposure is robustly negatively related to vacancies in the years following ChatGPT's release. It enters only weakly, and not yet significantly ($p=0.0565$), from December 2022 to October 2023, but from Year 2 onward it is significant with the theoretically predicted sign (mean exposure depressing demand) and remains so through Year 4 ($p=0.0001$, $p=0.0002$, and $p=0.0009$ in Years 2--4 respectively).\footnote{See Table \ref{tab:demand-headline} for raw coefficient estimates, and Table \ref{tab:demand-headline-std} for standardized impacts.} Dispersion's coefficient is positive in every year, consistent with the reallocation channel in \cite{HampoleEtAl2025AILaborMarket} partially offsetting the substitution channel, but it does not reach conventional significance in any single year ($p=0.1284$, $p=0.1832$, $p=0.6400$, and $p=0.0803$ in Years 1--4). 

\begin{figure}[!t]
  \centering
  \includegraphics[width=0.8\textwidth]{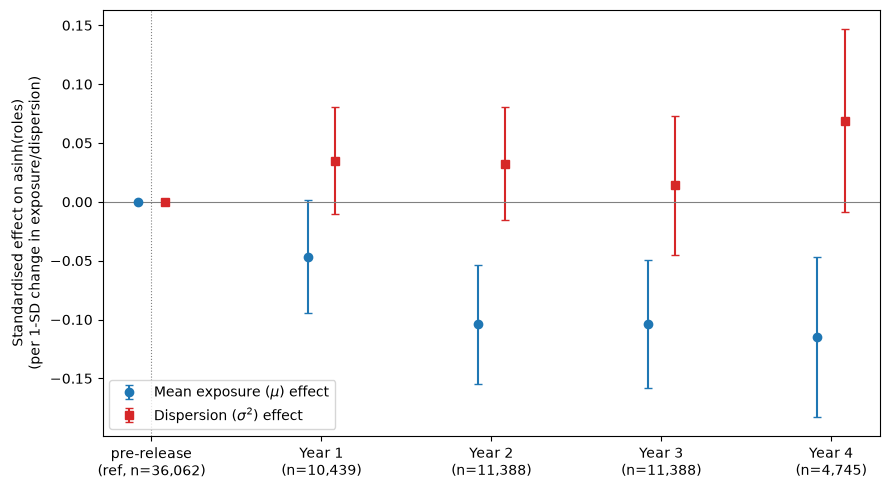}
  \caption{Standardized effect of mean exposure and dispersion by year}
  \label{fig:demand-event-study}
  \begin{minipage}{0.8\textwidth}
    \footnotesize
    \textit{Note.} Error bars show 95\% confidence intervals (clustered by occupation). Year 1: Dec 2022--Oct 2023; Year 2: Nov 2023--Oct 2024; Year 3: Nov 2024--Oct 2025; Year 4: Nov 2025--Mar 2026. Effects are computed as $\sinh\!\left(\operatorname{arcsinh}(n_0)+\hat\beta_{k}\cdot\mathrm{sd}\right)/n_0-1$, evaluated at the sample median occupation size ($n_0=3$ roles/month), using $\mathrm{sd}(\mu_i)=0.048$ and $\mathrm{sd}(\sigma_i^2)=0.0036$.
  \end{minipage}
\end{figure}

Figure \ref{fig:demand-event-study} presents the estimated standardized impacts from both channels. For example, for an occupation posting the sample median of 3 roles per month, a one-standard-deviation increase in mean exposure in Year 2 is associated with a $10.4\%$ reduction in role postings relative to the pre-release baseline (ranging from $11.1\%$ to $9.9\%$ across the 25th--75th percentile of occupation size), while a one-standard-deviation increase in task dispersion is associated with a $3.5\%$ increase that is not statistically distinguishable from zero.

The timing of the mean exposure's effect being only weak and not yet significant in the twelve months immediately after release (Year 1), then becoming clearly significant from Year 2 onward, is consistent with a delayed adoption effect. We believe that this is plausible given the lag between ChatGPT's release and the following AI integration and tooling, as well as the constraints on AI uptake within central government. It is not consistent with a single sharp trigger: Appendix~\ref{sec:AppendixDemandForLabor} tests the founding of the UK AI Safety Institute at the Bletchley Park AI Safety Summit (1 to 2 November 2023) and the Spring Budgets of 2023 and 2024, and finds no individual month with a disproportionate break, suggesting instead a gradual, diffuse shift concentrated in the second half of the sample period. Mean exposure's Year 2--4 result is robust to every check we run: it holds across minimum occupation-size thresholds from 5 to 150 pre-release postings, when the largest occupations are dropped, when an admin-grade confound is controlled for, when the pre-release reference window is narrowed to avoid the COVID-19 period, and also when estimated on its own without dispersion in the model. Dispersion is considerably more fragile: it fails to reach significance at most occupation-size thresholds, moves non-monotonically as the largest occupations are excluded, loses its already-marginal significance once the pre-release reference window is narrowed, and is never significant when estimated on its own.\footnote{A caveat is that both channels drift somewhat within the pre-release period itself, though only in its final third, closest to ChatGPT's release; as this drift predates the release, it cannot itself be an AI effect. Restricting the reference window to this same drift-affected period leaves the mean-exposure channel's estimates essentially unchanged, while the dispersion channel no longer reaches significance in any year, reinforcing the asymmetry between the two channels that we show above.} Full detail on this and all other robustness checks is given in Appendix~\ref{sec:AppendixDemandForLabor}.

\subsection{Occupational redesign}\label{sec:Redesign}

\subsubsection{Role redesign}\label{sec:ResultsRoleRedesign}

The model changes at least one task in 96\% of roles, but only 12\% of a role's original tasks are ever removed and replaced. We therefore measure exposure across the whole task list, weighting each task by the share of the working week it takes. Averaged across all 10,000 roles, exposure moves only slightly (Table~\ref{tab:RoleRedesignOverall}). It falls from 0.456 before redesign to 0.447 after the full redesign, and to 0.444 in the variant in which additional work must be something AI could not already do well. Exposure is lower after redesign in 65\% of roles (67\% in the variant).

\begin{table}[htbp]
    \centering
    \begin{threeparttable}
    \caption{Role-level exposure before and after redesign, across all 10,000 roles.}
    \label{tab:RoleRedesignOverall}
    \begin{tabular}{lrrr}
        \toprule
        & Before redesign & After redesign & After redesign (variant) \\
        \midrule
        Mean exposure & 0.456 & 0.447 & 0.444 \\
        Mean variance & 0.0277 & 0.0242 & 0.0239 \\
        Roles with lower exposure than before & --- & 64.7\% & 66.9\% \\
        Roles in the Augmentation quadrant & 100\% & 73.4\% & 71.9\% \\
        \bottomrule
    \end{tabular}
    \begin{tablenotes}[flushleft]
        \footnotesize
        \item \textit{Note.} Tasks are weighted by the share of the working week they take. The quadrant row adjusts the GPT-OSS scores to the scale of the ensemble used to define the quadrants (Appendix~\ref{sec:AppendixQuadrantCalibration}). In the variant, additional work must be something AI could not already do well.
    \end{tablenotes}
    \end{threeparttable}
\end{table}

The small overall change reflects two kinds of new work pulling in different directions. The tasks that replace removed work are less exposed than the tasks they replace, scoring 0.517 on average against 0.599, and 66\% explicitly describe reviewing, validating or checking AI-generated output. The scorer's reasoning for 87\% of these replacement tasks points to a part of the task that needs human judgment, approval or accountability, which suggests there is a limit to how far redesign can reduce a task's exposure. Although the model is not shown any exposure scores, it removes one of the highest-scoring tasks in the role in 63\% of roles, against 19\% by chance (Appendix~\ref{sec:AppendixRemovedTasks}). The model shows a strong tendency to replace a task with a requirement to oversee an AI system doing that task, regardless of which profession or domain the vacancy is working in. How far this reflects a change in the work and not just in the wording is discussed in Appendix~\ref{sec:AppendixReviewPattern}. 

The additional work proposed to fill freed time pulls the other way. Time is freed in 52\% of the roles where a task is replaced, averaging about 2.7 hours a week, and the model fills it mostly with routine reporting and monitoring work, such as keeping trackers, producing regular summaries and sampling cases for quality checks. This work scores 0.55 on average, higher than both the tasks retained in the role (0.41) and the replacement tasks. The model is not asked to avoid exposed work, and left to its own judgment it does not move the freed time toward less exposed tasks. 

When the model is told that additional work must be something AI could not already do well, the mean exposure of that work falls to 0.47, still above the retained tasks. Tasks involving direct contact, such as stakeholder liaison, court-user support and coaching, rise from about 23\% to 44\% of additional tasks, while written-output themes fall from about half to a third (Appendix~\ref{sec:AppendixBackfillClustering}). Many tasks still contain some routine checks and logs. These shifts have little effect on the redesigned roles overall, because additional work is generated for only 42\% of roles and accounts for about 3\% of total working time across all roles. 

As shown in Table \ref{tab:TaskCategoryShares}, across the nine categories of work, the shift is from administration toward assurance. Administration makes up 51\% of the tasks the model removes and falls from 30\% to 24.6\% of a role's working time after redesign, while assurance makes up 46\% of the replacements and rises from 8.7\% to 14.5\%. The other categories each move by less than two percentage points.

\begin{table}[htbp]
    \centering
    \small
    \begin{threeparttable}
    \caption{Share of a role's working time spent on each category of work}
    \label{tab:TaskCategoryShares}
    \begin{tabular}{lrrr}
        \toprule
        & Before redesign & After redesign & After redesign (variant) \\
        \midrule
        Administration & 30.0\% & 24.6\% & 24.4\% \\
        Enquiries & 10.2\% & 10.0\% & 10.0\% \\
        Casework & 7.6\% & 8.1\% & 8.2\% \\
        Stakeholder engagement & 10.5\% & 10.5\% & 10.8\% \\
        Analysis and reporting & 13.8\% & 13.5\% & 13.1\% \\
        Advice & 4.8\% & 5.0\% & 5.0\% \\
        Planning and delivery & 9.9\% & 9.5\% & 9.5\% \\
        People management & 2.5\% & 2.6\% & 2.7\% \\
        Assurance & 8.7\% & 14.5\% & 14.4\% \\
        Other & 2.0\% & 1.8\% & 1.8\% \\
        \bottomrule
    \end{tabular}
    \begin{tablenotes}[flushleft]
        \footnotesize
        \item \textit{Note.} Tasks are weighted by the share of the working week they take. In the variant, additional work must be something AI could not already do well.
    \end{tablenotes}
    \end{threeparttable}
\end{table}

Holding the thresholds in Table~\ref{tab:CategoryConditions} fixed, 73\% of redesigned roles remain in the Augmentation quadrant (72\% in the variant). The roles that leave Augmentation move mostly to Displacement (13\%) and Insulated (10\% to 12\%), and only about 4\% move to Concentrated. Redesign therefore lowers variance more than it lowers the mean, which moves roles out of Augmentation toward the low-variance quadrants.

Does AI-assisted redesign shift human effort toward less automatable work? Only at the margin. The tasks the model replaces do become less exposed, and nearly two-thirds of the replacements ask the role holder to review what AI now produces. But these tasks make up only about a tenth of a role, so the typical role barely moves, by about 0.01 across all roles. Using the freed time for additional work does not help when the model is left unguided, because that work is more exposed than the rest of the role. Telling the model to avoid work AI could already do lowers the exposure of the additional work, but does not make roles much less exposed overall.

\subsubsection{Team redesign}\label{sec:ResultsTeamRedesign}

Applying the redesign process within teams shows how team structures change once work can move between roles.
We focus on 80 relatively exposed teams, which together hold 1,256 posts in 603 distinct roles (median team size 10.5 posts, ranging from 5 to 158).
All figures describe a single model draw per team and should be read as indicative of what is feasible rather than as estimates.

Removing one post from each team takes 80 posts out of the 1,256 (6.4\%). The number of roles falls much further, from 603 to 496 (17.7\%), so the average role is held by 2.4 FTE rather than 2.1. In 66 of the 80 teams the model merges at least two roles. In the 14 teams where nothing is merged, the number of roles is unchanged, and in none of the 80 does it rise. Consolidation is therefore the model's usual response, but not always. Moreover, the structure of a team absorbs the loss of a post to a much greater degree than its headcount suggests: rather than thinning every role, the model bundles the remaining work into fewer, larger roles. This implies that the effects of automation within teams show up first in how work is grouped into roles, which an analysis holding role definitions fixed would miss. \footnote{A caveat is that some of the overlap the model finds may reflect near-identical advert templates rather than genuine duplication of work.}

Examining this role redesign and consolidation more closely, of the 496 new roles, 361 (73\%) are an original role carried over whole, and 120 combine two or more original roles, mostly peers (100 within a single seniority band\footnote{We group Civil Service grades into three bands so that the results can be read without knowledge of the grading system: junior (Administrative Officer, AO, and Executive Officer, EO), mid-level (Higher Executive Officer, HEO, and Senior Executive Officer, SEO) and senior (Grade~7, G7, and Grade~6, G6). No other grades appear in our teams.} and 20 across bands). Some consolidation is forced: in 20 teams, the chosen automated role has a single post and so its remaining work must be absorbed by another role elsewhere. But it is not only forced: of the 60 teams in which the post could simply come off a multi-post role, only 14 do so with every role kept one-to-one, while the other 46 also merge roles. Merged roles are as full as the roles carried over whole before any backfill, so they are not merely the remnants of roles that automation emptied. The share of roles that merge falls with team size, from 39\% in teams of five or six posts to about 21\% in teams of eleven or more, so larger teams absorb the cut with fewer changes to the roles around it, although most of them still consolidate somewhere. Taken together, this suggests that the model responds to a smaller team by pooling overlapping peer roles, leaving most of the structure in place, rather than by redistributing work widely.

The role chosen to lose a post is always at the junior or mid-level bands (Table~\ref{tab:RestructureGrades}): mid-level in 47 teams and junior in 33, and never senior. This is not because senior roles are absent: a senior role is present in 54 of the 80 teams, while a junior role is present in only 47, yet a junior role is the one chosen in 33 of them. The reduction decreases with seniority: junior posts fall by 13.6\% and mid-level posts by 6.7\%, while senior posts rise by 1.5\%. Junior roles hold 21\% of the posts but bear 45\% of the net loss, against 57\% of the posts and 60\% of the loss for mid-level roles. According to the model's rationale, the reason for this gradient by seniority is most often the role's lack of accountability or decision-making authority and the overlap of its duties with colleagues' roles. 

Posts also move between bands in three teams, always upward: in two, two junior posts are replaced by one mid-level post, and in the third, a very large team, five mid-level posts become senior posts. Collectively, the restructured teams are smaller, and also flatter at the bottom and weighted somewhat more towards seniority, with the senior share of posts rising from 22\% to 24\% and the junior share falling from 21\% to 19\%. The cut thus falls on the junior routes into the service, which has implications for how the remaining roles are staffed and for the pipeline of future senior staff, and the saving in pay terms will be smaller than the 6.4\% fall in headcount. Further results on how hours are redistributed within teams, and on the human work added to fill spare capacity, are reported in Appendix~\ref{sec:AppendixTeamRestructureResults}.

\begin{table}[!htb]
    \centering
    \newcommand{\hd}[1]{\begin{tabular}[b]{@{}c@{}}#1\end{tabular}}
    \begin{threeparttable}
    \caption{Posts by seniority band across the 80 viable teams, before and after the restructuring.}
    \label{tab:RestructureGrades}
    \begin{tabular}{lccccc}
        \toprule
        Band (grades) & \hd{Posts\\before} & \hd{Posts\\after} & Change & \% change & \hd{Teams where chosen\\role is in this band} \\
        \midrule
        Junior (AO, EO) & 264 & 228 & $-$36 & $-$13.6 & 33 \\
        Mid-level (SEO, HEO) & 718 & 670 & $-$48 & $-$6.7 & 47 \\
        Senior (G7, G6) & 274 & 278 & $+$4 & $+$1.5 & 0 \\
        \midrule
        Total & 1,256 & 1,176 & $-$80 & $-$6.4 & 80 \\
        \bottomrule
    \end{tabular}
    \begin{tablenotes}[flushleft]
        \footnotesize
        \item \textit{Note.} AO = Administrative Officer, EO = Executive Officer, HEO = Higher Executive Officer, SEO = Senior Executive Officer, G7 = Grade 7, G6 = Grade 6.
    \end{tablenotes}
    \end{threeparttable}
\end{table}
\section{Discussion}\label{section:Discussion}


We set out to understand how generative AI might reshape the job architecture of a large, multi-unit organization when tasks, rather than occupations, are taken as the unit of analysis. Using job descriptions from the UKCS and eight large language models to score the exposure of tasks, we documented substantial heterogeneity in AI exposure within and across roles. In this section, we discuss what these findings imply for research on AI exposure, AI and productivity, and job (re)design, and we outline the potential blockers and implications for organizations considering how to implement AI in practice.

\subsection{AI exposure as a design input rather than a risk indicator}
A first contribution of our study is to the task-based AI exposure literature. Prior work has constructed occupation-level exposure scores, typically to anticipate labor-market disruption at the economy-wide level \citep[e.g.][]{felten2023occupational, klein2025generative}, but these may be too coarse to guide organizational decisions about AI implementation. Even among roles with identical or similar titles, we find substantial variation in both average exposure and within-role dispersion, so exposure is a property of the task bundles that constitute jobs in particular organizational settings, not simply of occupations. Starting from actual job descriptions lets organizations treat AI exposure as a design input to their job architecture rather than solely as a risk indicator, identifying which tasks are most exposed and where there is scope for redesign rather than displacement.

\subsection{AI exposure and labor demand}
Mean exposure predicts weaker recruitment demand, while the dispersion coefficient has the theoretically predicted positive sign but is not robustly distinguishable from zero. This may partly reflect differences from the setting studied by \cite{HampoleEtAl2025AILaborMarket}, where productivity gains in publicly traded U.S. firms can contribute to firm growth and labor demand. In the UKCS, such gains may instead appear as increased output, improved service quality, or reduced staffing requirements, so the reallocation mechanism may be institutionally contingent, although without data on adoption or output this remains speculative. The findings also complement evidence that generative-AI adoption is followed by reduced junior employment, driven mainly by slower hiring \citep{HosseiniMaasoumLichtinger2025SeniorityBiased, BrynjolfssonChandarChen2025Canaries}, pointing to hiring as an important margin of organizational adjustment.

\subsection{AI-assisted redesign of roles and teams}
Our redesign exercises asked whether AI-assisted redesign moves human effort toward less automatable work, and how it changes the structure of teams. Within individual roles, the change is limited. Most replacement tasks ask the postholder to review or validate AI output, and freed time goes mostly to routine reporting and monitoring, so whole-role exposure barely changes. Replacement tasks also look alike across professions, mirroring evidence that generative AI can make outputs more similar and narrow differences in expertise \citep{DoshiHauser2024GenerativeAICreativityDiversity, DellAcquaEtAl2025CyberneticTeammate}, which suggests that unguided redesign may make roles more uniform as well as more focused on oversight. Within teams, the larger change is structural. The model merges overlapping roles into fewer, larger ones and takes the reduction in headcount from junior and middle grades, echoing the seniority-biased adjustment discussed above, with implications for the pipeline of future senior staff. Merging roles assumes work transfers without loss, yet knowledge is often specific to individuals and teams \citep{PloyhartMoliterno2011HumanCapitalResource, CampbellCoffKryscynski2012HumanCapitalAdvantage, WrightCoffMoliterno2014StrategicHumanCapital} and merged roles raise coordination demands \citep{PuranamRaveendranKnudsen2012}, while cutting junior posts may erode the domain experience that oversight of AI requires \citep{AllenChoudhury2022AlgorithmAugmentedWork}. In both cases, the model was not asked to avoid exposed work, which suggests that redesign has to be deliberate, with attention to what the organization wants from its roles and teams, rather than following from giving staff AI tools.

\subsection{Implications for the UKCS}\label{sec:UKCSImplications}
Arguably more than any other sector or organization, the UKCS has a number of enabling factors that must exist, and a number of blockers it must overcome, to widely adopt AI. There have been clear signals of political will to drive the adoption, with the appointment of the UK's first Minister of State for Artificial Intelligence in 2026. Political will is necessary but not sufficient. Even where tasks are technically exposed, adoption depends on data access and security classifications, legacy IT systems, procurement routes, staff capability to use and assure AI outputs, and the accountability requirements attached to public decisions. Workforce constraints add a further layer. The UKCS has a strong union presence,\footnote{The Public and Commercial Services Union is the eighth-largest trade union in the UK with almost 170,000 members in 2026.} and the Civil Service Redundancy Principles commit employers to ``taking appropriate measures to avoid compulsory redundancies'' \citep{CO2016}. In 2026, 74\% of Civil Service leavers came from either resignation or retirement, so any degree of role displacement in the UKCS as a result of AI or other technology will be long-term and require natural attrition.

These constraints shape how the four categories in Section \ref{sec:workforce-impact} should be read, indicating which lever is most relevant to each group rather than which roles will disappear. \textit{Displacement} roles are not immediately at risk, but are those where workforce planning and reskilling are most needed, with any reduction likely to come through not replacing leavers rather than through redundancy. The relative decline in postings for high-exposure roles (Section \ref{sec:DemandForLabor}) is consistent with this adjustment already being under way. \textit{Augmentation} roles are the strongest candidates for near-term job redesign, which will need to be carried out within the UKCS’s grading structures and consultation requirements. For \textit{Concentrated} roles, the opportunity lies in targeted tools for the small number of highly exposed tasks. Because these deployments leave roles and grades intact, they may face fewer of the barriers described above. \textit{Insulated} roles are likely to see limited change beyond the use of general-purpose assistants. Taken together, the near-term organizational impact of AI in the UKCS is more likely to come through gradual changes in hiring and task content than through large-scale restructuring.

\subsection{Transferability beyond the UK Civil Service}
A key takeaway from our analysis is that heterogeneity and context matter: exposure varies substantially between roles that share an occupational title, and occupation-level task lists can misstate it. Organizations seeking to understand how AI will affect their workforce are therefore likely to benefit from replicating this methodology to create a bespoke measure of AI exposure rather than relying on generalized occupation-level estimates. The requirements for doing so are modest: a corpus of job descriptions, workforce data against which to reweight them, and access to suitable models. Although our primary specification uses an ensemble exposure score calculated from eight different models, a local open-weight model closely aligned with our ensemble score would allow the scoring stage to be run within an organization's own infrastructure where data are sensitive.

Our point estimates are specific to the UKCS, but this setting also shows what applying the approach elsewhere might involve, both methodologically and in practice. Methodologically, the UKCS is an unusually large and complex case. In 2026, it reported total headcount of 557,825 (523,795 FTE) across more than 80 departments. This scale required us to process over 280,000 vacancies and reweight them by department, grade and profession in order to produce an accurate exposure profile. The volume of FTE is representative of the huge variation in the types of roles and tasks in the UKCS, which includes central government policymakers, but also frontline staff, such as UK Border Force, Prison Officers and Jobcentre staff. Applying this approach to a smaller organization, with fewer roles and less variation, should be considerably simpler and enable other researchers to perform further manual verification of each step. The main constraint is likely to be data rather than complexity. In practice, acting on the findings is also likely to be easier in the private sector. Private sector firms face fewer constraints on restructuring and can redeploy freed time into revenue-generating work. This is consistent with \citet{HampoleEtAl2025AILaborMarket} finding evidence of reallocation in US data that we do not detect in UKCS hiring. The same flexibility, however, means that displacement, as well as redesign, could happen more quickly.

\subsection{Limitations}
We caution that there are some limitations of our study. First, the approach taken throughout this paper relies on extracting tasks from job descriptions. While this task extraction approach builds upon the common approach of using standardized task classifications, this assumes that job descriptions are fully representative of the tasks that the role holder will undertake. This assumption is less likely to hold for the roles in the earlier years of our sample, where a role holder will have been in post for a number of years. In reality, job descriptions are promotional products aimed at attracting strong applicants. The tasks listed may therefore be aspirational and represent the most interesting and engaging elements of a job role. In other cases, they may not fully represent the realities of the job role, as hiring managers may reuse previous job descriptions or use AI to draft a standard job description. Our experience suggests that the types of tasks typically overlooked, particularly in UKCS job descriptions, are the more routine, admin-based tasks, which are likely to have higher exposure. 

In a public sector setting, changing Government priorities or the appointment of new Ministers will fundamentally change the responsibilities of at least a handful of job roles. Given the absence of time-use data, we weighted the tasks according to where they appear in the job description, assuming that tasks mentioned earlier are traditionally viewed as more central to a role and therefore occupy more of its time. The decay rate chosen is based on conversations with hiring managers, but has not been formally tested and should therefore not be considered to be a completely accurate figure. Nevertheless, our robustness checks, where we weight tasks more equally or place a larger emphasis on the first tasks, produce similar overall results, demonstrating that this assumption is not central to our conclusions.

Our redesign exercises also have limitations. Both are exploratory and describe what a model proposes, not changes the Civil Service has made, and neither has been validated by managers or postholders. Each rests on a single model run, and new tasks are scored by another language model, so some of the fall in exposure may reflect how tasks are worded rather than a change in the work. Team redesign also assumes work can move between people at the same productivity, ignores coordination costs, and treats the reduction in headcount as a counterfactual device rather than a recommendation. Because exposure depends on how tasks are bundled into roles in particular organizational settings, any redesign is specific to the task mix of an individual role or team, and validating it would require the judgment of the line managers responsible for each one. Validating a small number of teams would therefore only show that the redesign makes sense for those teams, not for the wider sample. Our data are also drawn from job descriptions rather than from postholders, and identifying the people in specific roles to review their redesigns would reduce the anonymity of the data. We therefore use an LLM to apply the same redesign consistently across a large number of roles and teams, so that the exercise describes the broad pattern of impact across the organization rather than recommendations for any single role.   \clearpage      

\bibliographystyle{apalike}   
\bibliography{references}

\clearpage
\appendix
\section*{Supplementary Materials}\label{sec:SuppementaryMaterials}
\section{Data description}\label{section:AppendixDatasetDescription}

\subsection{Government Recruitment Information Database Fields}\label{sec:GRID}

\begin{table}[!htb]
    \centering
    \small
    \begin{threeparttable}
    \caption{Variables in the GRID extract}
    \label{tab:GRIDVariables}
    \begin{tabularx}{\linewidth}{@{}p{3cm}X@{}}
        \toprule
        Variable & Description \\
        \midrule
        Vacancy ID & A unique identifier assigned to the job advert \\
        Vacancy title & Job title of the role \\
        Department & The Organization holding the vacancy. This could be a department, agency, arms-length body, or devolved administration. \\
        Job grade & The Civil Service grade associated with the role \\
        Job description & Normally a description of the role and the duties associated with it \\
        Job summary & Normally an overview of the department and team \\
        Posting date & The date and time the role was published on Civil Service Jobs \\
        Closing date & The date and time the role closed for applications on Civil Service Jobs \\
        Profession & The UKCS profession associated with the role — selected from 28 options or `Other' \\
        Jobs available & The number of FTE roles available for the particular vacancy \\
        Business unit & The business unit (free text) associated with the role \\
        \bottomrule
    \end{tabularx}
    \begin{tablenotes}[flushleft]
        \footnotesize
        \item \textit{Note.} GRID comprises a backend feed of Civil Service Jobs postings.
    \end{tablenotes}
    \end{threeparttable}
\end{table}

\subsubsection{GRID data cleaning}\label{sec:GRIDCleaning}

The raw GRID extract has no preprocessing and so requires us to clean and subset according to our needs. As job description is a free text field, we begin by dropping empty or duplicate jobs, removing adverts that are too brief, and stripping out boilerplate text or html encodings. We append the job summary onto the job description field in order to provide the LLM the role's context along with its substantive description. This also covers for a minority of cases in one department where the job summary and job description fields had been swapped. 

In order to be able to rake our vacancies and thereby assign weights, we filter GRID to include only those departments, grades, and professions that have a match in the UKCS statistics. We further remove any departments that would risk over-extrapolation by ensuring that there is a maximum of 12 FTEs per vacancy per department. 

We further remove all dummy/test vacancies by regex on the job title, being careful not to exclude valid roles such as ``test engineer''. Some roles are advertised simultaneously e.g. across both internal and external channels, so where multiple fields match and the Levenshtein distance is low for free text fields, we de-duplicate to avoid over-counting. Cleaning and de-duplication reduce the raw row count of 409,775 observations to 301,454 vacancies. We then drop vacancies from which fewer than three tasks could be extracted, together with those removed by the matching and over-extrapolation filters described above, leaving an analytical sample of 283,484 vacancies. As noted in Section \ref{section:DatasetDesc}, the 83 remaining departments in the dataset account for 94 percent of the total UKCS workforce. Detailed statistics on the sample's representativeness are further provided in Tables \ref{tab:department_coverage}, \ref{tab:grade_coverage}, and \ref{tab:profession_coverage} below. 

For any missing posting dates, we substitute the missing value with the closing date. For other vacancies this is typically a few weeks after the posting date. Implausible vacancy closing dates (stretching into 2027), often associated with very recent national recruitment campaigns are capped at the maximum observed posting date.

Finally, as mentioned in Section \ref{section:DatasetDesc}, the raw GRID contains a high share of ``Other'' profession vacancies. We pass each of these vacancies to Qwen 3 235b along with the prompt and schema in Appendix \ref{sec:AppendixProfessionReassignment} and have each assessed according to its likely profession given the job title and description. This reassigns the vast majority of ``Other'' profession vacancies: of the original 88,113, 1,080 remain ``Other'' while the other 88,050 are each reassigned. Those that are remain include vacancies such as broad-based graduate schemes that span several business areas, or genuinely niche roles. 

\subsection{Civil Service Statistics Bulletin}\label{sec:CivilServiceStatsBulletin}

Following the GRID cleaning and LLM batching, we join the resulting dataset onto aggregate statistics from the official UKCS.\footnote{See \href{https://www.gov.uk/government/collections/civil-service-statistics}{https://www.gov.uk/government/collections/civil-service-statistics}} These statistics are sourced from all UKCS organizations via the Annual Civil Service Employment Survey (ACSES). These statistics provide the target anchors across department, grade and profession which we rake our sample to. 

To get the level of granularity we require, we access the data via the Statistics Browser which un-bundles job grades.\footnote{\href{https://civil-service-statistics.jdac.service.cabinetoffice.gov.uk/}{https://civil-service-statistics.jdac.service.cabinetoffice.gov.uk/}} We access the FTEs in each of the marginal distributions of department, grade, and profession separately. Ideally, we would use the joint distribution of department x grade x profession, but due to the confidentiality masking and curse of dimensionality, we lose observations from a lot of smaller buckets. Therefore, we only retain the 3D marginals as a robustness check where required. 

\subsection{Department representation}\label{section:DeptDataDescAppendix}
In Table \ref{tab:department_coverage}, vacancies refer to row counts in our GRID extract. Roles correspond to the number of jobs available over all vacancies. FTE shows the headcount of Civil Servants in each department as per the 2026 annual UKCS Statistical Bulletin. The bracketed terms show the percentage to one decimal of each column total. Rows are ordered by 2026 FTE.

{\linespread{1}\small
\setlength{\tabcolsep}{4pt}
\begin{longtable}{@{}p{6.4cm}rrr@{}}
\caption{Coverage of vacancies, roles and FTE by department} \label{tab:department_coverage} \\
\toprule
Department & Vacancies & Roles & FTE \\
\midrule
\endfirsthead
\caption[]{Coverage of vacancies, roles and FTE by department} \\
\toprule
Department & Vacancies & Roles & FTE \\
\midrule
\endhead
\midrule
\multicolumn{4}{r}{Continued on next page} \\
\midrule
\endfoot
\bottomrule
\endlastfoot
Department for Work and Pensions (excl. agencies) & 8,341 (2.9\%) & 81,851 (10.9\%) & 84,110 (17.1\%) \\
HM Revenue and Customs (excl. agencies) & 33,215 (11.7\%) & 109,693 (14.7\%) & 66,415 (13.5\%) \\
HM Prison and Probation Service & 29,723 (10.5\%) & 116,849 (15.6\%) & 64,885 (13.2\%) \\
Home Office & 21,030 (7.4\%) & 77,598 (10.4\%) & 48,820 (9.9\%) \\
Ministry of Defence (excl. agencies) & 41,475 (14.6\%) & 56,744 (7.6\%) & 35,195 (7.2\%) \\
HM Courts and Tribunals Service & 6,636 (2.3\%) & 22,866 (3.1\%) & 14,940 (3.0\%) \\
Scottish Government (excl. agencies) & 7,101 (2.5\%) & 13,676 (1.8\%) & 8,640 (1.8\%) \\
Ministry of Justice (excl. agencies) & 9,285 (3.3\%) & 36,380 (4.9\%) & 8,190 (1.7\%) \\
Crown Prosecution Service & 3,505 (1.2\%) & 9,852 (1.3\%) & 7,800 (1.6\%) \\
Foreign, Commonwealth and Development Office (excl. agencies) & 9,847 (3.5\%) & 11,819 (1.6\%) & 7,500 (1.5\%) \\
Department for Education (excl. agencies) & 11,229 (4.0\%) & 16,984 (2.3\%) & 7,315 (1.5\%) \\
HM Land Registry & 1,968 (0.7\%) & 5,788 (0.8\%) & 6,410 (1.3\%) \\
Department for Environment, Food and Rural Affairs (excl. agencies) & 5,941 (2.1\%) & 10,527 (1.4\%) & 6,245 (1.3\%) \\
Department for Business and Trade (excl. agencies) & 6,382 (2.3\%) & 8,833 (1.2\%) & 5,810 (1.2\%) \\
National Crime Agency & 2,085 (0.7\%) & 9,372 (1.3\%) & 5,750 (1.2\%) \\
Welsh Government (excl. agencies) & 963 (0.3\%) & 1,601 (0.2\%) & 5,740 (1.2\%) \\
Driver and Vehicle Licensing Agency & 1,269 (0.4\%) & 5,524 (0.7\%) & 5,560 (1.1\%) \\
UK Health Security Agency & 6,661 (2.3\%) & 8,687 (1.2\%) & 5,560 (1.1\%) \\
Cabinet Office (excl. agencies) & 9,690 (3.4\%) & 21,412 (2.9\%) & 5,240 (1.1\%) \\
Driver and Vehicle Standards Agency & 2,210 (0.8\%) & 7,712 (1.0\%) & 4,990 (1.0\%) \\
Department for Energy Security and Net Zero & 2,210 (0.8\%) & 2,924 (0.4\%) & 4,725 (1.0\%) \\
Defence Science and Technology Laboratory & 1,993 (0.7\%) & 4,849 (0.6\%) & 4,605 (0.9\%) \\
Social Security Scotland & 551 (0.2\%) & 1,974 (0.3\%) & 4,275 (0.9\%) \\
Valuation Office Agency & 1,593 (0.6\%) & 4,731 (0.6\%) & 4,030 (0.8\%) \\
Ministry of Housing, Communities and Local Government (excl. agencies) & 4,349 (1.5\%) & 6,718 (0.9\%) & 3,665 (0.7\%) \\
Department for Science, Innovation and Technology (excl. agencies) & 1,498 (0.5\%) & 2,030 (0.3\%) & 3,570 (0.7\%) \\
Department for Transport (excl. agencies) & 4,197 (1.5\%) & 5,560 (0.7\%) & 3,305 (0.7\%) \\
Animal and Plant Health Agency & 2,521 (0.9\%) & 6,600 (0.9\%) & 3,225 (0.7\%) \\
Government Legal Department & 891 (0.3\%) & 3,501 (0.5\%) & 3,180 (0.6\%) \\
Department of Health and Social Care (excl. agencies) & 4,352 (1.5\%) & 8,397 (1.1\%) & 3,175 (0.6\%) \\
Health and Safety Executive & 2,590 (0.9\%) & 5,504 (0.7\%) & 3,110 (0.6\%) \\
Crown Office and Procurator Fiscal Service & 479 (0.2\%) & 945 (0.1\%) & 2,595 (0.5\%) \\
Met Office & 1,024 (0.4\%) & 1,597 (0.2\%) & 2,405 (0.5\%) \\
Rural Payments Agency & 412 (0.1\%) & 2,342 (0.3\%) & 2,400 (0.5\%) \\
Companies House & 1,428 (0.5\%) & 2,757 (0.4\%) & 2,310 (0.5\%) \\
Office of Gas and Electricity Markets & 2,328 (0.8\%) & 3,368 (0.4\%) & 2,275 (0.5\%) \\
Office for Standards in Education, Children's Services and Skills & 1,026 (0.4\%) & 2,089 (0.3\%) & 2,125 (0.4\%) \\
HM Treasury (excl. agencies) & 3,598 (1.3\%) & 4,454 (0.6\%) & 2,095 (0.4\%) \\
The Insolvency Service & 1,489 (0.5\%) & 2,767 (0.4\%) & 1,975 (0.4\%) \\
Office of the Public Guardian & 489 (0.2\%) & 2,317 (0.3\%) & 1,730 (0.4\%) \\
Intellectual Property Office & 1,305 (0.5\%) & 3,479 (0.5\%) & 1,675 (0.3\%) \\
Medicines and Healthcare Products Regulatory Agency & 1,977 (0.7\%) & 2,578 (0.3\%) & 1,570 (0.3\%) \\
Government Commercial Organization & 1,733 (0.6\%) & 2,587 (0.3\%) & 1,555 (0.3\%) \\
Food Standards Agency & 1,798 (0.6\%) & 2,371 (0.3\%) & 1,535 (0.3\%) \\
Maritime and Coastguard Agency & 1,583 (0.6\%) & 2,621 (0.4\%) & 1,270 (0.3\%) \\
Legal Aid Agency & 568 (0.2\%) & 1,041 (0.1\%) & 1,255 (0.3\%) \\
FCDO Services & 1,066 (0.4\%) & 1,549 (0.2\%) & 1,155 (0.2\%) \\
Registers of Scotland & 259 (0.1\%) & 456 (0.1\%) & 1,130 (0.2\%) \\
Department for Culture, Media and Sport & 2,812 (1.0\%) & 3,781 (0.5\%) & 1,125 (0.2\%) \\
Forestry and Land Scotland & 454 (0.2\%) & 736 (0.1\%) & 1,055 (0.2\%) \\
Advisory, Conciliation and Arbitration Service & 756 (0.3\%) & 1,482 (0.2\%) & 1,050 (0.2\%) \\
Planning Inspectorate & 578 (0.2\%) & 965 (0.1\%) & 1,010 (0.2\%) \\
Competition and Markets Authority & 806 (0.3\%) & 1,415 (0.2\%) & 945 (0.2\%) \\
UK Hydrographic Office & 1,011 (0.4\%) & 1,838 (0.2\%) & 910 (0.2\%) \\
Crown Commercial Service & 876 (0.3\%) & 1,147 (0.2\%) & 705 (0.1\%) \\
The National Archives & 475 (0.2\%) & 558 (0.1\%) & 630 (0.1\%) \\
Centre for Environment, Fisheries and Aquaculture Science & 327 (0.1\%) & 420 (0.1\%) & 605 (0.1\%) \\
Serious Fraud Office & 485 (0.2\%) & 1,008 (0.1\%) & 595 (0.1\%) \\
UK Export Finance & 1,022 (0.4\%) & 1,245 (0.2\%) & 580 (0.1\%) \\
Water Services Regulation Authority & 115 (0.0\%) & 136 (0.0\%) & 470 (0.1\%) \\
Government Property Agency & 787 (0.3\%) & 1,057 (0.1\%) & 465 (0.1\%) \\
Charity Commission & 548 (0.2\%) & 743 (0.1\%) & 420 (0.1\%) \\
Government Internal Audit Agency & 332 (0.1\%) & 931 (0.1\%) & 420 (0.1\%) \\
Office of Rail and Road & 500 (0.2\%) & 553 (0.1\%) & 370 (0.1\%) \\
Criminal Injuries Compensation Authority & 34 (0.0\%) & 135 (0.0\%) & 355 (0.1\%) \\
Office of Qualifications and Examinations Regulation & 385 (0.1\%) & 482 (0.1\%) & 320 (0.1\%) \\
UK Space Agency & 742 (0.3\%) & 810 (0.1\%) & 310 (0.1\%) \\
National Savings and Investments & 326 (0.1\%) & 388 (0.1\%) & 280 (0.1\%) \\
Vehicle Certification Agency & 306 (0.1\%) & 451 (0.1\%) & 260 (0.1\%) \\
Scottish Forestry & 117 (0.0\%) & 175 (0.0\%) & 245 (0.0\%) \\
Building Digital UK & 199 (0.1\%) & 292 (0.0\%) & 245 (0.0\%) \\
Government Actuary's Department & 109 (0.0\%) & 218 (0.0\%) & 225 (0.0\%) \\
Veterinary Medicines Directorate & 244 (0.1\%) & 312 (0.0\%) & 200 (0.0\%) \\
Northern Ireland Office & 416 (0.1\%) & 490 (0.1\%) & 180 (0.0\%) \\
Estyn & 58 (0.0\%) & 107 (0.0\%) & 130 (0.0\%) \\
Welsh Revenue Authority & 98 (0.0\%) & 112 (0.0\%) & 120 (0.0\%) \\
Active Travel England & 116 (0.0\%) & 147 (0.0\%) & 100 (0.0\%) \\
Wilton Park & 77 (0.0\%) & 87 (0.0\%) & 80 (0.0\%) \\
Attorney General's Office & 155 (0.1\%) & 184 (0.0\%) & 65 (0.0\%) \\
UK Supreme Court & 86 (0.0\%) & 105 (0.0\%) & 55 (0.0\%) \\
Office for Budget Responsibility & 76 (0.0\%) & 88 (0.0\%) & 55 (0.0\%) \\
Queen Elizabeth II Centre & 114 (0.0\%) & 130 (0.0\%) & 50 (0.0\%) \\
Office of the Secretary of State for Wales & 79 (0.0\%) & 88 (0.0\%) & 45 (0.0\%) \\
\midrule
Total & 283,484 (100.0\%) & 748,690 (100.0\%) & 491,710 (100.0\%) \\
\end{longtable}
}

\subsection{Grade}\label{section:GradeDataDescAppendix}
In Table \ref{tab:grade_coverage}, vacancies refer to row counts in our GRID extract. Roles correspond to the number of jobs available over all vacancies. FTE shows the headcount of civil servants in each grade as per the 2026 annual UKCS Statistical Bulletin. The bracketed terms show the percentage to one decimal of each column total.
\begin{table}[!htb]
    \centering
    \small
    \begin{threeparttable}
    \caption{Grade coverage}
    \label{tab:grade_coverage}
    \begin{tabular}{@{}lrrr@{}}
        \toprule
        Grade & Vacancies & Roles & FTE \\
        \midrule
        AA & 6,386 (2.3\%) & 36,751 (4.9\%) & 12,894 (2.6\%) \\
        AO & 35,085 (12.4\%) & 202,749 (27.1\%) & 121,483 (24.7\%) \\
        EO & 39,124 (13.8\%) & 163,221 (21.8\%) & 123,749 (25.2\%) \\
        HEO & 62,185 (21.9\%) & 125,633 (16.8\%) & 79,725 (16.2\%) \\
        SEO & 65,127 (23.0\%) & 110,891 (14.8\%) & 66,066 (13.4\%) \\
        G7 & 49,201 (17.4\%) & 77,043 (10.3\%) & 58,834 (12.0\%) \\
        G6 & 18,939 (6.7\%) & 24,179 (3.2\%) & 21,752 (4.4\%) \\
        SCS level & 7,437 (2.6\%) & 8,223 (1.1\%) & 7,207 (1.5\%) \\
        \midrule
        Total & 283,484 (100.0\%) & 748,690 (100.0\%) & 491,710 (100.0\%) \\
        \bottomrule
    \end{tabular}
    \end{threeparttable}
\end{table}

\subsection{Profession}\label{section:ProfessionDataDescAppendix}
In Table \ref{tab:profession_coverage}, vacancies refer to row counts in our GRID extract. Roles correspond to the number of jobs available over all vacancies. FTE shows the headcount of civil servants in each profession as per the 2026 annual UKCS Statistical Bulletin. The bracketed terms show the percentage to one decimal of each column total.
{\linespread{1}\small
\setlength{\tabcolsep}{4pt}
\begin{longtable}{@{}p{6.4cm}rrr@{}}
\caption{Profession coverage} \label{tab:profession_coverage} \\
\toprule
Profession & Vacancies & Roles & FTE \\
\midrule
\endfirsthead
\caption[]{Profession coverage} \\
\toprule
Profession & Vacancies & Roles & FTE \\
\midrule
\endhead
\midrule
\multicolumn{4}{r}{Continued on next page} \\
\midrule
\endfoot
\bottomrule
\endlastfoot
Operational Delivery & 87,200 (30.8\%) & 396,846 (53.0\%) & 264,278 (53.7\%) \\
Policy & 38,292 (13.5\%) & 56,656 (7.6\%) & 35,589 (7.2\%) \\
Digital and Data & 36,546 (12.9\%) & 64,062 (8.6\%) & 28,814 (5.9\%) \\
Other & 847 (0.3\%) & 8,270 (1.1\%) & 20,732 (4.2\%) \\
Project Delivery & 21,551 (7.6\%) & 35,927 (4.8\%) & 17,960 (3.7\%) \\
Tax & 3,981 (1.4\%) & 18,690 (2.5\%) & 15,853 (3.2\%) \\
Counter Fraud & 3,815 (1.3\%) & 15,491 (2.1\%) & 12,660 (2.6\%) \\
Finance & 10,469 (3.7\%) & 14,667 (2.0\%) & 12,075 (2.5\%) \\
Science and Engineering & 10,956 (3.9\%) & 18,693 (2.5\%) & 11,419 (2.3\%) \\
Human Resources & 13,078 (4.6\%) & 18,105 (2.4\%) & 11,289 (2.3\%) \\
Legal & 3,749 (1.3\%) & 14,876 (2.0\%) & 10,639 (2.2\%) \\
Security & 5,889 (2.1\%) & 10,402 (1.4\%) & 9,488 (1.9\%) \\
Property & 5,681 (2.0\%) & 8,939 (1.2\%) & 7,606 (1.5\%) \\
Commercial & 8,083 (2.9\%) & 13,971 (1.9\%) & 5,805 (1.2\%) \\
Communications & 8,250 (2.9\%) & 9,908 (1.3\%) & 4,939 (1.0\%) \\
Intelligence Analysis & 3,353 (1.2\%) & 7,731 (1.0\%) & 3,613 (0.7\%) \\
Statistics & 2,331 (0.8\%) & 4,282 (0.6\%) & 2,877 (0.6\%) \\
Economics & 2,655 (0.9\%) & 4,473 (0.6\%) & 2,822 (0.6\%) \\
Knowledge and Information Management & 6,036 (2.1\%) & 7,437 (1.0\%) & 2,742 (0.6\%) \\
Clinical & 665 (0.2\%) & 2,180 (0.3\%) & 2,337 (0.5\%) \\
Operational Research & 940 (0.3\%) & 1,819 (0.2\%) & 1,781 (0.4\%) \\
Social Research & 1,865 (0.7\%) & 2,985 (0.4\%) & 1,746 (0.4\%) \\
Inspector of Education and Training & 241 (0.1\%) & 1,081 (0.1\%) & 1,096 (0.2\%) \\
Internal Audit & 872 (0.3\%) & 1,673 (0.2\%) & 691 (0.1\%) \\
Veterinarian & 330 (0.1\%) & 1,070 (0.1\%) & 560 (0.1\%) \\
Geography & 134 (0.0\%) & 558 (0.1\%) & 530 (0.1\%) \\
Risk Management & 1,878 (0.7\%) & 2,107 (0.3\%) & 485 (0.1\%) \\
Planning Inspectors & 56 (0.0\%) & 138 (0.0\%) & 470 (0.1\%) \\
Planning & 1,016 (0.4\%) & 1,364 (0.2\%) & 385 (0.1\%) \\
Corporate Finance & 2,494 (0.9\%) & 3,818 (0.5\%) & 195 (0.0\%) \\
Actuary & 44 (0.0\%) & 143 (0.0\%) & 130 (0.0\%) \\
Occupational Psychology & 187 (0.1\%) & 328 (0.0\%) & 100 (0.0\%) \\
\midrule
Total & 283,484 (100.0\%) & 748,690 (100.0\%) & 491,710 (100.0\%) \\
\end{longtable}
}

\subsection{Task processing error rates}\label{sec:AppendixTaskErrorRates}

\begin{table}[htbp]
  \centering
  \small
  \begin{threeparttable}
  \caption{Post-processing bucket-snap events by model (\texttt{automation\_potential} coerced to the nearest 0.1)}
  \label{tab:coercion-audit}
  \begin{tabular}{llrrrr}
    \toprule
    Run & Model & Records replayed & Refused & Snapped & \% snapped \\
    \midrule
    Extraction & sonnet-4-6       & 320{,}207 & 297 & 0     & 0.00\% \\
    Scoring    & nova-2-lite      & 317{,}613 & 60  & 1{,}126 & 0.35\% \\
    Scoring    & nemotron-super   & 317{,}592 & 81  & 95    & 0.03\% \\
    Scoring    & glm-5            & 317{,}634 & 39  & 115   & 0.04\% \\
    Scoring    & qwen-3-235b      & 317{,}668 & 5   & 2     & 0.00\% \\
    Scoring    & llama-maverick   & 317{,}592 & 81  & 0     & 0.00\% \\
    Scoring    & openai-oss-120b  & 317{,}665 & 8   & 0     & 0.00\% \\
    Scoring    & ministral-14b-3  & 317{,}614 & 59  & 0     & 0.00\% \\
    \bottomrule
  \end{tabular}
  \end{threeparttable}
\end{table}

\begin{figure}[!htb]
    \centering
    \includegraphics[width=1\linewidth]{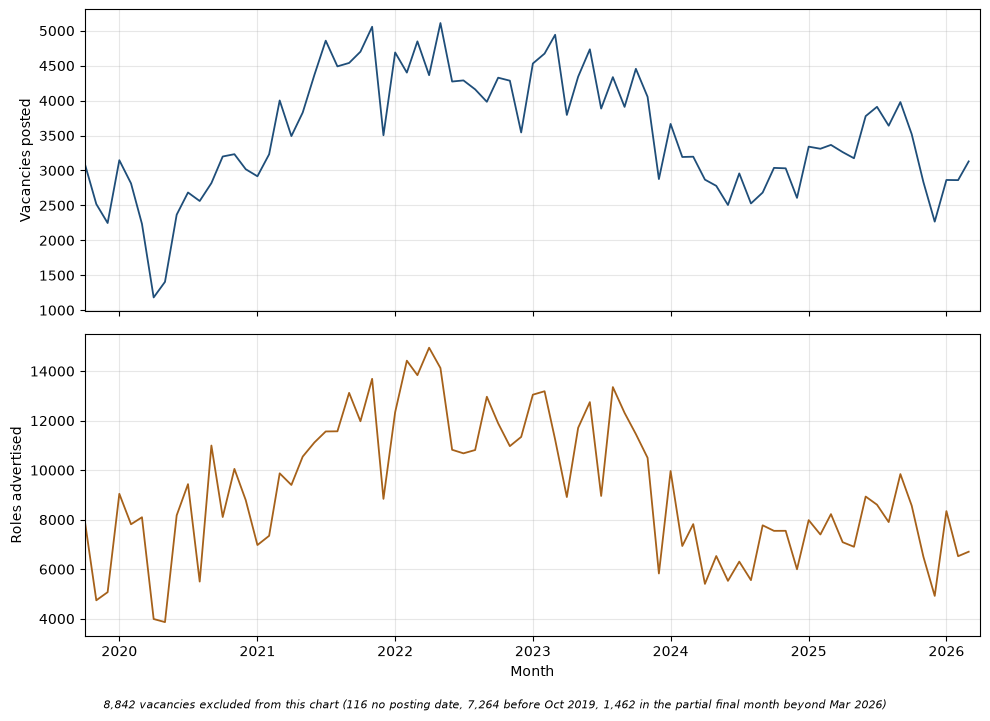}
    \caption{Vacancies (top) and roles (bottom) over time}
    \label{fig:VacanciesOverTime}
\end{figure} \clearpage
\section{Prompts and output schema}\label{AppendixLLMPrompts}

This appendix reproduces the prompt used at each step of the pipeline, in the order the steps are run. Stages 1 and 2 refer to the two stages of the job redesign exercise (Section~\ref{sec:MethodologyRedesign}), the profession, extraction and scoring prompts come before it, and the task category prompt follows it. Text in braces is filled in at run time. Prompts are shown as they are sent, including their original spelling. In every case the model is required to return structured output with a short reasoning field placed before the answer fields, and the fields are described under each prompt.

\subsection{Profession reassignment prompt}\label{sec:AppendixProfessionReassignment}

The model returns one sentence of reasoning followed by a single profession, which must match the closed list in the prompt exactly.

\begin{lstlisting}[language={}]
SYSTEM:
You are an expert on UK Civil Service job roles and the official profession
categories used in UK Civil Service Statistics. Your job is to read a job
advert's title and description and determine which single official profession
category the role most plausibly belongs to, based ONLY on the substantive
content of the advert itself.

USER:
Read the job title and description carefully. Decide which ONE profession
category the substantive content of this role most plausibly belongs to.

If the advert genuinely does not contain enough information to determine a
specific profession (e.g. it describes a generic recruitment campaign covering
multiple functions, graduate schemes that rotate, a highly specialised role not
adequately described by the set of professions, or provides no real detail
about day-to-day duties), select "Other" rather than guessing. Do not force an
assignment to a named profession if the evidence does not support it.

First, briefly explain your reasoning (1 sentence) referring to specific
content in the job title/description. Then state your final profession
category choice.

Job title: {vacancy_title}

Job description:
{job_description}

Valid profession categories (choose EXACTLY ONE, verbatim, from this list):
"Operational Delivery", "Policy", "Digital and Data", "Project Delivery",
"Finance", "Corporate Finance", "Commercial", "Human Resources",
"Communications", "Legal", "Security", "Property", "Science and Engineering",
"Intelligence Analysis", "Counter Fraud", "Statistics", "Economics",
"Social Research", "Operational Research",
"Knowledge and Information Management", "Internal Audit", "Planning",
"Planning Inspectors", "Tax", "Veterinarian", "Actuary", "Clinical",
"Geography", "Inspector of Education and Training",
"Occupational Psychology", "Risk Management", "Other".

You MUST give your answer using the tool provided.
\end{lstlisting}

\subsection{Extraction and scoring prompt}\label{sec:AppendixExtractAndScorePrompt}

This prompt extracts the tasks from each job description and scores them. For each task the model returns the task text, one sentence of reasoning, and an automation potential score. Scores are rounded to the nearest 0.1. The scoring definition and rubric below are the same in every scoring prompt in this appendix.

\begin{lstlisting}[language={}]
SYSTEM:
You are an expert on UK Civil Service work. Your job is to extract the tasks
from a job advert and assess the automation potential of each with current AI
systems.

USER:
Extract each work task from the job description below and score it for its
AUTOMATION POTENTIAL from current AI systems, using the provided format.

TASK EXTRACTION RULES:
1. Return tasks in the order they appear in the job description. Order
   matters: downstream processing preserves the list order.
2. Exclude tasks that describe this role's own recruitment, onboarding,
   working hours, or working conditions. If the role involves recruiting or
   onboarding others as part of the job, retain those tasks.
3. Extract the individual role's actionable tasks (what the person does), not
   skills (what the person knows) or team goals.
4. Do NOT extract items that describe vague outcomes or accountability, not
   actions (e.g. "meet KPIs", "deliver on strategic priorities"). Test: if you
   cannot answer "what specifically does the postholder DO on a typical day?",
   it is not a task worth extracting.

SCORING DEFINITION (automation_potential):
Your estimate of the FRACTION OF THE TASK'S SUBSTANTIVE CONTENT that could be
performed by current AI systems (LLMs, multimodal models, AI-powered software,
transcription tools, retrieval tools). The rubric below describes how AI and
human involvement combine at different points on the scale.

- 0.0 means no substantive part of the task can be performed by current AI
  systems.
- 1.0 means the entire substantive content can be performed by current AI
  systems with negligible human involvement.
- Score in DISCRETE 0.1 intervals. Return only values in
  {0.0, 0.1, 0.2, 0.3, 0.4, 0.5, 0.6, 0.7, 0.8, 0.9, 1.0}.

RUBRIC:

| Score     | Share automatable | Character of AI involvement           |
|-----------|-------------------|---------------------------------------|
| 0.0       | None              | No substantive part is automatable.   |
|           |                   | Task requires physical presence,      |
|           |                   | physical action, in-person live       |
|           |                   | judgement, or capabilities beyond     |
|           |                   | current AI systems.                   |
| 0.1-0.2   | Small share       | The human performs the substantive    |
|           |                   | work. AI can support with background  |
|           |                   | preparation, information retrieval,   |
|           |                   | or partial drafting, but the core of  |
|           |                   | the task remains with the human.      |
| 0.3-0.4   | Meaningful share  | AI can produce initial drafts,        |
|           |                   | intermediate outputs, or process      |
|           |                   | structured inputs. The human still    |
|           |                   | performs core reasoning, adaptation,  |
|           |                   | and substantive decision-making.      |
| 0.5-0.6   | Roughly half      | Shared execution. AI processes and    |
|           |                   | drafts a substantial share; the human |
|           |                   | directs, contextualises, and verifies |
|           |                   | with materially more than spot-check  |
|           |                   | involvement.                          |
| 0.7-0.8   | Majority share    | AI produces the substantive output    |
|           |                   | (applying codified rules, synthesising|
|           |                   | evidence, or processing structured    |
|           |                   | information). The human reviews,      |
|           |                   | adjusts, and handles exceptions.      |
| 0.9-1.0   | Near-complete     | AI performs the substantive work      |
|           |                   | end-to-end. Human role is limited to  |
|           |                   | setup, sampling, spot-checks, or      |
|           |                   | handling residual edge cases.         |

Score continuously across the eleven buckets; the bands above are for
orientation only. Choose the bucket that best matches your
substantive-component decomposition.

OUTPUT INSTRUCTIONS:
For each extracted task, provide in this order:
1. task (str): the task text as extracted from the job description. Reproduce
   the substantive activity; do not paraphrase heavily.
2. reasoning (one sentence): before scoring, name the substantive
   component(s) of the task current AI could perform, and the substantive
   component(s) that require a human. This decomposition should determine
   your score, not justify a pre-committed number.
3. automation_potential: based on your reasoning, the fraction of the task's
   substantive content AI could perform. Return exactly one of the eleven
   allowed values.

<job_description>{job_description}</job_description>
\end{lstlisting}

\subsection{Scoring prompts}\label{sec:AppendixScorePrompt}\label{sec:AppendixRescoringPrompt}

A second prompt scores a fixed list of tasks that have already been extracted. It is used to score the original tasks and, unchanged, the new tasks written in the job redesign (Stage 1 and Stage 2, in the main redesign and the variant), so the two are directly comparable. The prompt does not say that any task is new or what it replaced, and the model sees only the job description and the bare task text. The system prompt is ``You are an expert on UK Civil Service work. Your job is to assess the automation potential of tasks with current AI systems.'' The model returns, for each task, its number, one sentence of reasoning and a score, and the task numbers must be unique.

\begin{lstlisting}[language={}]
You are given a job description and a NUMBERED LIST OF TASKS that have already
been extracted from it. Your job is to score EACH task for its AUTOMATION
POTENTIAL from current AI systems, using the provided format.

IMPORTANT - DO NOT RE-EXTRACT:
- Score exactly the tasks listed below, no more and no fewer.
- Do NOT merge, split, reorder, add, or drop tasks.
- Return one entry per task, echoing its task_number so your scores align to
  the list. Preserve the given order.
- Use the full job description only as CONTEXT to understand each task; do not
  score anything not in the numbered list.

[SCORING DEFINITION and RUBRIC: identical to the extraction prompt]

OUTPUT INSTRUCTIONS:
For each task, provide in this order:
1. task_number (int): the number of the task from the list below. Echo it
   exactly so your score can be matched to the task.
2. reasoning (one sentence): [as in the extraction prompt]
3. automation_potential: [as in the extraction prompt]

<job_description>{job_description}</job_description>

<tasks_to_score>
{numbered_tasks}
</tasks_to_score>
\end{lstlisting}

\subsection{Job redesign prompt}\label{sec:AppendixRedesignPrompt}

This is the prompt for Stage 1 of the job redesign, as used in the full 10,000-role run, with exposure scores withheld. The pilot condition that shows scores adds the current score and reasoning beside each task and tells the model not to react to them. The model returns the numbers of the tasks it removes (possibly none) and, for each, one sentence of reasoning and the replacement task. The numbers of removed and replaced tasks must match one for one.

\begin{lstlisting}[language={}]
SYSTEM:
You are an expert on UK Civil Service work, advising on how individual roles
should evolve as AI tools become more capable. Your job is to redesign a
role's task list to maximise its productivity -- not to make the role look
less automatable.

USER:
You are redesigning ONE role's task list to make the postholder more
productive, now that AI tools can take on more of the routine work.

<job_description>{job_description}</job_description>

<current_tasks>
{numbered_tasks}
</current_tasks>

YOUR JOB:
Decide whether removing any of this role's current tasks -- and replacing each
removed task with exactly one new task -- would make the role MORE PRODUCTIVE,
given that AI can now handle more of the routine work described above. You are
not given any automation-potential estimate for these tasks -- decide purely
on your own judgement of what this role most needs, given what current AI
tools are generally capable of.

CRITICAL RULES:
1. Optimise for PRODUCTIVITY ONLY. Do not choose new tasks because they look
   "less automatable" or "more human" -- we will independently measure how
   automatable your new tasks are afterwards, in a separate process you cannot
   see; your job here is only to make the role better.
2. n (the number of tasks you remove) MAY BE ZERO, and often should be. Most
   roles need little or no redesign. Removing tasks is not rewarded -- you are
   not required to find something to change. Only remove a task if removal is
   clearly justified: e.g. its substantive content is now handled by AI and
   its small residual human element can be absorbed elsewhere without harming
   the role.
3. Each removed task must be replaced by EXACTLY ONE new task (headcount and
   scope are conserved -- this is a swap, not a cut). Do not remove a task
   unless you have a genuinely good replacement for it.
4. Each new task must be a CONCRETE, role-appropriate action -- you should be
   able to answer "yes" if asked "does the postholder literally DO this on a
   working day?" Not a vague aspiration, skill, or accountability statement
   (not "provide strategic leadership"; instead, a specific thing this person
   now spends time on).
5. New tasks must fit the seniority and grade this job description implies. Do
   not invent tasks that belong to a different, more senior (or more junior)
   role than the one described.

OUTPUT INSTRUCTIONS:
List the task_number of every task you remove in `removed_task_numbers` (this
list may be empty).

For each removal, add ONE entry to `new_tasks` with:
- replaces_task_number: the task_number it replaces. The set of
  replaces_task_number values must exactly match `removed_task_numbers`,
  one-to-one.
- reasoning: one sentence, decided BEFORE new_task_text -- why this task
  raises the role's productivity now that AI handles more of the routine work,
  and why it fits this role's seniority/grade.
- new_task_text: the concrete new task, informed by the reasoning above.

If no redesign is justified, return an empty list for both fields.
\end{lstlisting}

\subsection{Task-insertion ranking prompt}\label{sec:AppendixTaskInsertionPrompt}

This prompt places each new task within the role's retained tasks, which are shown in their real advert order. The model returns, for each new task, one sentence of reasoning and the number of the existing task it should follow (0 if it takes more time than all of them). Each new task's position in the merged order is converted to a rank fraction, (rank $-$ 1)/(n $-$ 1), where n is the number of tasks in the merged list. Ties between new tasks with the same placement are broken by the order in which they are returned.

\begin{lstlisting}[language={}]
SYSTEM:
You are an expert on UK Civil Service work. Your job is to judge where NEW
tasks fit, by time priority, within a role's EXISTING task list.

USER:
Below is a role's EXISTING task list, already ordered from the task that takes
up the MOST of the postholder's working time to the task that takes up the
LEAST. Treat this order as fixed and correct -- do not reorder it, and do not
second-guess it.

<job_description>{job_description}</job_description>

<existing_tasks_in_time_priority_order>
{existing_tasks}
</existing_tasks_in_time_priority_order>

One or more NEW tasks have been added to this role, and are not part of the
ordered list above. Your job is to decide where EACH new task fits into that
existing time-priority order -- based only on your own judgement of how much of
the postholder's time each new task is likely to take, relative to the
existing tasks shown.

<new_tasks>
{new_tasks}
</new_tasks>

OUTPUT INSTRUCTIONS:
For each new task listed above, add ONE entry to `placements` with:
- new_task_ref: the "New-X" label for this task.
- reasoning: one sentence, decided BEFORE insert_after -- how much of the
  postholder's time this task is likely to take relative to the existing list.
- insert_after: the NUMBER of the existing task this new task's time priority
  sits immediately after (use 0 if this new task takes up MORE time than every
  existing task, i.e. it belongs before all of them).
\end{lstlisting}

\subsection{Job description cleaning prompt}\label{sec:AppendixCleanJDPrompt}

Between Stage 1 and Stage 2 of the job redesign, this prompt removes any text describing the replaced tasks from the job description, so that the Stage 2 model cannot propose something close enough to undo a Stage 1 removal. The model returns one or two sentences of reasoning and the edited job description.

\begin{lstlisting}[language={}]
SYSTEM:
You are an expert editor of UK Civil Service job adverts. Your job is to
remove references to ONE specific task from a job description, without
changing anything else about it.

USER:
A role's task list has just been redesigned: the task(s) below have been
REMOVED from this role because AI now handles that work.

<removed_tasks>
{removed_tasks}
</removed_tasks>

Here is the job description this role was originally advertised with, which
may still describe this now-removed work in its own prose:

<job_description>{job_description}</job_description>

YOUR JOB:
Identify which sentences or phrases in the job description describe the
removed task(s) above, then return the job description with ONLY those parts
removed.

CRITICAL RULES:
1. Remove ONLY content that describes the removed task(s). Leave every other
   sentence, duty, and detail in the job description completely unchanged --
   same wording, same order.
2. Do NOT add, summarise, or rephrase anything. This is a deletion pass, not a
   rewrite -- if a sentence covers the removed task AND something else, keep
   the other-topic sentence but drop or edit only the removed-task portion of
   it, no more.
3. If nothing in the job description clearly corresponds to the removed
   task(s) (some tasks are extracted/generalised and don't map to a specific
   sentence), return the job description completely unchanged rather than
   guessing at a cut.
4. The result must still read as a coherent, natural job description -- fix
   simple stitching issues (stray connectives, empty bullet points) left
   behind by a removed sentence, but do not restructure anything beyond that.

OUTPUT INSTRUCTIONS:
- reasoning: one or two sentences -- which part(s) of the job description
  describe the removed task(s), or a note that nothing corresponds to them,
  decided BEFORE you edit.
- cleaned_job_description: the full job description text with only the
  identified parts removed (or unchanged, per rule 3).
\end{lstlisting}

\subsection{Stage 2 task generation prompt}\label{sec:AppendixBackfillPrompt}

This prompt generates the new tasks in Stage 2 of the job redesign, and is used in both the main redesign and the variant. The model sees the cleaned job description, the role's current task list (the retained tasks plus the Stage 1 replacements) and the hours freed by Stage 1. For each new task it returns one sentence of reasoning, the task, and an estimated number of hours per week (at least 1.0). An empty list is allowed. The variant adds one further rule, shown after the listing, and is otherwise identical.

\begin{lstlisting}[language={}]
SYSTEM:
You are an expert on UK Civil Service work, advising on how individual roles
should evolve as AI tools become more capable. Your job is to propose
genuinely valuable new work for a postholder whose role has freed up time, not
to pad the role with filler.

USER:
This role has recently been redesigned so that AI tools now handle more of the
routine work. Based on that redesign, an estimated {hours_freed} hours per week
of this postholder's time is now free and not accounted for by any task
currently on their list.

<job_description>{cleaned_job_description}</job_description>

<current_tasks>
{numbered_tasks}
</current_tasks>

YOUR JOB:
Propose NEW tasks that make good use of this newly freed time -- work that
raises this role's value and fits its seniority and profession, not filler
invented just to look busy.

CRITICAL RULES:
1. Each new task must be CONCRETE -- you should be able to answer "yes" if
   asked "does the postholder literally DO this on a working day?" Not a vague
   aspiration, skill, or accountability statement.
2. Each new task must represent AT LEAST 1 hour of the postholder's time per
   week. Do not propose small, trivial, or filler tasks just to hit a number
   -- a shorter list of substantive tasks is better than a long list of thin
   ones.
3. New tasks must fit the seniority and grade this job description implies,
   and must not duplicate or overlap with anything already in <current_tasks>.
4. There is no fixed number of tasks to propose -- propose as many genuinely
   valuable tasks as you can concretely justify, and no more. It is fine to
   propose few tasks, or even none, if you cannot think of enough good uses
   for this time -- do not invent weak tasks just to use up all {hours_freed}
   hours.
5. Your estimated hours across all new tasks combined should not greatly
   exceed the {hours_freed} hours available -- a modest undershoot is fine and
   expected; a large overshoot is not.

OUTPUT INSTRUCTIONS:
For each new task, add ONE entry to `new_tasks` with:
- reasoning: one sentence, decided BEFORE new_task_text -- why this task is a
  good use of this role's newly freed time and fits this role's
  seniority/grade.
- new_task_text: the concrete new task, informed by the reasoning above.
- estimated_hours_per_week: your best estimate of how many hours per week this
  task will take (at least 1.0), informed by new_task_text.

If no new task is justified, return an empty list.
\end{lstlisting}

The variant adds the following as rule 6, directly after rule 5.

\begin{lstlisting}[language={}]
6. Each new task must also be one that an AI system could not already do well
   on its own. Assume the AI has access to the same documents, data and tools
   as the postholder. Propose only tasks that raise this role's value AND that
   a current large language model could not already carry out without the
   postholder.
\end{lstlisting}

\subsection{Task category classification prompt}\label{sec:AppendixTaskCategoryPrompt}

This prompt assigns every original task and every new task from the job redesign (Stage 1 and Stage 2, in the main redesign and the variant) to a category of work (Appendix~\ref{sec:AppendixTaskCategories}). The same prompt is used for all tasks, it contains only the category definitions and the task text, and tasks from all groups are shuffled together within each request, 20 to a request. For each task the model returns the task number and the number of the best-fitting category, with 0 for a task that fits none of them.

\begin{lstlisting}[language={}]
SYSTEM:
You classify tasks from UK Civil Service job adverts into categories of work.

USER:
Classify each task below into the one category that best describes the main work it
involves. Judge each task on its own wording.

<categories>
1. Administration: filing, data entry, record keeping, post, diaries, booking,
   minute taking, processing purchase orders, and routine upkeep of systems,
   equipment and supplies
2. Enquiries: answering questions and requests from the public, customers or
   colleagues by phone, email, webchat or in person, including correspondence, FOI
   and parliamentary questions, and taking payments
3. Casework: assessing and deciding individual cases, claims or applications,
   carrying out interviews and checks on individuals, and front-line operational
   duties such as enforcement, searches or supervising people
4. Stakeholder engagement: building and keeping up relationships with partners,
   colleagues and external bodies, liaising, acting as a point of contact, and
   representing the team
5. Analysis and reporting: analysing data and information, producing reports,
   statistics, intelligence and briefings, and monitoring performance
6. Advice: giving expert advice or guidance to managers, staff, customers or
   ministers on policy, legal, technical, financial or operational matters
7. Planning and delivery: planning, managing and delivering projects, programmes and
   strategies, including milestones, budgets and benefits, and leading on
   initiatives
8. People management: line managing staff, managing performance, recruitment,
   induction, training, coaching and development
9. Assurance: audits, quality checks, compliance and policy checks, health and
   safety, and identifying and escalating risk
0. Other: does not fit any of the categories above
</categories>

<tasks>
{numbered_tasks}
</tasks>

Return a label for every task.
\end{lstlisting}

\subsection{Team restructuring: step 1 role choice prompt}\label{sec:AppendixTeamStage1Prompt}

In Step 1 of the team restructuring exercise (Appendix~\ref{sec:AppendixTeamRestructure}), this prompt asks the model to choose the one role in a team that loses one post and to identify the tasks of that role that cannot be automated away. The model sees every role in the team with its grade, number of posts, job description and tasks with hours per week for one post-holder, but not our automation scores. It returns the chosen role, a short rationale, and, for each protected task, a one-sentence reason. Each team is run once.

\begin{lstlisting}[language={}]
SYSTEM:
You are advising on how a UK Civil Service team should evolve as AI tools
become more capable. You are given a whole team: every role, its job
description, how many posts (people) currently hold it, and its task list
with hours/week per post-holder. Automation-potential scores are
deliberately withheld from you.

YOUR JOB, IN TWO PARTS:
1. Choose EXACTLY ONE role in this team to lose exactly ONE post (a whole
post, not a fraction). All else equal, prefer the role whose tasks look
most amenable to being done by current AI systems, based on what you read
in its task list and job description -- you are not given this project's
own automation-potential scores, so form this judgement yourself from the
material above. You may also weigh other substantive reasons (e.g.
redundancy, consolidation, the temporary or seasonal nature of a post) if
they justify the cut more strongly than automatability alone.
2. For THAT CHOSEN ROLE ONLY, go through its own task list and identify
which tasks CANNOT be automated away (these will be protected from removal
later). A task CANNOT be automated away if it still needs a human to do it:
it requires a human to physically be present, requires a specific person's
accountability or legal authority, requires trust built through an ongoing
human relationship, or requires judgement on a genuinely contested value
where there is no single correct answer. A task CAN be automated away if a
current AI system could substantially complete it without a human doing the
substantive work themselves. It is fine for this list to include all, some,
or none of the role's tasks.

Do not evaluate any other role's tasks here.

USER:
Team: {team} ({department}). Roles in this team:

For each role:
{role_title} ({grade}, role_id={role_id}, {n_posts} post(s)):
  Job description: {job_description}
  task_number={n}: [~{hours} hrs/wk per post-holder] {task}
  (one line per task)

OUTPUT INSTRUCTIONS:
- chosen_role_id: the role_id of the ONE role losing exactly one post.
- rationale: why this role, and not another, loses a post.
- set_aside: a list, possibly empty, with one entry per task of the chosen
role that cannot be automated away. Each entry has role_id (the chosen
role), task_number, and reasoning (one sentence, decided before concluding:
why this task cannot be automated away).
\end{lstlisting}

\subsection{Team restructuring: step 2 automation triage prompt}\label{sec:AppendixTeamStage2Prompt}

In Step 2, this prompt asks the model to decide which of the remaining tasks in the team can be automated away. The tasks it sees are those of every role, excluding the tasks protected in Step 1 and the share of the chosen role's other work that leaves with the departing post. Our automation scores are again withheld. The model lists only the tasks it removes, each with a one-sentence reason, and tasks it does not list stay in the team. Each team is run once.

\begin{lstlisting}[language={}]
SYSTEM:
You are triaging a UK Civil Service team's tasks for automation, after one
role has already been chosen to lose a post (already decided, not your job
here). You are given every task still needing attention across the team,
EXCLUDING the tasks already set aside as un-automatable and excluding the
one post's worth of work already being removed with the chosen role.
Automation-potential scores are withheld.

For each task, decide whether it CAN be automated away. A task CANNOT be
automated away if it still needs a human to do it: it requires a human to
physically be present, requires a specific person's accountability or legal
authority, requires trust built through an ongoing human relationship, or
requires judgement on a genuinely contested value where there is no single
correct answer. A task CAN be automated away if a current AI system could
substantially complete it without a human doing the substantive work
themselves.

List ONLY the tasks you are removing (anything not listed stays unchanged).
It is normal, and often correct, to remove few or none -- only remove a task
when removal is clearly justified.

USER:
Team: {team} ({department}). Tasks to triage, by role:

For each role:
{role_title} ({grade}, role_id={role_id}):
  task_number={n}: [~{hours} hrs/wk per post-holder] {task}
  (one line per task)

OUTPUT INSTRUCTIONS:
- removed: a list, possibly empty, with one entry per task removed. Each
entry has role_id, task_number, and reasoning (one sentence, decided before
concluding: why this task can be automated away).
\end{lstlisting}

\subsection{Team restructuring: step 3 regrouping prompt}\label{sec:AppendixTeamStage3Prompt}

In Step 3, this prompt asks the model to regroup the work that still needs a human into a new set of roles. The model sees every remaining task (including the protected tasks from Step 1) with its original role and its hours per week across the whole team, and is told the exact number of posts the new roles must sum to. It returns a short rationale for the headcount and, for each new role, a title, grade, number of posts, the original tasks assigned to it, and a rationale. Where a new role's workload exceeds the capacity of its posts, or the posts do not sum to the required total, the same prompt is sent once more with the feedback shown at the end of the prompt appended. In the prompt below, {total\_posts} is the team's original number of posts minus one.

\begin{lstlisting}[language={}]
SYSTEM:
You are redesigning a UK Civil Service team's roles around its remaining
workload, after some tasks have been removed/automated. You are given every
task that STILL NEEDS A HUMAN, with hours/week ACROSS THE WHOLE TEAM
(already multiplied by headcount). Group this into a new set of roles, each
with a title, grade, and an INTEGER number of posts. If an existing role's
shape (its title, grade, and task mix) already suits the remaining workload
well, you may keep it essentially unchanged rather than inventing something
new -- only redesign a role where redesign is actually warranted. The posts
across ALL your new roles MUST SUM TO EXACTLY {total_posts} -- the team's new
total headcount (one less post than before). Each role's workload must
realistically fit its posts: a role with P posts has 37*P hours/week of
capacity, and must not be given materially more work than that. Every
remaining task must end up in exactly one new role. You are not told the
posts are correct in advance -- get the sum exactly right.

USER:
Team: {team} ({department}). Remaining workload tasks, by original role:

For each original role:
Originally '{role_title}' ({grade}, role_id={role_id}):
  task_number={n}: [~{hours} hrs/wk team-wide] {task}
  (one line per task)

OUTPUT INSTRUCTIONS:
- headcount_rationale: how the posts were allocated across the new roles.
- new_roles: a list of new roles. Each has new_role_id (a short unique id
such as R1), title, grade (one of AA, AO, EO, HEO, SEO, G7, G6, SCS),
posts (an integer of at least 1), task_assignments (a list of role_id and
task_number pairs, at least one), and rationale.

FEEDBACK ADDED ON RETRY (only if the first plan breaks a constraint):
Your previous attempt had problems -- please fix and resubmit a COMPLETE
new plan:
- {title} ({new_role_id}): {workload}h workload > {capacity}h capacity
({posts} posts)
- Total posts across your new roles was {n}, must be exactly {total_posts}.
\end{lstlisting}

\subsection{Team restructuring: step 4 backfill prompt}\label{sec:AppendixTeamStage4Prompt}

In Step 4, this prompt is sent once for each new role with spare capacity, asking the model to propose human-only work to fill it. The model sees the role's title, grade and posts, its spare hours per post-holder, the tasks that have been automated in the team (so that it does not re-create them under a new label) and the role's existing task list. For each addition it returns a description, one of ten reasons a human is needed, a one- or two-sentence explanation, and weekly hours per post-holder, and an addition can instead expand the hours on a numbered existing task. Additions are capped at the role's spare capacity after the model responds. Afterwards, a second model (GPT-OSS-120B) scores every task in the role, retained and new alike and without labels, using the corpus scoring prompt in Appendix~\ref{sec:AppendixScorePrompt} with the system prompt shown at the end of this section.

\begin{lstlisting}[language={}]
SYSTEM:
You are adding new work to a just-redesigned UK Civil Service team role. It
has spare capacity (slack hours) because of task removals elsewhere in the
team. Propose new tasks that make the role MORE PRODUCTIVE. Every new task
must be CONCRETE (the postholder literally does this on a working day), at
least 1 hour/week per post-holder (no filler), must not duplicate the
existing task list, and must be tagged with exactly one of ten reasons a
human is needed rather than AI:
- accountability_or_legal_authority (a named person must take legal/formal
responsibility)
- physical_presence
- trust_and_relationships
- judgement_on_contested_values
- verification_and_exception_handling (checking AI output for errors and
handling cases it cannot)
- tacit_or_contextual_knowledge (local/organisational/domain knowledge not
available to AI)
- ethics_safety_duty_of_care
- democratic_or_political_legitimacy (decisions that must be visibly owned
by a person for public/political acceptability)
- creativity_or_novel_problem_solving
- coordination_or_negotiation (brokering between people or bodies)

CRITICAL: write each task's description as JUST the human-only sliver -- the
specific act a human must perform, be accountable for, be present for, or
hold the relationship for -- not the full end-to-end process including
AI-automatable substance. For example, write 'Reviews the AI-drafted
determination and formally approves or rejects it', NOT 'Determines
eligibility for X'. Size estimated_hours_per_week to reflect only the
human's own time on that sliver (review/sign-off/presence/relationship
time), not the full task's end-to-end execution time. It is fine, and often
correct, to propose nothing. Do not re-propose, under a new label, any task
already removed/automated for this team -- it will be independently scored
afterwards for its automation potential.

As well as proposing brand-new tasks, you may instead choose to EXPAND the
hours on an EXISTING task already on this role's task list, if deepening
that existing responsibility (e.g. more review/sign-off/presence time on
something already happening) is a more productive use of the spare capacity
than inventing a new discrete task. To do this, set
expands_existing_task_number to that task's number from the list below, and
set estimated_hours_per_week to the ADDITIONAL hours/week only, not the
task's new total.

USER:
{role_title} ({grade}), {posts} post(s), ~{slack_per_post} hrs/wk spare
capacity PER POST-HOLDER.

Already automated -- do not re-create any of this under a new label:
  - {automated_task}
  (one line per automated task in the team, if any)

Existing task list (team-wide hrs/wk; numbered so you can reference one to
expand its hours):
  [{i}] [~{hours} hrs/wk] {task}
  (one line per task)

OUTPUT INSTRUCTIONS:
- new_tasks: a list, possibly empty. Each entry has description (just the
human-only sliver), reason (one of the ten tags above), reason_explanation
(one or two sentences applying the reason), estimated_hours_per_week (at
least 1.0, per post-holder; the additional hours if expanding), and
expands_existing_task_number (empty for a new task, otherwise the number of
the existing task being expanded).

SYSTEM PROMPT FOR THE FINAL SCORING CALL (user prompt as in the corpus
scoring prompt):
You are an expert on UK Civil Service work. You are given a job description
and a fixed, pre-extracted list of a role's tasks. Score each task for its
automation potential with current AI systems, exactly as the rest of this
project's task corpus was scored. Do not extract, add, or remove tasks.
\end{lstlisting}
 \clearpage
\section{Prompt ablation study}\label{sec:AppendixPromptAblation}

\subsection{Baseline prompt}\label{sec:AppendixPromptAblationPrompt}

\begin{lstlisting}[language={}]
def task_scoring_prompt(job_description: str, tasks: list[str]) -> str:
    """Score extracted tasks for AI exposure."""
    task_list = "".join(f"- {t}" for t in tasks)
    return f"""
    Score each of the following tasks for AI exposure using the provided format.
    The tasks were extracted from the job description below, which is provided 
    for context only - do not extract new tasks.

    SCORING RUBRIC (exposure_score):
    For each task, assess the potential for current AI systems (LLMs, multimodal 
    models, AI-powered software, machine learning) to reduce the time required to 
    complete the task without a loss in quality. Score on a continuous 0.0-1.0 scale 
    using these anchors:

    - 0.0-0.2 (No/Low Exposure): The task is physical, requires deep empathy, 
      complex negotiation, or high-stakes moral judgement. AI cannot meaningfully 
      reduce time spent. E.g., conducting a sensitive staff welfare conversation.

    - 0.2-0.4 (Minor Augmentation): AI assists with peripheral sub-components 
      (retrieval, templates, spell-checking) but the core work requires human 
      judgement. Time savings are modest. E.g., chairing a cross-departmental 
      working group.

    - 0.4-0.6 (Moderate Augmentation): AI can draft components or process 
      underlying data, but significant human input is needed to contextualise, 
      interpret, or adapt outputs. Time savings are notable but the human 
      remains the primary driver. E.g., writing a policy briefing on a novel topic.

    - 0.6-0.8 (High Augmentation): AI can draft initial outputs, process data, 
      or handle routine interactions. The human shifts from creator to reviewer. 
      Most of the time spent on this task can be saved. E.g., summarising 
      consultation responses.

    - 0.8-1.0 (Near/Full Automation): The task is routine information processing, 
      standard text generation, or data entry. AI can perform it near-autonomously. 
      Almost all time spent can be saved. E.g., converting a spreadsheet into a 
      formatted table.

    REASONING INSTRUCTION:
    For each task, first reason about how it maps to the thresholds above, then 
    assign the score. Provide your reasoning as exactly one sentence.

    <job_description>{job_description}</job_description>

    <tasks>
    {task_list}
    </tasks>
    """
\end{lstlisting}

\subsection{Sensitivity variants summary}\label{sec:AppendixPromptAblationTable}
\begin{table}[!htb]
    \centering
    \small
    \begin{threeparttable}
    \caption{Prompt sensitivity variants}
    \label{tab:PromptVariants}
    \begin{tabularx}{\linewidth}{@{}p{3.2cm}X>{\centering\arraybackslash}p{1.6cm}>{\centering\arraybackslash}p{1.6cm}@{}}
        \toprule
        Variant & What changes & Within $\pm$0.1 & Noise floor \\
        \midrule
        \multicolumn{4}{@{}l}{\textbf{Baseline and control}} \\
        0. Baseline & Initial scoring prompt as defined in Section \ref{sec:AppendixPromptAblationPrompt} above to provide a baseline score to compare to & --- & --- \\
        1. Control rerun & An identical copy of the baseline to establish the noise floor. Any variant whose divergence is comparable to the control rerun's is within stochastic variation and should not be interpreted as meaningful. & 94.6\% & 1.0$\times$ \\
        \addlinespace
        \multicolumn{4}{@{}l}{\textbf{Essential components}} \\
        2. No numeric bands & Keep the descriptions and examples, remove the numeric anchors of 0.0--0.2 / 0.2--0.4 / \dots\ & 85.6\% & 27.2$\times$ \\
        3. No reasoning & Remove the instruction to first provide reasoning before assigning the exposure score & 89.7\% & 12.4$\times$ \\
        4. No rubric & Remove the entire scoring rubric; keep only ``Score on a scale of 0.0--1.0'' & 83.5\% & 9.5$\times$ \\
        \addlinespace
        \multicolumn{4}{@{}l}{\textbf{Moderate impact}} \\
        5. No examples & Remove the worked examples (e.g.\ ``conducting a sensitive staff welfare conversation'') but keep band descriptions and numeric anchors & 93.3\% & 8.3$\times$ \\
        6. Three bands & Collapse to three bands: Low (0.0--0.3), Medium (0.3--0.7), High (0.7--1.0) & 92.3\% & 6.4$\times$ \\
        \addlinespace
        \multicolumn{4}{@{}l}{\textbf{Near noise floor}} \\
        7. No band descriptions & Keep the numeric bands and examples, remove the descriptive text & 91.9\% & 4.8$\times$ \\
        8. No exposure type & Remove the augmentation/automation field & 94.6\% & 4.5$\times$ \\
        9. System prompt persona & Add a system prompt persona ``You are an expert occupational analyst\dots'' & 94.3\% & 4.2$\times$ \\
        10. Percentage thresholds & Replace qualitative language (modest, notable, most, almost all) with explicit percentages of the potential time savings (10--30\%, 30--50\%, 50\%+, 80\%+) & 94.2\% & 2.9$\times$ \\
        \bottomrule
    \end{tabularx}
    \begin{tablenotes}[flushleft]
        \footnotesize
        \item \textit{Note.} Variants are grouped by the size of their effect and ordered within each group by noise floor multiple. Within $\pm$0.1 reports the share of tasks scored within 0.1 of the baseline (Variant 0) score. Noise floor expresses each variant's distributional shift as a multiple of the control rerun (Variant 1), which sets the run-to-run noise baseline.
    \end{tablenotes}
    \end{threeparttable}
\end{table}

We applied all essential components to our final prompt for task extraction and scoring (see Appendix \ref{sec:AppendixExtractAndScorePrompt}). For the moderately important components, we removed specific tasks as examples to avoid anchoring too heavily, but included types of work in the rubric instead (e.g. ``synthesizing evidence''). We also reduced the number of bands from 21 to 11 (intervals of 0.1 rather than 0.05) to allow the model adequate scoring flexibility without spurious precision. For the least important components, we decided to keep the system prompt given its low cost and overhead and remove the others. \clearpage
\section{Task extraction sensitivity}\label{sec:AppendixExtraction}

\begin{table}[!htbp]
    \centering
    \small
    \begin{threeparttable}
    \caption{Job description with clearly defined tasks.}
    \label{tab:task_extract_defined}
    \begin{tabular}{@{}>{\raggedright\arraybackslash}p{0.48\linewidth}>{\raggedright\arraybackslash}p{0.48\linewidth}@{}}
        \toprule
        \multicolumn{2}{@{}p{\dimexpr\linewidth\relax}@{}}{
        The role of the PMO Support Officer covers a diverse range of activities to support the delivery of the programme objectives. The PMO Support Officer enables the smooth running of the programme by supporting the PMO Manager through the operation of project/ programme management processes, and the co-ordination of business management actions and activities on their behalf. This includes:

        •	Scheduling meetings; taking accurate minutes with clearly attributable actions, monitoring and prompting the timely completion of actions.  
        •	Drafting skeleton packs for governance board and liaising across the team to ensure team members complete sections attributed to them on time and to quality.  
        •	Support in maintaining programme controls and in producing project reports.  
        •	Making sure programme documents such as delivery plans and risk and issue registers are regularly updated and used across the programme to support delivery.  
        •	Maintaining the stakeholder log for internal and external stakeholders.  
        •	Programme file management using robust version control.  
        •	Support the monitoring of programme spend and contribute to the compilation of budgets.
        } \\
        \midrule
        LLM inferred tasks: & Manually inferred tasks: \\
        \midrule
        Scheduling meetings and taking accurate minutes with clearly attributable actions, monitoring, and prompting the timely completion of actions. 
        & 
        Scheduling meetings; taking accurate minutes with clearly attributable actions, monitoring and prompting the timely completion of actions. 
        \\
        \addlinespace
        Drafting skeleton packs for governance board and coordinating with team members to ensure timely and quality completion of their sections. 
        & 
        Drafting skeleton packs for governance board and liaising across the team to ensure team members complete sections attributed to them on time and to quality. 
        \\
        \addlinespace
        Supporting the maintenance of programme controls and producing project reports. 
        & 
        Support in maintaining programme controls and in producing project reports. 
        \\
        \addlinespace
        Ensuring regular updates and usage of programme documents such as delivery plans and risk and issue registers to support delivery. 
        & 
        Making sure programme documents such as delivery plans and risk and issue registers are regularly updated and used across the programme to support delivery. 
        \\
        \addlinespace
        Maintaining the stakeholder log for internal and external stakeholders. 
        & 
        Maintaining the stakeholder log for internal and external stakeholders. 
        \\
        \addlinespace
        Managing programme files using robust version control. 
        & 
        Programme file management using robust version control. 
        \\
        \addlinespace
        Supporting the monitoring of programme spend and contributing to the compilation of budgets. 
        & 
        Support the monitoring of programme spend and contribute to the compilation of budgets. 
        \\
        \bottomrule
    \end{tabular}
    \end{threeparttable}
\end{table}

\begin{table}[!htbp]
    \centering
    \small
    \begin{threeparttable}
    \caption{Job description with generally defined tasks.}
    \label{tab:task_extract_undefined}
    \begin{tabular}{@{}>{\raggedright\arraybackslash}p{0.48\linewidth}>{\raggedright\arraybackslash}p{0.48\linewidth}@{}}
        \toprule
        \multicolumn{2}{@{}p{\dimexpr\linewidth\relax}@{}}{
        Within the directorate, the vision is to build a community of highly skilled and motivated project delivery professionals to establish and transform our Test and Trace services. Part of this will include establishing a pool of talented project delivery professionals who will be forward-deployed across the change portfolio based on the demands of Test and Trace/UKHSA. Members of the pool will be assigned to projects and programmes to manage, support, guide, direct and intervene at programme and project levels where necessary. Flexibility will be key as we will seek to retain and redeploy resources as demand shifts in line with our portfolio.

        Senior Project Manager: The role of the Senior Project Manager is to lead / manage the project and the project team on a day-to-day basis. The Senior Project Manager is responsible for driving and overseeing the delivery of the project to ensure that the objectives are clearly defined and achieved within the agreed time, cost and quality constraints. The Senior Project Manager has a key role in project governance and working with stakeholders, to ensure the agreed project outputs are delivered to enable benefits to be realised.
        } \\
        \midrule
        LLM inferred tasks: & Manually inferred tasks: \\
        \midrule
        Lead and manage the project team on a daily basis. 
        & 
        \textit{No matching task extracted} 
        \\
        \addlinespace
        Drive and oversee the delivery of the project to ensure objectives are met within agreed time, cost, and quality constraints. 
        & 
        The Senior Project Manager is responsible for driving and overseeing the delivery of the project to ensure that the objectives are clearly defined and achieved within the agreed time, cost and quality constraints. 
        \\
        \addlinespace
        Play a key role in project governance. 
        & 
        \textit{No matching task extracted} 
        \\
        \addlinespace
        Work with stakeholders to ensure project outputs are delivered, enabling benefits realization. 
        & 
        The Senior Project Manager has a key role in project governance and working with stakeholders, to ensure the agreed project outputs are delivered to enable benefits to be realised. 
        \\
        \bottomrule
    \end{tabular}
    \begin{tablenotes}[flushleft]
        \footnotesize
        \item \textit{Note.} Job descriptions have been manually selected by the authors to represent the contrasting quality and clarity of task information in job descriptions in our sample dataset. For the manually inferred tasks, an analyst was given the full job description text and asked to extract each unique task.
    \end{tablenotes}
    \end{threeparttable}
\end{table} \clearpage
\section{Sample reweighting through raking}\label{app:Raking}

Next, we performed a process of iterative proportional fitting (IPF). Given that each vacancy \( i \) in our sample is characterized by department \( d_i \in D \), grade \( g_i \in G \) and profession \( p_i \in P \), we defined the initial weight \( w_i^{(0)} \) of each vacancy and the known FTE totals over the estimation population\footnote{$D, G, P$ are described in Section \ref{section:DatasetDesc} and listed in Tables \ref{tab:department_coverage}, \ref{tab:grade_coverage}, \ref{tab:profession_coverage} respectively.} for each department, grade, and profession as:
\[
w_i^{(0)} = \text{Number of roles available}_{i}, \qquad
N_d \; (d \in D), \quad
N_g \; (g \in G), \quad
N_p \; (p \in P).
\]
We then converted known totals into target marginal proportions, and compare these against the current weighted proportion of each category in the sample:
\[
\pi_c = \frac{N_c}{N}, \qquad
s_c(w) = \frac{\sum_{i : c_i = c} w_i}{\sum_i w_i}
\]
where \( c \in \{d, g, p\} \) indicates the dimension being adjusted (department, grade, or profession), and \( c_i \) denotes the category membership of vacancy \( i \) on that dimension. At each iteration \( k \), the IPF procedure updated the weight by multiplying the previous weight by the ratio of the target marginal proportion to the current sample proportion for the department, grade, or profession:
\[
w_i^{(k+1)}
\;\leftarrow\;
w_i^{(k)}
\cdot
\Biggl(
  \frac{\pi_{\,c_i}}{\,s_{\,c_i}(w^{(k)})}
\Biggr),
\qquad
c \in \{d, g, p\},
\]
The weights were iteratively updated once per dimension, in sequence: department, grade, then profession. The process continued until convergence, defined as:
\[
\max_i
\Biggl|\,
\frac{w_i^{(k+1)} - w_i^{(k)}}{w_i^{(k)}}
\Biggr|
< 10^{-6},
\]
or until 50 iterations had been completed. After convergence, the resulting weights \( w_i^{(*)} \) sum to an arbitrary constant, so we rescale them to the estimation population's total published FTE \( N \):
\[
w_i = w_i^{(*)} \cdot \frac{N}{\sum_i w_i^{(*)}}, \qquad \text{so that} \quad \sum_i w_i = N
\] \clearpage
\section{Aggregating exposure scores}\label{app:Aggregation}

We next derived statistics using exposure scores and weights. Let $i = 1, \ldots, N$ index vacancies, $j = 1, \ldots, n_i$ index the tasks extracted from vacancy $i$ (vacancies vary in how many tasks they contain), and $k=1, \ldots K$ index the foundation models. Then, we denote task $j$ of vacancy $i$ scored by model $k$ as $x_{ijk} \in \mathcal{B}, \qquad \mathcal{B} = \{0.0,\, 0.1,\, \ldots,\, 1.0\}$. This represents the model's assessment of the share of that task's substantive content performable by current AI systems. Then, the task-level ensemble score is simply the average (mean\footnote{See Appendix \ref{AppendixMeanMedianMode} which demonstrates this choice of average function is immaterial.}) of all model assessments:

$$\bar x_{ij} = \frac{1}{K}\sum_{k=1}^{K} x_{ijk} \quad\text{(task-level ensemble score)}$$

Tasks are extracted from each job description in the order they appear in the advert. We assume this order carries information about time allocation: responsibilities listed first in a vacancy's advert plausibly occupy more of the role than those listed last. Accordingly, each task receives a geometric decay weight based on its position within the vacancy:

\begin{equation}
    \tilde\tau_{ij} = \delta^{\,j-1}, \qquad \tau_{ij} = \frac{\tilde\tau_{ij}}{\sum_{k=1}^{n_i} \tilde\tau_{ik}}, \qquad \text{so that} \quad \sum_{j=1}^{n_i} \tau_{ij} = 1,
    \label{eq:task-decay-weight}
\end{equation}

for every vacancy $i$. The parameter $\delta \in (0,1]$ governs the rate of decay: $\delta = 1$ recovers uniform weighting ($\tau_{ij} = 1/n_i$ for every task, equivalent to assuming task order carries no information about time allocation), while smaller values concentrate weight on earlier-listed tasks. Our primary specification sets $\delta = 0.75$, a non-trivial decay consistent with the intuition that job adverts foreground their more significant duties; we also compute every statistic below $\delta \in \{1.0, 0.9, 0.5\}$ as sensitivity checks, and find our results are not materially affected by this choice.

Thus for vacancy $i$, its mean exposure is given by the task-decay-weighted average of its task-level ensemble scores:

\begin{equation} 
    \mu_i = \sum_{j=1}^{n_i} \tau_{ij}\, \bar x_{ij}
    \label{eq:mean} 
\end{equation} 

and its exposure dispersion is given by its (task-decay-weighted) variance:
\begin{equation} 
    \sigma_{i}^{2} = \sum_{j=1}^{n_i} \tau_{ij}\left(\bar x_{i,j}-\mu_{i}\right)^2
    \label{eq:variance} 
\end{equation}

Finally, for each task $j$ of vacancy $i$, we further define the sample variance of the $K$ individual model scores around that task's own ensemble mean $\bar x_{ij}$:

\begin{equation}
    d_{ij}^2 = \frac{1}{K-1}\sum_{k=1}^{K}\left(x_{ijk}-\bar x_{ij}\right)^2. \label{eq:task-var}
\end{equation}

This can be thought of as a measure of how much the foundation models disagree on a task's automation potential, with a maximum possible standard deviation of $d_{ij}=0.5$ (half the models at each extreme). Normalizing $d_{ij}$ by this bound puts task-level disagreement on the same $[0,1]$ scale as $\mu_i$ and $\sigma_i^2$: $\tilde d_{ij} = \frac{d_{ij}}{0.5} = 2\,d_{ij} \;\in\; [0,1].$

Aggregating to vacancy level using the same task-decay weights $\tau_{ij}$ as $\mu_i$ and $\sigma_i^2$ gives a vacancy-level measure of model disagreement:

\begin{equation}
    D_i = \sum_{j=1}^{n_i} \tau_{ij}\,\tilde d_{ij}. \label{eq:vacancy-disagreement}
\end{equation}

Note that while $\sigma_i^2$ summarizes how heterogeneous a role's own tasks are, $D_i$ focuses on how reliable the exposure score is regardless of which foundation model produced it. \clearpage
\section{Role heterogeneity illustrative examples}\label{sec:AppendixRoleHeterogeneity}

\begin{figure}[htbp]
  \centering
  \begin{subfigure}[b]{0.32\textwidth}
    \centering
    \includegraphics[width=\textwidth]{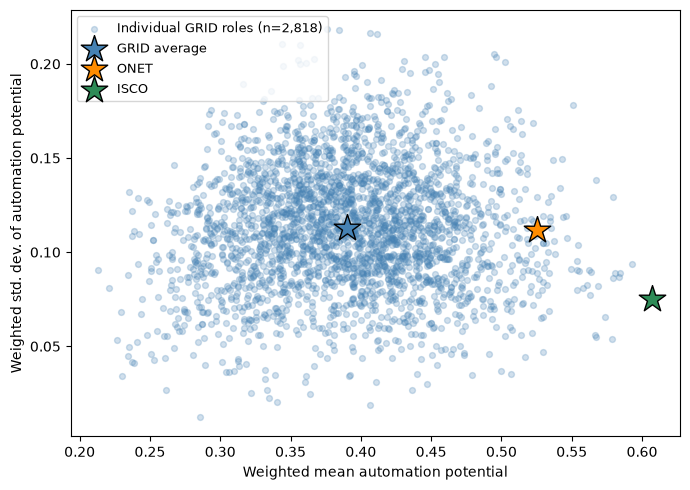}
    \caption{Economists}
    \label{fig:onet-isco-economists}
  \end{subfigure}
  \hfill
  \begin{subfigure}[b]{0.32\textwidth}
    \centering
    \includegraphics[width=\textwidth]{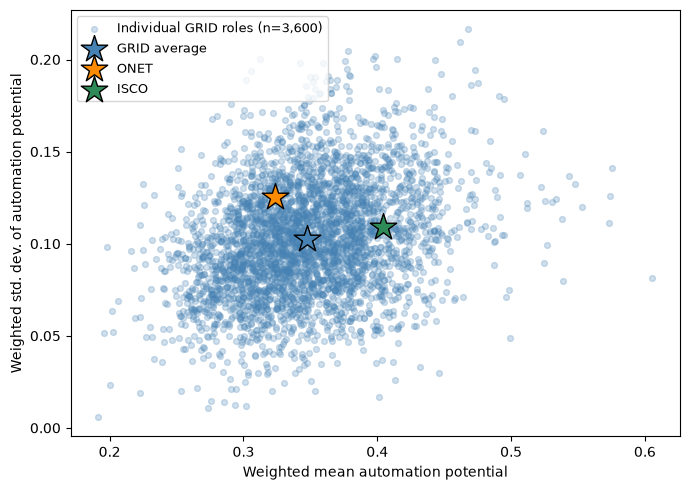}
    \caption{HR Managers}
    \label{fig:onet-isco-hr}
  \end{subfigure}
  \hfill
  \begin{subfigure}[b]{0.32\textwidth}
    \centering
    \includegraphics[width=\textwidth]{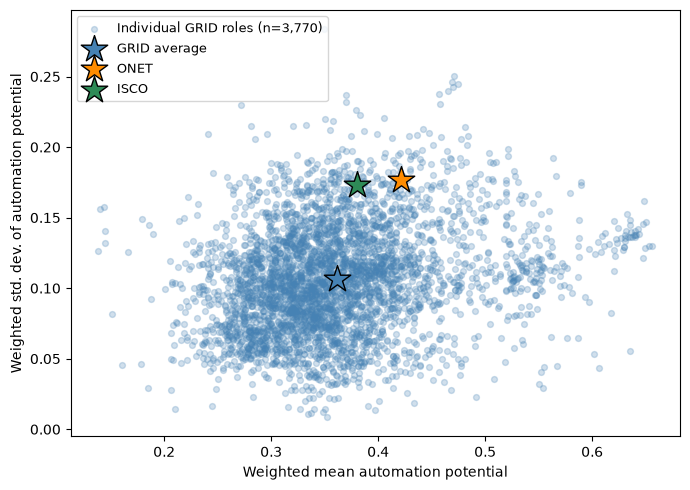}
    \caption{Lawyers}
    \label{fig:onet-isco-lawyers}
  \end{subfigure}
  \caption{GRID roles vs. O*NET vs. ISCO-08, by profession.}
  \label{fig:onet-isco-comparison}
\end{figure} \clearpage
\providecommand{\hd}[1]{\begin{tabular}[b]{@{}c@{}}#1\end{tabular}}

\section{Choice of ensemble aggregation}
\label{AppendixMeanMedianMode}

Each task is scored independently by up to eight foundation models, giving up to eight
automation-potential judgments per task that must be combined into a single score. We
aggregate these judgments using the (unweighted) \textbf{mean} across the available model
scores for each task.

Table~\ref{tab:mean-median-mode} compares this choice against three natural alternatives --
the median, the mode, and the mean rounded to the nearest point on the underlying 0.1-spaced
scoring scale -- computed over all 2,729,344 scored tasks in the corpus. All four measures
produce materially identical estimates of population-level exposure: the raking-weighted
headline exposure figure varies by at most 0.003 (0.3 percentage points) across the four
measures, both overall and within every major profession. Switching to the median does not,
in general, resolve cases where model judgments are split between two extremes, since mean
and median coincide whenever disagreement is symmetric; only the mode differs in that case,
and it is undefined for roughly one in five tasks in our corpus, where models are equally
split between two or more scores. We therefore retain the mean as our primary aggregation
rule, and treat cross-model disagreement as a question of dispersion (Appendix
\ref{AppendixModelConsistencyModelAgreement}), not of which point estimate to report.

\begin{table}[htbp]
    \centering
    \small
    \begin{threeparttable}
    \caption{Population-level exposure under four candidate ensemble aggregation rules.}
    \label{tab:mean-median-mode}
\begin{tabular}{lrrrr}
    \toprule
    Aggregation rule & Task-level & Population exposure & On-bucket & Corr.\ with \\
                      & average (unweighted) & (weight-1d weighted) & tasks (\%) & mean \\
    \midrule
    Mean (primary)               & 0.392 & 0.363 & 12.6  & -- \\
    Median                       & 0.393 & 0.364 & 72.2  & 0.984 \\
    Mode                         & 0.391 & 0.360 & 100.0  & 0.939 \\
    Mean, rounded to nearest 0.1 & 0.392 & 0.363 & 100.0  & 0.983 \\
    \bottomrule
  \end{tabular}
    \end{threeparttable}
\end{table}
\clearpage

\section{Exposure score distributions}\label{AppendixExposureScoreDistributions}

\begin{figure}[!htb]
    \centering
    \includegraphics[width=1\linewidth]{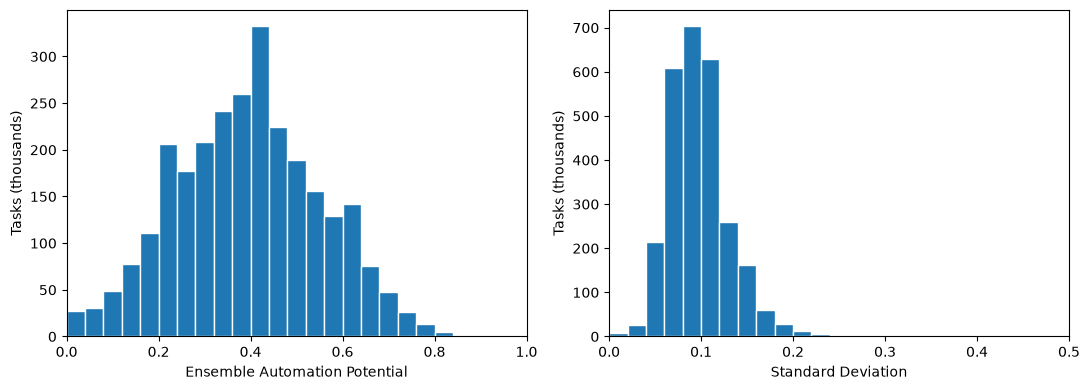}
    \caption{Ensemble Exposure Score Moment Distributions}
    \label{fig:EnsembleExposureScoreMomentDist}
\end{figure}
\clearpage

\section{Model consistency}\label{AppendixModelConsistency}

\subsection{Summary statistics}\label{AppendixModelConsistencySummaryStats}

\begin{table}[!htbp]
    \centering
    \small
    \begin{threeparttable}
    \caption{Per-model descriptive statistics of task-level automation scores.}
    \label{tab:cross-model-desc}
\begin{tabular}{lrrrrrrr}
        \toprule
         & mean & median & std & min & max & p\_zero & p\_high \\
        \midrule
        llama-maverick & 0.438 & 0.400 & 0.187 & 0.000 & 1.000 & 0.016 & 0.138 \\
        qwen-3-235b & 0.438 & 0.400 & 0.197 & 0.000 & 1.000 & 0.024 & 0.158 \\
        ministral-14b-3 & 0.426 & 0.400 & 0.186 & 0.000 & 1.000 & 0.041 & 0.110 \\
        nova-2-lite & 0.426 & 0.400 & 0.165 & 0.000 & 1.000 & 0.027 & 0.090 \\
        openai-oss-120b & 0.387 & 0.400 & 0.187 & 0.000 & 1.000 & 0.040 & 0.074 \\
        nemotron-super & 0.385 & 0.400 & 0.153 & 0.000 & 1.000 & 0.011 & 0.046 \\
        glm-5 & 0.365 & 0.400 & 0.175 & 0.000 & 1.000 & 0.025 & 0.064 \\
        sonnet-4-6 & 0.274 & 0.300 & 0.169 & 0.000 & 1.000 & 0.068 & 0.029 \\
        \bottomrule
    \end{tabular}
    \begin{tablenotes}[flushleft]
        \footnotesize
        \item \textit{Note.} p\_high refers to the fraction of scores above 0.7. Likewise, p\_zero refers to the fraction of scores assigned 0.
    \end{tablenotes}
    \end{threeparttable}
\end{table}

Per-model mean automation-potential ranges from 0.274 (Sonnet 4.6, most conservative) to 0.438 (Llama Maverick and Qwen, most generous), showing an absolute spread of 0.165 and a 1.60× ratio. Sonnet is a clear low outlier (0.09 below the next-lowest, GLM-5), while the remaining seven cluster within about 0.07 of each other.
\clearpage
\begin{figure}[!htb]
    \centering
    \includegraphics[width=1\linewidth]{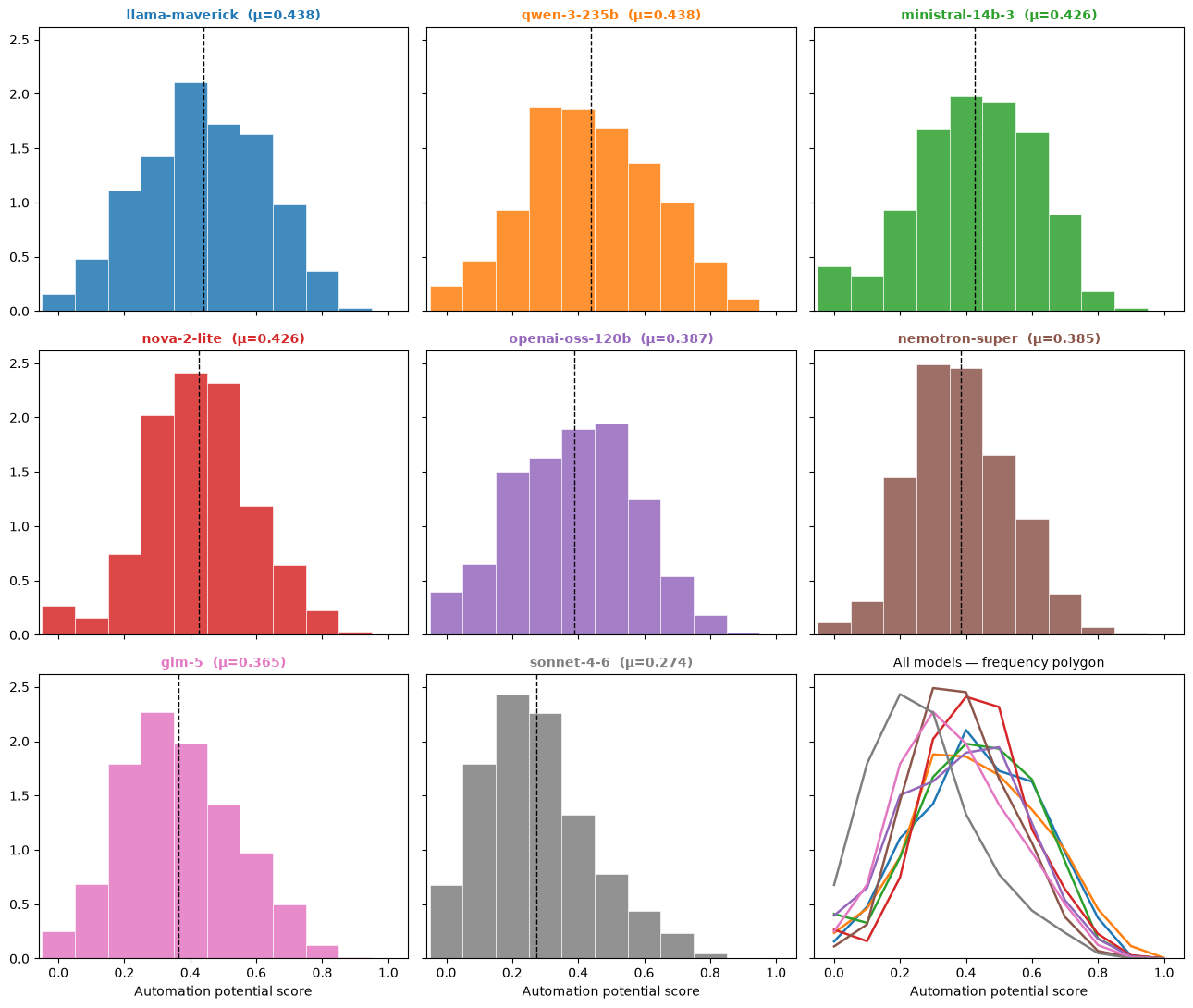}
    \caption{Per-model task AI exposure score distributions}
    \label{fig:FMScoreDistributions}
\end{figure}

All eight models produce broadly uni-modal, right-skewed distributions peaking around 0.3-0.4, but Sonnet 4.6 is visibly shifted left (peak at 0.2, near-zero mass above 0.6) while Llama Maverick and Qwen are shifted right with heavier tails toward 0.6–0.7. The frequency polygon shows this as a consistent leftward offset for Sonnet rather than a difference in shape.

\subsection{Model agreement}\label{AppendixModelConsistencyModelAgreement}

Here we calculate several statistics to characterize the degree of cross-model agreement, correcting for chance. First, we use the quadratic-weighted Cohen's $\kappa$. We use the quadratic-weighting in order to penalize disagreements in proportion to the squared distance between categories, so that a disagreement between two models of 0.2 vs 0.8 is treated more severely than 0.5 and 0.6, for example. 

\begin{table}[!htbp]
    \centering
    \footnotesize
    \setlength{\tabcolsep}{4pt}
    \begin{threeparttable}
    \caption{Pairwise quadratic-weighted Cohen's $\kappa$ between models.}
    \label{tab:kappa}
\begin{tabular}{@{}lrrrrrrrr@{}}
    \toprule
     & \rotatebox{90}{sonnet-4-6} & \rotatebox{90}{openai-oss-120b} & \rotatebox{90}{qwen-3-235b} & \rotatebox{90}{nova-2-lite} & \rotatebox{90}{llama-maverick} & \rotatebox{90}{nemotron-super} & \rotatebox{90}{glm-5} & \rotatebox{90}{ministral-14b-3} \\
    \midrule
    sonnet-4-6 &  &  &  &  &  &  &  &  \\
    openai-oss-120b & 0.668 &  &  &  &  &  &  &  \\
    qwen-3-235b & 0.576 & 0.770 &  &  &  &  &  &  \\
    nova-2-lite & 0.524 & 0.702 & 0.745 &  &  &  &  &  \\
    llama-maverick & 0.511 & 0.692 & 0.755 & 0.717 &  &  &  &  \\
    nemotron-super & 0.617 & 0.719 & 0.715 & 0.707 & 0.679 &  &  &  \\
    glm-5 & 0.759 & 0.799 & 0.755 & 0.706 & 0.694 & 0.756 &  &  \\
    ministral-14b-3 & 0.529 & 0.719 & 0.751 & 0.723 & 0.725 & 0.679 & 0.715 &  \\
    \bottomrule
\end{tabular}
    \end{threeparttable}
\end{table}

The quadratic-weighted $\kappa_{w}$ averages 0.693 (substantial, per \cite{LandisKoch1977}). Sonnet's kappa agreement is uneven rather than uniformly weak: it pairs worst with Llama Maverick, Nova, and Ministral (0.51–0.53) but best of all with GLM-5 (0.758). Its weakest kappa partners are the same generous, high-mean models it also correlates least strongly with in rank terms.

Next, we look at Krippendorff's ordinal $\alpha$ which evaluates all eight foundation models simultaneously rather than compute a set of pairwise statistics. In this case, the ordinal distance is built from
the observed distribution of categories, not their raw numeric spacing, so it
doesn't assume 0.9-vs-1.0 is the same magnitude of disagreement as 0.4-vs-0.5
unless the data category frequencies support that.

\begin{table}[!htbp]
    \centering
    \small
    \begin{threeparttable}
    \caption{Krippendorff's ordinal $\alpha$: all 8 models vs. leave-one-out.}
    \label{tab:krippendorff}
\begin{tabular}{lr}
    \toprule
     & alpha \\
    \midrule
    drop sonnet-4-6 & 0.719 \\
    drop llama-maverick & 0.676 \\
    drop nova-2-lite & 0.674 \\
    all 8 & 0.673 \\
    drop ministral-14b-3 & 0.672 \\
    drop nemotron-super & 0.667 \\
    drop qwen-3-235b & 0.659 \\
    drop openai-oss-120b & 0.658 \\
    drop glm-5 & 0.654 \\
    \bottomrule
\end{tabular}
    \end{threeparttable}
\end{table}

The headline Krippendorff's $\alpha=0.673$ which under conventional thresholds dictates that our models show ``tentative'' agreement. Dropping Sonnet 4.6 lifts $\alpha$ from 0.673 to 0.719, the largest single-model effect (the next largest, dropping Llama Maverick, barely moves it to 0.676). This confirms that Sonnet is disproportionately responsible for the ``tentative'' verdict; even without it, though, $\alpha$ stays below the 0.800 ``good'' threshold, so it is a meaningful contributor rather than the sole cause, and the remaining seven models show only moderate agreement among themselves.

The final statistic we use to measure model agreement is the intraclass correlation coefficient $ICC(2,1)$. The (2,1) notation follows \cite{ShroutFleiss1979}. Model 2 means that we take both tasks and raters as random samples from a larger population and raters are systematically distinguishable across all units. The one means that the unit of interest is a single rater's score, not an average across raters. $ICC(2,1)$ provides a variance decomposition, treating the score matrix (tasks $\times$ models) as a two-way random-effects ANOVA. We report $ICC(2,1)=0.687$, placing agreement in the ``moderate'' band \cite{KooLi2016}, closely matching the value and corresponding band of our Krippendorff's $\alpha$ result. 

Next, we consider the extent to which the models agree on their rank ordering i.e. their relative scoring of tasks. First, we calculate the Spearman's rank correlation $\rho$. Table \ref{tab:spearman} summarizes the results over all pairwise model comparisons:

\begin{table}[!htbp]
    \centering
    \footnotesize
    \setlength{\tabcolsep}{4pt}
    \begin{threeparttable}
    \caption{Pairwise Spearman's $\rho$ between models.}
    \label{tab:spearman}
\begin{tabular}{@{}lrrrrrrrr@{}}
    \toprule
     & \rotatebox{90}{sonnet-4-6} & \rotatebox{90}{openai-oss-120b} & \rotatebox{90}{qwen-3-235b} & \rotatebox{90}{nova-2-lite} & \rotatebox{90}{llama-maverick} & \rotatebox{90}{nemotron-super} & \rotatebox{90}{glm-5} & \rotatebox{90}{ministral-14b-3} \\
    \midrule
    sonnet-4-6 &  &  &  &  &  &  &  &  \\
    openai-oss-120b & 0.815 &  &  &  &  &  &  &  \\
    qwen-3-235b & 0.820 & 0.793 &  &  &  &  &  &  \\
    nova-2-lite & 0.746 & 0.717 & 0.754 &  &  &  &  &  \\
    llama-maverick & 0.736 & 0.709 & 0.749 & 0.718 &  &  &  &  \\
    nemotron-super & 0.771 & 0.732 & 0.770 & 0.731 & 0.723 &  &  &  \\
    glm-5 & 0.871 & 0.806 & 0.817 & 0.753 & 0.750 & 0.768 &  &  \\
    ministral-14b-3 & 0.739 & 0.725 & 0.749 & 0.721 & 0.722 & 0.713 & 0.760 &  \\
    \bottomrule
\end{tabular}
    \end{threeparttable}
\end{table}

The mean pairwise $\rho=0.756$. The highest pair is between Sonnet 4.6 and GLM 5 at $0.871$, and the lowest is between OpenAI OSS 120B and Llama Maverick at $0.709$. These pairwise values indicate that the models rank tasks' automation potential fairly consistently. 

Given that we restrict the output domain to only eleven possible values, we also use Kendall's $\tau_b$ statistic which helps us account for ties within each annotator's scores. 

\begin{table}[!htbp]
    \centering
    \footnotesize
    \setlength{\tabcolsep}{4pt}
    \begin{threeparttable}
    \caption{Pairwise Kendall's $\tau_b$ between models.}
    \label{tab:kendall}
\begin{tabular}{@{}lrrrrrrrr@{}}
    \toprule
     & \rotatebox{90}{sonnet-4-6} & \rotatebox{90}{openai-oss-120b} & \rotatebox{90}{qwen-3-235b} & \rotatebox{90}{nova-2-lite} & \rotatebox{90}{llama-maverick} & \rotatebox{90}{nemotron-super} & \rotatebox{90}{glm-5} & \rotatebox{90}{ministral-14b-3} \\
    \midrule
    sonnet-4-6 &  &  &  &  &  &  &  &  \\
    openai-oss-120b & 0.710 &  &  &  &  &  &  &  \\
    qwen-3-235b & 0.715 & 0.679 &  &  &  &  &  &  \\
    nova-2-lite & 0.644 & 0.609 & 0.646 &  &  &  &  &  \\
    llama-maverick & 0.623 & 0.591 & 0.631 & 0.607 &  &  &  &  \\
    nemotron-super & 0.669 & 0.622 & 0.661 & 0.628 & 0.612 &  &  &  \\
    glm-5 & 0.781 & 0.697 & 0.708 & 0.648 & 0.635 & 0.662 &  &  \\
    ministral-14b-3 & 0.629 & 0.611 & 0.635 & 0.614 & 0.606 & 0.605 & 0.650 &  \\
    \bottomrule
\end{tabular}
    \end{threeparttable}
\end{table}

These results are similar to the Spearman's $\rho$, just at a lower absolute magnitude. Sonnet and GLM remains the most concordant pair. This is particularly interesting as it is Sonnet that, when considering absolute task exposure scores, is most clearly the outlier. 

We can also inspect instances where model disagreement is high: below we show three typical examples.

\begin{table}[htbp]
    \centering
    \small
    \begin{threeparttable}
    \caption{Cross-model disagreement example: photocopying and scanning documents.}
    \label{tab:task-example-photocopying}
\begin{tabularx}{\textwidth}{@{}lrX@{}}
\toprule
\multicolumn{3}{@{}p{\textwidth}@{}}{\textbf{Task:} Photocopying and scanning documents} \\
\midrule
Model & Score & Reasoning \\
\midrule
qwen-3-235b & 0.00 & Photocopying and scanning are physical tasks requiring manual operation of equipment, though digitized document workflows can be automated once scanned; the physical act remains non-automatable. \\
openai-oss-120b & 0.00 & Photocopying and scanning require physical equipment operation, which AI cannot do directly. \\
glm-5 & 0.00 & Photocopying and scanning require physical operation of equipment to handle documents, which cannot be performed by current AI systems without robotics integration. \\
sonnet-4-6 & 0.10 & This is a physical, manual task requiring in-person operation of equipment; AI cannot perform the physical action, though it could assist with organising or naming scanned files. \\
nemotron-super & 0.60 & AI can control photocopiers and scanners via voice or automation for routine tasks, but handling jams, special formats, and quality checks require human intervention. \\
nova-2-lite & 0.90 & AI can automate photocopying and scanning documents through integrated hardware and software solutions, with minimal human intervention needed for setup and troubleshooting. \\
ministral-14b-3 & 0.90 & Photocopying and scanning documents can be fully automated with AI-powered software, but human oversight may still be required for quality control or handling sensitive documents. \\
llama-maverick & 1.00 & Photocopying and scanning are routine tasks that can be automated with existing technology, not necessarily AI. \\
\bottomrule
\end{tabularx}
    \begin{tablenotes}[flushleft]
        \footnotesize
        \item \textit{Note.} Department: HM Revenue and Customs (excl. agencies) \textbar{} Profession: Operational Delivery \textbar{} Grade: AO. Mean automation score = 0.44 (SD = 0.46); normalized disagreement $\tilde s_{ij}$ = 0.91.
    \end{tablenotes}
    \end{threeparttable}
\end{table}

\begin{table}[htbp]
    \centering
    \small
    \begin{threeparttable}
    \caption{Cross-model disagreement example: reading aloud to a judge.}
    \label{tab:task-example-reading-aloud}
\begin{tabularx}{\textwidth}{@{}lrX@{}}
\toprule
\multicolumn{3}{@{}p{\textwidth}@{}}{\textbf{Task:} Read aloud quickly and accurately to the judge} \\
\midrule
Model & Score & Reasoning \\
\midrule
nemotron-super & 0.00 & AI can read aloud text with high accuracy and speed using text-to-speech systems, but the task requires a human to perform the vocal delivery in real-time as specified, making automation irrelevant to the role's intent. \\
ministral-14b-3 & 0.00 & AI cannot replace the need for human reading aloud with the appropriate tone, pace, and clarity required for assisting a visually impaired judge. \\
glm-5 & 0.10 & AI can convert text to speech, but reading aloud quickly and accurately to a judge in real-time during proceedings requires human presence, the ability to respond to immediate needs, and contextual understanding of courtroom dynamics. \\
sonnet-4-6 & 0.20 & Text-to-speech AI can read text aloud with high accuracy, but the role requires a physically present human to guide the judge, select the right passages in real time, and respond to the judge's live instructions in a courtroom environment. \\
openai-oss-120b & 0.70 & Advanced text-to-speech can read aloud quickly and accurately, though a human may still need to monitor for errors or adjust pacing. \\
nova-2-lite & 0.70 & AI text-to-speech systems can read aloud accurately, but may lack the nuanced tone and pace required for effective communication with a visually impaired judge. \\
llama-maverick & 0.80 & Reading aloud to the judge is a task that can be performed by a human, with AI potentially supporting with text-to-speech functionality. \\
qwen-3-235b & 1.00 & AI can convert text to speech accurately and at speed, fully automating the reading function for a visually impaired judge. \\
\bottomrule
\end{tabularx}
    \begin{tablenotes}[flushleft]
        \footnotesize
        \item \textit{Note.} Department: HM Courts and Tribunals Service \textbar{} Profession: Other \textbar{} Grade: EO. Mean automation score = 0.44 (SD = 0.40); normalized disagreement $\tilde s_{ij}$ = 0.81, domain-adjusted $\hat s_{ij}$ = 0.76.
    \end{tablenotes}
    \end{threeparttable}
\end{table}

\begin{table}[htbp]
    \centering
    \small
    \begin{threeparttable}
    \caption{Cross-model disagreement example: recording parliamentary divisions.}
    \label{tab:task-example-divisions}
\begin{tabularx}{\textwidth}{@{}lrX@{}}
\toprule
\multicolumn{3}{@{}p{\textwidth}@{}}{\textbf{Task:} Take divisions (record votes) on a rota basis} \\
\midrule
Model & Score & Reasoning \\
\midrule
sonnet-4-6 & 0.00 & Recording votes is a procedural, in-person task requiring physical presence in the Chamber; AI has no substantive role in this activity. \\
openai-oss-120b & 0.00 & Recording votes requires physical presence and real-time verification, which current AI cannot perform. \\
llama-maverick & 0.00 & Taking divisions involves physical presence and real-time recording, which is not automatable by current AI systems. \\
glm-5 & 0.00 & AI cannot physically record votes in the division lobbies, verify members' identities, or manage the real-time logistics of a division, which requires physical presence and direct human oversight. \\
ministral-14b-3 & 0.20 & AI can automate the recording of votes using digital tools, but taking divisions requires physical presence, managing human interactions, and ensuring procedural accuracy, which cannot be fully automated. \\
qwen-3-235b & 0.80 & Recording votes is a procedural, rule-based task that can be automated through digital systems; however, current rota-based human involvement ensures accuracy and accountability in live settings, limiting full automation. \\
nova-2-lite & 0.90 & AI can automate vote recording systems and maintain division logs, requiring minimal human intervention beyond technical oversight and verification of system outputs. \\
nemotron-super & 0.90 & AI can accurately record votes from official feeds and generate division reports, which is a structured, rule-based task with minimal need for interpretation or judgment. \\
\bottomrule
\end{tabularx}
    \begin{tablenotes}[flushleft]
        \footnotesize
        \item \textit{Note.} Department: Cabinet Office (excl. agencies) \textbar{} Profession: Other \textbar{} Grade: HEO. Mean automation score = 0.35 (SD = 0.43); normalized disagreement $\tilde s_{ij}$ = 0.87.
    \end{tablenotes}
    \end{threeparttable}
\end{table}
\clearpage

\subsection{Sensitivity to model vintage}\label{sec:vintage}

Table~\ref{tab:vintage} reports task-level agreement between the older and current versions of the two models, alongside their mean scores.

\begin{table}[!htbp]
    \centering
    \small
    \begin{threeparttable}
    \caption{Agreement between older and current versions of two ensemble models}
    \label{tab:vintage}
\begin{tabular}{lrr}
    \toprule
    Measure & \hd{Llama 3 70B\\vs Maverick} & \hd{Nova Lite vs\\Nova 2 Lite} \\
    \midrule
    Tasks compared & 93,301 & 93,562 \\
    Mean task score, older vs current & 0.34 vs 0.44 & 0.31 vs 0.42 \\
    Workforce-weighted mean exposure, older vs current & 0.32 vs 0.41 & 0.30 vs 0.40 \\
    Task-level correlation & 0.62 & 0.53 \\
    Tasks given the same score & 25\% & 24\% \\
    Tasks scored within 0.1 & 60\% & 61\% \\
    \bottomrule
  \end{tabular}
    \begin{tablenotes}[flushleft]
        \footnotesize
        \item \textit{Note.} Mean task scores are unweighted averages across tasks. Workforce-weighted means apply the raking weights to role-level scores.
    \end{tablenotes}
    \end{threeparttable}
\end{table}

\section{AI exposure and the demand for labor}
\label{sec:AppendixDemandForLabor}

This appendix reports the full estimation results underlying
Section~\ref{sec:DemandForLabor}, and presents eight robustness checks in
turn: sensitivity to the occupation-size threshold (Appendix~\ref{app:B1}),
sensitivity to the largest occupations (Appendix~\ref{app:B2}), an
admin-grade confound (Appendix~\ref{app:B3}), a disclosed limitation
regarding drift within the pre-release period itself
(Appendix~\ref{app:B4}), a cross-check against published Civil Service
headcount statistics (Appendix~\ref{app:B5}), a test of each channel in
isolation (Appendix~\ref{app:B6}), sensitivity to the task-decay rate
used to construct $\mu_i$ and $\sigma_i^2$ (Appendix~\ref{app:B7}), and a
check of specific named dates (Appendix~\ref{app:named-dates}). As in
Section~\ref{sec:DemandForLabor}, every check below uses
$\operatorname{arcsinh}(n_{it})$ rather than $\log(n_{it})$, retains every
occupation-month including those with zero postings ($74{,}022$ in total,
rather than the $44{,}783$ occupation-months with at least one posting that
an earlier, $\log$-based version of this analysis was restricted to), and
constructs $\mu_i$/$\sigma_i^2$ using the primary geometric task-decay rate
($\delta=0.75$), except Appendix~\ref{app:B5}, which uses annual headcount
data that is never zero and so is unaffected by the first of these choices,
and Appendix~\ref{app:B7}, which varies the decay rate directly. We begin
with the headline results.

\begin{table}[!htbp]
    \centering
    \small
    \begin{threeparttable}
    \caption{Year-by-year effects of AI exposure on labor demand (Equation~\ref{eq:event-study})}
    \label{tab:demand-headline}
\begin{tabular}{lrrrrr}
    \toprule
    Period & $N$ & \hd{$\hat\beta_{1k}$\\(mean exp.)} & $p$ &
      \hd{$\hat\beta_{2k}$\\(dispersion)} & $p$ \\
    \midrule
    Pre-release (reference) & 36,062 & -- & -- & -- & -- \\
    Year 1 (Dec 2022--Oct 2023) & 10,439 & -0.977 & 0.057 & +9.665 & 0.128 \\
    Year 2 (Nov 2023--Oct 2024) & 11,388 & -2.183 & 0.000 & +8.967 & 0.183 \\
    Year 3 (Nov 2024--Oct 2025) & 11,388 & -2.183 & 0.000 & +3.873 & 0.640 \\
    Year 4 (Nov 2025--Mar 2026) & 4,745 & -2.408 & 0.001 & +19.062 & 0.080 \\
    \bottomrule
  \end{tabular}
    \begin{tablenotes}[flushleft]
        \footnotesize
        \item \textit{Note.} $N$ = occupation-months, including occupation-months with zero postings. Standard errors clustered by occupation ($949$ occupations, $74{,}022$ occupation-months total). $R^2$ = 0.0022; $R^2$ (incl.\ FE) = 0.495; $F$-test for the joint significance of occupation and time fixed effects: $67.83$ ($p<0.001$).
    \end{tablenotes}
    \end{threeparttable}
\end{table}

The within-$R^2$ is small (0.002), which is expected given the identification strategy: entity and time fixed effects absorb the majority of variation in $\operatorname{arcsinh}(\mathrm{roles})$, and the $\mu_i$/$\sigma_i^2$ interaction terms are estimated only from the residual occupation-by-time variation that remains after this level and common-time variation is removed. 

\begin{table}[!htbp]
    \centering
    \small
    \begin{threeparttable}
    \caption{Standardized effects: a one-standard-deviation increase in mean exposure or dispersion, relative to the pre-release baseline}
    \label{tab:demand-headline-std}
\begin{tabular*}{0.7\linewidth}{@{\extracolsep{\fill}}lrr}
    \toprule
    Period & Mean exposure & Dispersion \\
    \midrule
    Year 1 & $-4.8\%$ & $+3.8\%$ \\
    Year 2 & $-10.4\%$ & $+3.5\%$ \\
    Year 3 & $-10.4\%$ & $+1.5\%$ \\
    Year 4 & $-11.5\%$ & $+7.5\%$ \\
    \bottomrule
  \end{tabular*}
    \begin{tablenotes}[flushleft]
        \footnotesize
        \item \textit{Note.} Because $\operatorname{arcsinh}$ is not exactly log-linear, effects are computed exactly rather than approximated, as $\sinh\!\left(\operatorname{arcsinh}(n_0)+\hat\beta_{k}\cdot\mathrm{sd}\right)/n_0-1$, evaluated at the sample median occupation size ($n_0=3$ roles/month), using $\mathrm{sd}(\mu_i)=0.048$ and $\mathrm{sd}(\sigma_i^2)=0.0036$. Effects vary modestly with occupation size; see Figure~\ref{fig:demand-event-study} in the main text for the full 25th--75th percentile range.
    \end{tablenotes}
    \end{threeparttable}
\end{table}

\begin{figure}[!htbp]
    \centering
    \includegraphics[width=1\linewidth]{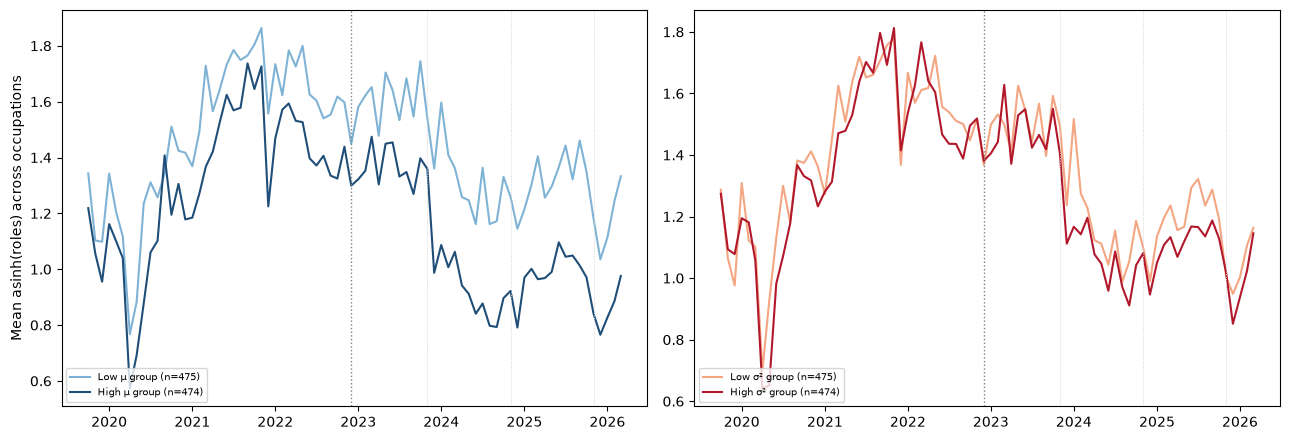}
    \caption{Raw group averages: arcsinh(roles) by month, split at the median exposure/dispersion}
    \label{fig:TreatmentAverages}
    \caption*{\footnotesize\textit{Note: Dotted lines correspond to ChatGPT's release, then each subsequent year boundary)}}
\end{figure}
\clearpage

\subsection{Occupation-size threshold sensitivity}
\label{app:B1}
We re-estimate Equation~\ref{eq:event-study} at six minimum-pre-release-role thresholds for occupation inclusion (5, 10, 20, 40, 80, 150 - the main text uses 40), focusing on Year 2, the first year in which the effect emerges. Mean exposure remains significant, and grows in magnitude, at every threshold tested, from the smallest (5 pre-release roles) to the largest (150); dispersion is far less robust. In Year 2 it is significant only at a threshold of 80 pre-release roles ($p=0.012$), falls short at the largest threshold (150, $p=0.087$), and is nowhere near significance at the smaller thresholds (5--40, $p\ge0.18$).

\begin{table}[htbp]
    \centering
    \small
    \begin{threeparttable}
    \caption{Threshold sensitivity of Year-2 coefficients}
    \label{tab:app-b1}
\begin{tabular}{lrrrrr}
    \toprule
    Min. pre-release roles & $N$ occ. & $\hat\beta_{1,2}$ & $p$ &
      $\hat\beta_{2,2}$ & $p$ \\
    \midrule
    5 & 3,061 & -0.559 & 0.000 & +0.990 & 0.497 \\
    10 & 2,117 & -0.907 & 0.000 & +2.396 & 0.327 \\
    20 & 1,392 & -1.504 & 0.000 & +5.120 & 0.253 \\
    40 & 949 & -2.183 & 0.000 & +8.967 & 0.183 \\
    80 & 563 & -3.396 & 0.000 & +25.113 & 0.012 \\
    150 & 340 & -3.553 & 0.006 & +24.480 & 0.087 \\
    \bottomrule
  \end{tabular}
    \end{threeparttable}
\end{table}

The same threshold sweep for the other three years shows a similarly asymmetric pattern. Year 1 is null or marginal for both channels at low thresholds: mean exposure reaches conventional significance only at a threshold of 80 pre-release roles ($p=0.018$), remaining marginal at 40 ($p=0.057$) and 150 ($p=0.063$); dispersion is significant at both 80 ($p=0.0095$) and 150 ($p=0.025$). Year 3's mean-exposure coefficient is significant at every threshold from 5 to 80 ($p\le0.001$) and marginal at 150 ($p=0.066$); its dispersion coefficient is never significant at any threshold, and is wrong-signed (negative) at the three lowest thresholds (5, 10, 20) before turning positive, though still insignificant, from 40 upward. Year 4's mean-exposure coefficient is significant at every threshold ($p=0.000$--$0.046$); dispersion is significant only at 80 ($p=0.0496$, just under the 5\% level), marginal at 40 ($p=0.080$), and not significant at the smallest (5, 10, 20) or largest (150, $p=0.208$) thresholds. Taken together, mean exposure's Year 2--4 result is robust across the full range of thresholds tested. Dispersion is markedly less so: at thresholds of 5--40 pre-release roles it is not significant in any year (smallest $p=0.080$, Year 4 at 40), and it reaches the 5\% level only at 80 or 150 roles (Year 1 at both, Year 2 and Year 4 at 80 only, Year 3 never). Its significance therefore depends on restricting the sample to the largest occupations.

\subsection{Leverage of the largest occupations}
\label{app:B2}

To rule out the result being driven by a small number of high-volume occupations, we re-estimate Equation~\ref{eq:event-study} after dropping the largest occupations by pre-release posting volume.

\begin{table}[htbp]
    \centering
    \small
    \begin{threeparttable}
    \caption{Leverage check: dropping the largest occupations (Year 2)}
    \label{tab:app-b2}
\begin{tabular}{lrrrrr}
    \toprule
    Occupations dropped & $N$ occ. & $\hat\beta_{1,2}$ & $p$ &
      $\hat\beta_{2,2}$ & $p$ \\
    \midrule
    0 & 949 & -2.183 & 0.000 & +8.967 & 0.183 \\
    5 & 944 & -2.421 & 0.000 & +8.067 & 0.231 \\
    10 & 939 & -2.611 & 0.000 & +11.007 & 0.059 \\
    20 & 929 & -2.526 & 0.000 & +9.664 & 0.093 \\
    50 & 899 & -2.479 & 0.000 & +8.145 & 0.193 \\
    \bottomrule
  \end{tabular}
    \end{threeparttable}
\end{table}

Mean exposure remains significant throughout, and if anything strengthens as the largest occupations are dropped -- the result is not driven by a handful of high-volume occupations. Dispersion is not significant under any configuration tested: it is furthest from significance with none or the five largest occupations removed ($p=0.183$, $p=0.231$), closest -- though still not significant at conventional levels -- once ten or twenty are removed ($p=0.059$, $p=0.093$), and recedes again once fifty are removed ($p=0.193$). Dispersion's result is therefore sensitive to the specific composition of large occupations retained in the sample in a way that mean exposure is not, and under the primary task-decay specification it does not reach conventional significance at any point in this check.

\subsection{An admin-grade confound?}
\label{app:B3}

Since a post-COVID headcount-reduction target \citep{bbc2022civilservicecuts} included administrative grades, and admin-grade occupations happen to be somewhat higher-exposure on average\footnote{$25.6\%$ of the qualifying sample are AA/AO/EO grades; mean $\mu_i = 0.417$ for admin occupations vs.\ $0.374$ for others;
$\mathrm{corr}(\mu_i,\mathrm{Admin}_i)=0.388$}, we test whether an
admin-specific Year-2 effect explains the result:
\begin{equation}
  \operatorname{arcsinh}(n_{it}) = \alpha_i + \gamma_t +
    \sum_{k=1}^{4} \beta_{1k}\left(\mu_i \cdot D_t^k\right) +
    \sum_{k=1}^{4} \beta_{2k}\left(\sigma_i^2 \cdot D_t^k\right) +
    \delta\left(\mathrm{Admin}_i \cdot D_t^{2}\right) + \varepsilon_{it},
  \label{eq:app-b3}
\end{equation}
where $\mathrm{Admin}_i=1$ for AA/AO/EO-grade occupations. Estimated:
$\hat\beta_{1,2}=-2.00$ ($p=0.001$), $\hat\beta_{2,2}=+10.44$
($p=0.113$), $\hat\delta=-0.066$ ($p=0.150$). The admin-grade interaction is not significant, and $\hat\beta_{1,2}$ remains close to its value in Table~\ref{tab:demand-headline} and strongly significant; $\hat\beta_{2,2}$ also remains close to its (already non-significant) headline value, so the confound is ruled out for both channels.

\subsection{Drift within the pre-release window}
\label{app:B4}

We use the entire pre-release period (October 2019--November 2022), rather
than a shorter window immediately preceding ChatGPT's release, for two
reasons. First, due to the pandemic GRID-wide vacancy volumes were unusually volatile in the
early part of this period (see Figure \ref{fig:VacanciesOverTime}): postings fell sharply during the initial
COVID-19 lockdown (April/May 2020, to roughly 1{,}200--1{,}400 vacancies
against $2{,}900$--$3{,}300$ in surrounding months) before rebounding
strongly in the following months. A short window overlapping this
disruption would estimate each occupation's mean exposure and dispersion
from a thin and potentially unrepresentative set of tasks, while
averaging over the full 38-month period reduces this sampling noise.
Second, a longer window increases the number of qualifying occupations
(from 607 under a narrower Jul 2021 to Nov 2022 window to 949 here), since
more occupations clear the minimum-posting-volume threshold. The checks
below test whether this choice of window is itself stable.

\textit{(a) Drift within the pre-release period.} We split pre-release rows
into three sub-periods -- $\mathrm{Pre}_A$ (before Oct 2020, $N=11{,}388$,
reference), $\mathrm{Pre}_B$ (Oct 2020--Oct 2021, $N=11{,}388$), and
$\mathrm{Pre}_C$ (Oct 2021--Nov 2022, $N=13{,}286$) -- and estimate, on the
pre-release rows only:
\begin{equation}
  \operatorname{arcsinh}(n_{it}) = \alpha_i + \gamma_t +
    \beta_{1B}\left(\mu_i \cdot \mathrm{Pre}_B\right) +
    \beta_{1C}\left(\mu_i \cdot \mathrm{Pre}_C\right) +
    \beta_{2B}\left(\sigma_i^2 \cdot \mathrm{Pre}_B\right) +
    \beta_{2C}\left(\sigma_i^2 \cdot \mathrm{Pre}_C\right) +
    \varepsilon_{it}.
  \label{eq:app-b4a}
\end{equation}
We find $\hat\beta_{1B}=-0.757$ ($p=0.101$), $\hat\beta_{1C}=-1.251$
($p=0.016$), $\hat\beta_{2B}=+6.283$ ($p=0.318$), $\hat\beta_{2C}=+12.776$
($p=0.074$): drift is concentrated almost entirely in the final third of the pre-release
period ($\mathrm{Pre}_C$) for both channels, though only marginally so for
dispersion; neither channel drifts significantly in the middle third
($\mathrm{Pre}_B$).

\textit{(b) Does narrowing the reference period change the result?}
We re-estimate Equation~\ref{eq:event-study} restricting the reference period to Jul 2021--Nov 2022 only (dropping Oct 2019--Jun 2021 pre-release rows; reference $N$ falls from $36{,}062$ to $16{,}133$).

\begin{table}[htbp]
    \centering
    \small
    \begin{threeparttable}
    \caption{Narrowed pre-release reference window (Jul 2021--Nov 2022 only)}
    \label{tab:app-b4b}
\begin{tabular}{lrrrr}
    \toprule
    Period & $\hat\beta_{1k}$ (mean exp.) & $p$ &
      $\hat\beta_{2k}$ (dispersion) & $p$ \\
    \midrule
    Year 1 & -0.576 & 0.224 & +3.676 & 0.509 \\
    Year 2 & -1.782 & 0.001 & +2.978 & 0.650 \\
    Year 3 & -1.782 & 0.002 & -2.116 & 0.798 \\
    Year 4 & -2.008 & 0.005 & +13.074 & 0.206 \\
    \bottomrule
  \end{tabular}
    \begin{tablenotes}[flushleft]
        \footnotesize
        \item \textit{Note.} $F$-test for the joint significance of occupation and time fixed effects: $56.36$ ($p<0.001$).
    \end{tablenotes}
    \end{threeparttable}
\end{table}

Mean exposure remains significant in every year under this narrower reference, essentially unchanged from Table~\ref{tab:demand-headline}; dispersion is not significant in a single year -- a stronger version of the asymmetry already visible in Table~\ref{tab:demand-headline} itself.

One plausible driver of this drift is that $\mathrm{Pre}_A$, our reference
sub-period, itself spans the COVID-19 disruption,
including both the trough and the subsequent rebound; the apparent shift
in $\mathrm{Pre}_C$ relative to this unusual baseline may
reflect COVID-era recovery dynamics rather than a genuine, ongoing trend.
In any case, since this drift occurs entirely within the pre-release period,
it cannot itself explain the Year 2 divergence in the main result, but it
is disclosed here as a potential drawback of using a long, fixed reference
window.

\subsection{An independent cross-check against real headcount}
\label{app:B5}

Using published Civil Service headcount statistics (2021--2026, annual),
matched by occupation, we estimate a coarser version of
Equation~\ref{eq:event-study} with a single post-2023 dummy in place of the
four annual bins (only 6 annual observations per occupation are available):
\begin{equation}
  \log(\mathrm{FTE}_{it}) = \alpha_i + \gamma_t +
    \beta_1\left(\mu_i \cdot \mathrm{Post}_t\right) +
    \beta_2\left(\sigma_i^2 \cdot \mathrm{Post}_t\right) + \varepsilon_{it},
  \label{eq:app-b5}
\end{equation}
where $\mathrm{Post}_t=1$ for 2023--2026. $897/949$ qualifying occupations
have $\geq 1$ year of ACSES coverage, $709$ have $\geq 3$ years, and $437$
have all 6 years.

\begin{table}[htbp]
    \centering
    \small
    \begin{threeparttable}
    \caption{ACSES headcount cross-check}
    \label{tab:app-b5}
\begin{tabular}{lrrrrr}
    \toprule
    Coverage & $N$ occ. & \hd{$\hat\beta_1$\\(mean only)} & $p$ &
      \hd{$\hat\beta_1$\\(joint)} & \hd{$\hat\beta_2$\\(joint)} \\
    \midrule
    $\geq 3$ years & 709 & $-0.890$ & $0.165$ & $-0.900\ (p=0.211)$ &
      $+0.196\ (p=0.981)$ \\
    All 6 years    & 437 & $-0.880$ & $0.212$ & $-1.009\ (p=0.227)$ &
      $+2.286\ (p=0.817)$ \\
    \bottomrule
  \end{tabular}
    \end{threeparttable}
\end{table}

Mean exposure is correctly signed but not significant in either coverage
group; dispersion is correctly signed but essentially zero and, with
$\mathrm{corr}(\mu_i\cdot\mathrm{Post}_t,\sigma_i^2\cdot\mathrm{Post}_t)$
of $0.92$--$0.93$, too collinear with mean exposure to separate from it.
With only six annual data points per occupation, this check can neither
confirm nor contradict the main result.

\subsection{Does each channel hold up on its own?}
\label{app:B6}

For each year $k$, we drop one moment from Equation~\ref{eq:event-study}
and restrict the sample to pre-release rows plus year $k$ only:
\begin{equation}
  \operatorname{arcsinh}(n_{it}) = \alpha_i + \gamma_t + \beta_{1k}\left(\mu_i \cdot D_t^k\right)
    + \varepsilon_{it}
  \quad\text{and}\quad
  \operatorname{arcsinh}(n_{it}) = \alpha_i + \gamma_t + \beta_{2k}\left(\sigma_i^2 \cdot D_t^k\right)
    + \varepsilon_{it}.
  \label{eq:app-b6}
\end{equation}

\begin{table}[htbp]
    \centering
    \small
    \begin{threeparttable}
    \caption{Single-moment comparison, by year}
    \label{tab:app-b6}
\begin{tabular}{lrrrr}
    \toprule
    Period & $\hat\beta_{1k}$ (mean only) & $p$ &
      $\hat\beta_{2k}$ (dispersion only) & $p$ \\
    \midrule
    Year 1 & -0.662 & 0.141 & +4.171 & 0.455 \\
    Year 2 & -1.891 & 0.000 & -3.312 & 0.578 \\
    Year 3 & -2.057 & 0.000 & -8.409 & 0.239 \\
    Year 4 & -1.788 & 0.003 & +5.514 & 0.547 \\
    \bottomrule
  \end{tabular}
    \end{threeparttable}
\end{table}

Mean exposure alone is a significant predictor from Year 2 onward (missing significance only in Year 1, where it is also not significant in the joint model); dispersion alone
is not significant in any year, and in Years 2 and 3 even takes the opposite
sign to its coefficient in the joint model (Table~\ref{tab:demand-headline}).
The dispersion channel's already-marginal role in
Table~\ref{tab:demand-headline} therefore relies on jointly controlling for
mean exposure to an even greater extent than under an earlier, $\log$-based
version of this check, consistent with the two moments' cross-sectional
correlation ($0.36$--$0.43$ across the occupation-size thresholds tested in
Appendix~\ref{app:B1}).

\subsection{Sensitivity to the task-decay rate}
\label{app:B7}

The primary specification constructs $\mu_i$ and $\sigma_i^2$ using
geometric task-decay weighting with $\delta=0.75$, under the assumption
that job adverts tend to list a role's more important, more
time-consuming duties first. We test whether the headline result is
sensitive to this choice by refitting Equation~\ref{eq:event-study} with
$\mu_i$/$\sigma_i^2$ recomputed under each of the four decay rates used
throughout this project ($\delta\in\{0.5,0.75,0.9,1.0\}$), holding the
panel construction, occupation set, and year bins fixed.

\begin{table}[htbp]
    \centering
    \small
    \begin{threeparttable}
    \caption{Sensitivity of the primary labor-demand specification to the task-decay rate used to construct $\mu_i$ and $\sigma_i^2$}
    \label{tab:app-b7}
\begin{tabular}{@{}lrrrr@{}}
    \toprule
    Period & $\delta=0.5$ & $\delta=0.75$ & $\delta=0.9$ & $\delta=1.0$ \\
    \midrule
    \multicolumn{5}{@{}l}{Mean exposure ($\hat\beta_{1k}$, $p$)} \\
    Year 1 & $-0.86$ (0.080) & $-0.98$ (0.056) & $-1.00$ (0.056) & $-0.95$ (0.069) \\
    Year 2 & $-1.70$ (0.001) & $-2.18$ (0.000) & $-2.38$ (0.000) & $-2.37$ (0.000) \\
    Year 3 & $-1.92$ (0.001) & $-2.18$ (0.000) & $-2.25$ (0.000) & $-2.21$ (0.000) \\
    Year 4 & $-1.97$ (0.004) & $-2.41$ (0.001) & $-2.51$ (0.001) & $-2.37$ (0.002) \\
    \addlinespace
    \multicolumn{5}{@{}l}{Dispersion ($\hat\beta_{2k}$, $p$)} \\
    Year 1 & $7.43$ (0.408) & $9.66$ (0.128) & $9.49$ (0.088) & $8.45$ (0.107) \\
    Year 2 & $1.02$ (0.923) & $8.97$ (0.183) & $10.65$ (0.077) & $9.43$ (0.104) \\
    Year 3 & $-0.34$ (0.977) & $3.87$ (0.640) & $4.54$ (0.528) & $3.25$ (0.621) \\
    Year 4 & $16.21$ (0.273) & $19.06$ (0.080) & $19.07$ (0.039) & $16.14$ (0.055) \\
    \bottomrule
  \end{tabular}
    \begin{tablenotes}[flushleft]
        \footnotesize
        \item \textit{Note.} $\delta=0.75$ is the primary specification (identical to Table~\ref{tab:demand-headline}); the other three columns recompute $\mu_i$/$\sigma_i^2$ under each alternative decay rate and refit the identical equation.
    \end{tablenotes}
    \end{threeparttable}
\end{table}

Mean exposure is a significant predictor of reduced role postings from Year
2 onward under every decay rate tested ($p<0.01$ in each case), and
marginal in Year 1 under all four ($p=0.056$--$0.080$); point estimates
move only modestly across $\delta$ (Year 2: $-1.70$ at $\delta=0.5$ to
$-2.38$ at $\delta=0.9$). Dispersion is not significant in Years 1--3 under
any decay rate, and crosses conventional significance in Year 4 only once,
at $\delta=0.9$ ($p=0.039$); at $\delta=0.5$, the most aggressive decay
tested, dispersion is not significant in any year at all. The choice of
decay rate therefore does not change which channel is robust: mean
exposure holds up regardless of $\delta$, and dispersion remains the
fragile channel under every specification tested in this appendix.

\subsection{Named-date check: Bletchley Summit / AISI, and the Spring
Budgets}
\label{app:named-dates}

Rather than assume the year-since-release binning is the right resolution,
we estimate a monthly-resolution event study over Jan 2023--Jun 2024 (18
months), with the whole pre-release period as reference and all other
post-release months absorbed into a single ``other'' bucket. This allows us
to probe whether any particular dates are relevant to changes in UKCS hiring:
\begin{equation}
  \begin{split}
  \operatorname{arcsinh}(n_{it}) = {} & \alpha_i + \gamma_t +
    \sum_{m=1}^{18} \beta_{1m}\left(\mu_i \cdot D_t^m\right) +
    \beta_{1,\mathrm{other}}\left(\mu_i \cdot \mathrm{Other}_t\right) \\
    & +
    \sum_{m=1}^{18} \beta_{2m}\left(\sigma_i^2 \cdot D_t^m\right) +
    \beta_{2,\mathrm{other}}\left(\sigma_i^2 \cdot \mathrm{Other}_t\right) +
    \varepsilon_{it},
  \end{split}
  \label{eq:app-named}
\end{equation}
where $D_t^m$ indexes each calendar month from January 2023 to June 2024.

\begin{table}[htbp]
    \centering
    \small
    \begin{threeparttable}
    \caption{Named dates: founding of AISI / Bletchley Summit, and the Spring Budgets}
    \label{tab:app-named}
\begin{tabular}{lrrrr}
    \toprule
    Date & $\hat\beta_{1m}$ & $p$ & $\hat\beta_{2m}$ & $p$ \\
    \midrule
    Spring Budget 2023 (Mar 2023)       & -0.124 & 0.891 & +15.147 & 0.195 \\
    Bletchley Summit / AISI (Nov 2023)  & +1.330 & 0.192 & -9.027 & 0.384 \\
    Spring Budget 2024 (Mar 2024)       & -1.487 & 0.133 & +3.810 & 0.720 \\
    \bottomrule
  \end{tabular}
    \end{threeparttable}
\end{table}

None of the three dates shows a disproportionate break. Notably, the
month in which the UK AI Safety Institute was founded and the Bletchley
Declaration was signed is itself null, despite sitting almost exactly at
the start of Year 2. Across the full 18-month scan, six individual
months reach conventional significance for mean exposure (Aug 2023,
$p=0.003$; Oct 2023, $p=0.048$; Dec 2023, $p=0.041$; Jan 2024, $p=0.002$; Feb 2024,
$p=0.004$; Apr 2024, $p=0.013$), clustered in the second half of the
window rather than tied to any single date, and none does for dispersion
(closest: Apr 2024, $p=0.106$). The
transition to Year 2's effect therefore appears gradual and diffuse across
late 2023 and early 2024, rather than attributable to any single named
event. \clearpage
\section{Exposure scores: human reviewers}\label{section:AppendixHumanReview}

\begin{figure}[htbp]
    \centering
    \begin{subfigure}[t]{0.48\textwidth}
        \centering
        \includegraphics[width=\linewidth]{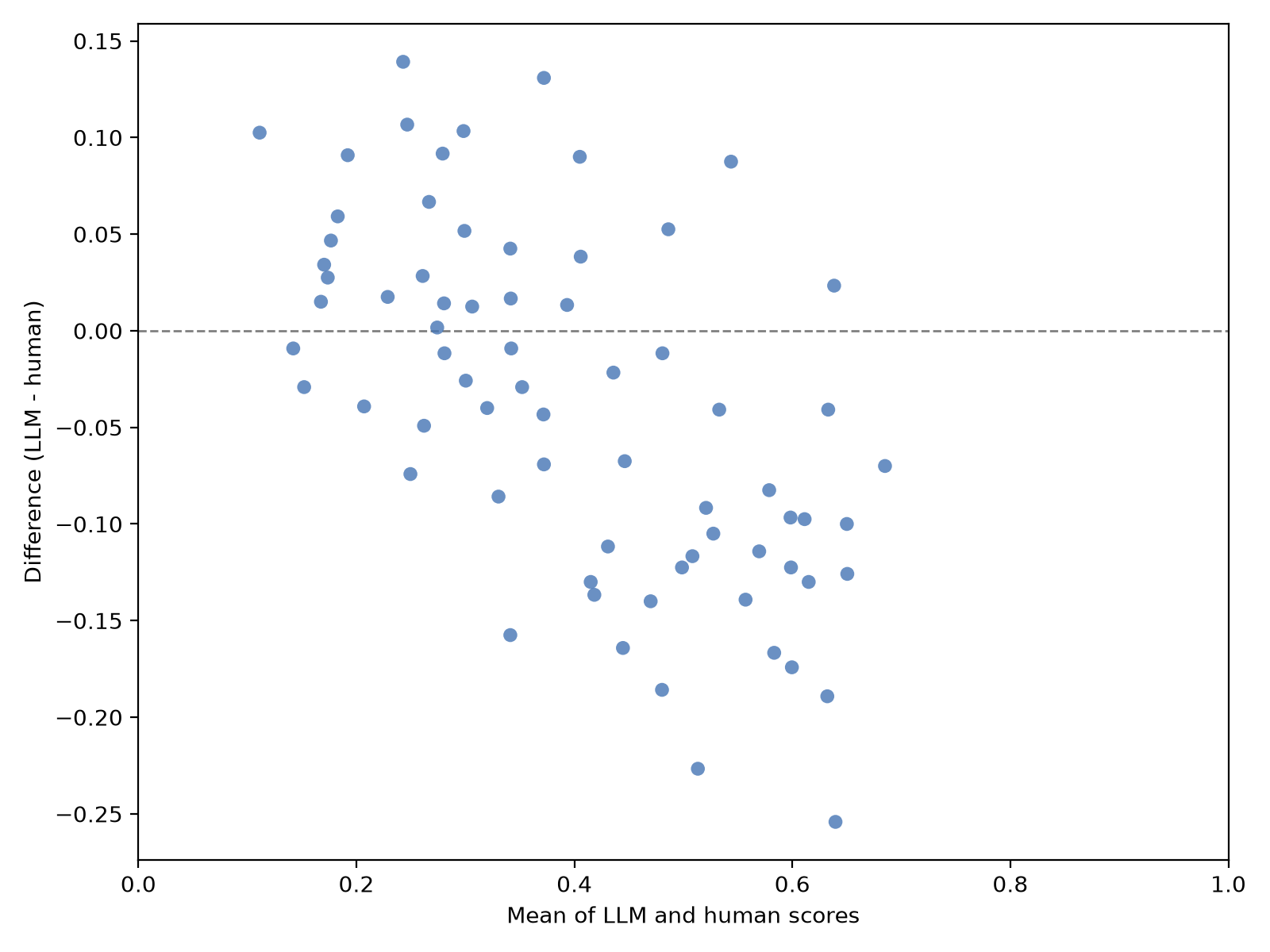}
        \caption{Task level, Bland-Altman plot (n = $71$).}
        \label{fig:human-task-level}
    \end{subfigure}
    \hfill
    \begin{subfigure}[t]{0.48\textwidth}
        \centering
        \includegraphics[width=\linewidth]{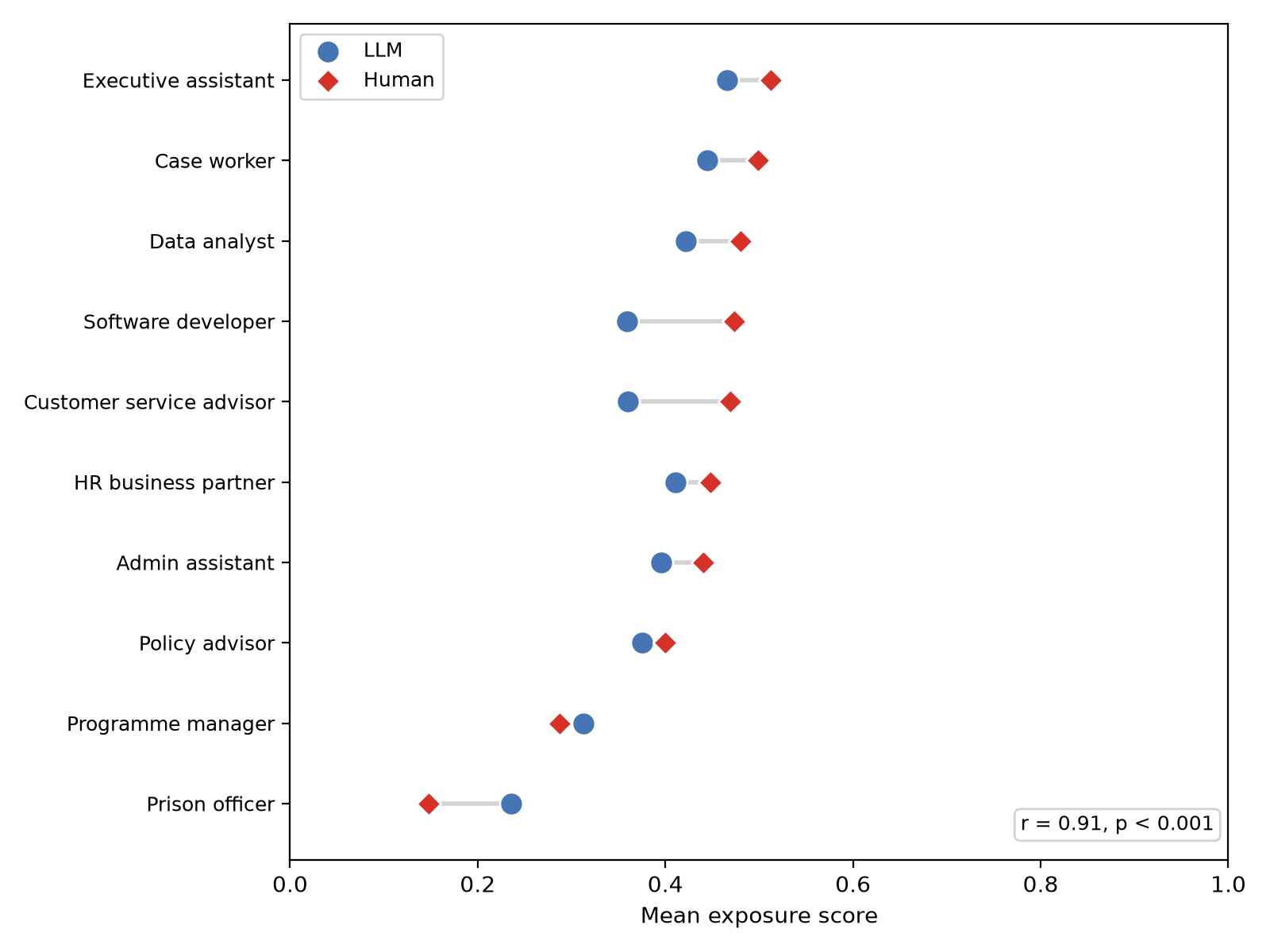}
        \caption{Role level, mean exposure by role (n = $10$).}
        \label{fig:human-role-level}
    \end{subfigure}
    \caption{Ensemble versus human-reviewer agreement on task exposure.}
    \label{fig:human_review_ba_role}
\end{figure}

Figure \ref{fig:human_review_ba_role} plots the ensemble versus human-reviewer comparison at task level and role level. Panel \ref{fig:human-task-level} plots each task's score difference (ensemble minus human mean) against the mean of the two, the standard method for comparing two measurement methods rather than treating one as ground truth. We find a negative trend, with the ensemble increasingly under-scoring relative to the human mean as the average score rises ($r = -0.618$ between the mean and the difference, $p < 0.001$). 
Panel \ref{fig:human-role-level} shows the role-by-role comparison between the average LLM- and human-assigned scores, showing in most cases the LLMs score more conservatively. The human average at this level is computed as each reviewer's own mean score within a role, averaged across reviewers. The ensemble scores below the human consensus in 8 of the 10 roles; the two exceptions are Prison Officer, where the ensemble scores $0.088$ points higher than the human mean, and Programme Manager, where the ensemble scores $0.025$ points higher.

\begin{table}[htbp]
    \centering
    \small
    \begin{threeparttable}
    \caption{Individual models compared to human reviewers}
    \label{tab:model_comparison}
\begin{tabular}{lcccc}
\toprule
Model & $r$ & $\rho$ & Bias & MAE \\
\midrule
\textbf{Ensemble} & \textbf{0.894} & \textbf{0.901} & \textbf{-0.039} & \textbf{0.081} \\
sonnet-4-6 & 0.874 & 0.875 & -0.171 & 0.171 \\
qwen-3-235b & 0.868 & 0.869 & +0.029 & 0.081 \\
llama-maverick & 0.813 & 0.820 & +0.048 & 0.099 \\
glm-5 & 0.812 & 0.846 & -0.068 & 0.104 \\
openai-oss-120b & 0.787 & 0.790 & -0.052 & 0.103 \\
nemotron-super & 0.717 & 0.732 & -0.078 & 0.116 \\
ministral-14b-3 & 0.714 & 0.729 & +0.001 & 0.113 \\
nova-2-lite & 0.676 & 0.691 & -0.020 & 0.119 \\
\bottomrule
\end{tabular}
    \end{threeparttable}
\end{table}

Table \ref{tab:model_comparison} documents how each model's scores align with the human-reviewer mean across tasks. Each of the eight individual scoring models is scored on the same 71 reviewed tasks, and each model's task-level score is correlated against the human-reviewer mean for that same task. The near-tie between the top two models masks a substantial difference in calibration. Qwen-3-235b's bias and MAE are small ($+0.029$ and $0.081$ respectively), while sonnet-4-6's are roughly double any other model's ($-0.171$ and $0.171$). It therefore ranks tasks in almost the same order as qwen-3-235b but scores them systematically lower than the human panel. This is consistent with the same low-scoring pattern this model exhibits in Appendix \ref{AppendixModelConsistencyModelAgreement}. \clearpage
\section{Exposure scores: NLP exposure}\label{section:AppendixNLPExposure}

\begin{figure}[htbp]
    \centering
    \begin{subfigure}[t]{0.48\textwidth}
        \centering
        \includegraphics[width=\linewidth]{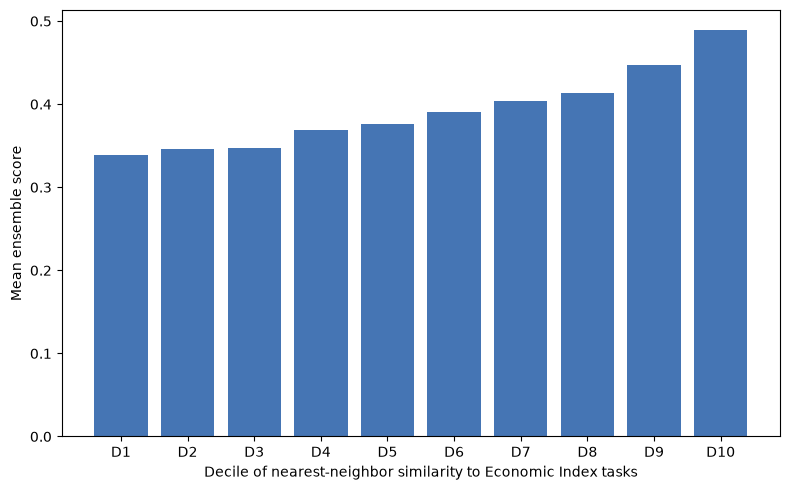}
        \caption{Similarity by task ($n = 2{,}739{,}619$)}
        \label{fig:nlp_tasks_similarity}
    \end{subfigure}
    \hfill
    \begin{subfigure}[t]{0.48\textwidth}
        \centering
        \includegraphics[width=\linewidth]{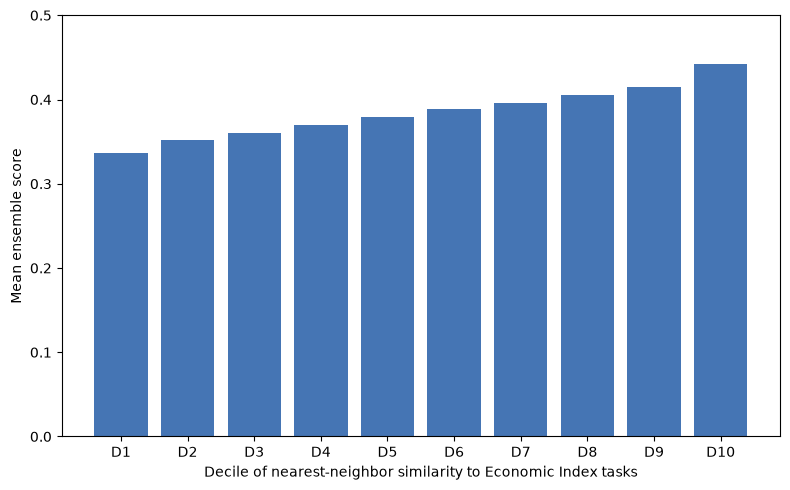}
        \caption{Similarity by vacancy ($n = 289{,}504$)}
        \label{fig:nlp_roles_similarity}
    \end{subfigure}
    \caption{Ensemble exposure scores by decile of NLP similarity}
    \label{fig:nlp_similarity}
\end{figure}

This appendix provides supporting detail for the NLP-based exposure validation described in Section \ref{sec:NLPScores}. Both the task- and role-level results reported use the same underlying similarity measure. Each task's cosine similarity to its closest match in the 680-task reference corpus is computed a single time.\footnote{As a robustness check, we replace nearest-neighbor similarity with the mean cosine similarity to the five nearest reference tasks, weighted by each task’s share of Claude usage. The relationship with ensemble exposure is essentially unchanged, and weighting by usage makes no difference compared with the unweighted mean of the same five tasks.} For task $j$ of vacancy $i$, denote this similarity $c_{ij}$. Task-level figures use $c_{ij}$ directly, alongside the task-level ensemble score. Vacancy-level figures are the average of $c_{ij}$ across all $J$ tasks belonging to vacancy $i$:
\[
C_i = \frac{1}{J}\sum_j c_{ij}.
\]

\begin{table}[h]
    \centering
    \small
    \begin{threeparttable}
    \caption{Individual models compared to Anthropic Economic Index}
    \label{tab:model_sim_comparison}
\begin{tabular}{lcc}
\toprule
Model & Task $r$ & Vacancy $r$ \\
\midrule
sonnet-4-6 & 0.290 & 0.392 \\
\textbf{Ensemble} & \textbf{0.284} & \textbf{0.385} \\
glm-5 & 0.271 & 0.367 \\
nova-2-lite & 0.251 & 0.314 \\
ministral-14b-3 & 0.248 & 0.336 \\
qwen-3-235b & 0.248 & 0.324 \\
nemotron-super & 0.244 & 0.321 \\
llama-maverick & 0.240 & 0.333 \\
openai-oss-120b & 0.230 & 0.278 \\
\bottomrule
\end{tabular}
    \end{threeparttable}
\end{table}

Table \ref{tab:model_sim_comparison} reports the same comparison broken out by individual scoring model, correlating each model's own exposure score against the similarity measure directly rather than through the ensemble average. Sonnet-4-6 ranks highest at both levels of aggregation, with the ensemble close behind in second place throughout. The other models cluster closely together at task level, all falling between $r = 0.230$ (openai-oss-120b) and $r = 0.271$ (glm-5), but this clustering loosens considerably once aggregated to vacancy level. 
 \clearpage
\section{AI exposure comparison: O*NET and ISCO databases}\label{section:AppendixOnetIsco}

\begin{table}[!htb]
    \centering
    \footnotesize
    \begin{threeparttable}
    \caption{Full list of tasks for Economist roles from O*NET and ISCO-08 databases}
    \label{tab:onet_isco_mapping}
    \begin{tabular}{@{}>{\centering\arraybackslash}p{0.8cm}>{\raggedright\arraybackslash}p{7cm}>{\raggedright\arraybackslash}p{7cm}@{}}
        \toprule
        \# & O*NET Task Description & ISCO-08 Task Description \\
        \midrule
        1 & Study economic and statistical data in area of specialization, such as finance, labor, or agriculture 
          & Forecasting changes in the economic environment for short-term budgeting, long-term planning and investment evaluation \\
        \addlinespace
        2 & Compile, analyze, and report data to explain economic phenomena and forecast market trends, applying mathematical models and statistical techniques 
          & Formulating recommendations, policies and plans for the economy, corporate strategies and investment, and undertaking feasibility studies for projects \\
        \addlinespace
        3 & Study the socioeconomic impacts of new public policies, such as proposed legislation, taxes, services, and regulations 
          & Monitoring economic data to assess the effectiveness, and advise on the appropriateness, of monetary and fiscal policies \\
        \addlinespace
        4 & Explain economic impact of policies to the public 
          & Forecasting production and consumption of specific products and services based on records of past production and consumption and general economic and industry-specific conditions \\
        \addlinespace
        5 & Review documents written by others 
          & Preparing forecasts of income and expenditure, interest rates and exchange rates \\
        \addlinespace
        6 & Provide advice and consultation on economic relationships to businesses, public and private agencies, and other employers 
          & Analysing factors that determine labour-force participation, employment, wages, unemployment and other labour-market outcomes \\
        \addlinespace
        7 & Formulate recommendations, policies, or plans to solve economic problems or to interpret markets 
          & Applying mathematical formulae and statistical techniques to test economic theories and devise solutions to economic problems \\
        \addlinespace
        8 & Supervise research projects and students’ study projects 
          & Compiling, analysing and interpreting economic data using economic theory and a variety of statistical and other techniques \\
        \addlinespace
        9 & Conduct research on economic issues, and disseminate research findings through technical reports or scientific articles in journals 
          & Evaluating the outcome of political decisions concerning public economy and finances and advising on economic policy and possible courses of action in light of past, present and projected economic factors and trends \\
        \addlinespace
        10 & Develop economic guidelines and standards, and prepare points of view used in forecasting trends and formulating economic policy 
           & Preparing scholarly papers and reports \\
        \addlinespace
        11 & Teach theories, principles, and methods of economics 
           & Examining problems related to the economic activities of individual companies \\
        \addlinespace
        12 & Testify at regulatory or legislative hearings concerning the estimated effects of changes in legislation or public policy, and present recommendations based on cost-benefit analyses 
           & Conducting research on market conditions in local, regional or national areas to set sales and pricing levels for goods and services, to assess market potential and future trends and to develop business strategies \\
        \addlinespace
        13 & Provide litigation support, such as writing reports for expert testimony or testifying as an expert witness 
           & N/A \\
        \bottomrule
    \end{tabular}
    \begin{tablenotes}[flushleft]
        \footnotesize
        \item \textit{Note.} O*NET tasks are taken from the O*NET 29.3 Database, SOC code 19-3011.00. O*NET tasks are ordered by their data value on the importance scale ('IM'). ISCO-08 tasks are listed for ISCO-08 Code 2631. They are unordered by importance so are simply assigned equal weighting.
    \end{tablenotes}
    \end{threeparttable}
\end{table} \clearpage
\section{Role impact sensitivity}\label{sec:RoleSensitivity}

The quantification of potential organizational impact outlined in Section \ref{sec:workforce-impact} relies upon two main assumptions: the task-level decay weight and the choice of thresholds to classify roles into the four categories using the mean and variance. This appendix tests those assumptions, by comparing different assumption values to the chosen baseline values.

\subsection{Decay weight sensitivity}

\begin{figure}[htbp]
  \centering
  \includegraphics[width=0.8\textwidth]{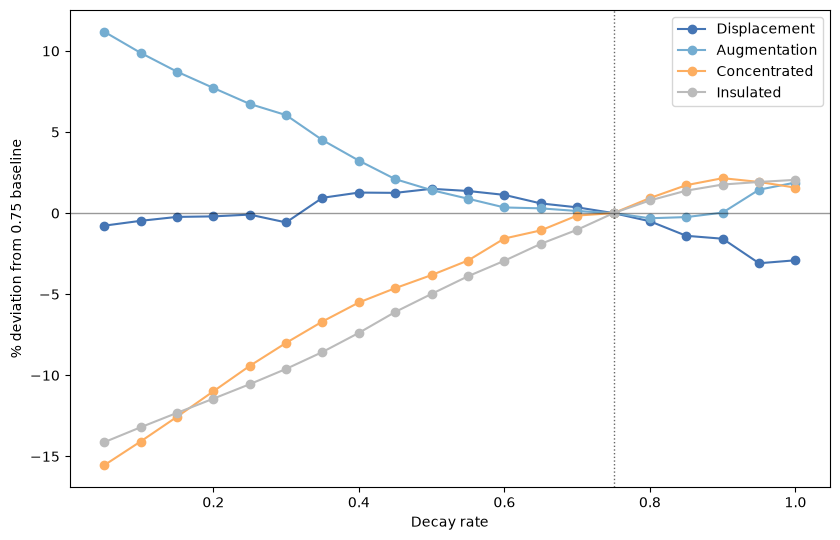}
  \caption{Impact of decay weight on estimated scale of opportunity.}
  \label{fig:decay-sensitivity}
\end{figure}

Using the same thresholds for mean and variance, we recalculate the value of the technical opportunity in each quadrant under different decay weight parameters.

\subsection{Threshold sensitivity}\label{sec:AppendixThresholdSensitivityTable}

\begin{table}[!htb]
    \centering
    \small
    \begin{threeparttable}
    \caption{Share of roles in each category under alternative quartile thresholds}
    \label{tab:ThresholdSensitivityShares}
    \begin{tabularx}{\linewidth}{@{}>{\centering\arraybackslash}p{1.6cm}>{\centering\arraybackslash}p{2.0cm}>{\centering\arraybackslash}X>{\centering\arraybackslash}X>{\centering\arraybackslash}X>{\centering\arraybackslash}X@{}}
        \toprule
        $\mu$ threshold & $\sigma^2$ threshold ($P$ / $1-P$) & Insulated (\%) & Concentrated (\%) & Augmentation (\%) & Displacement (\%) \\
        \midrule
        $\mu_{0.90}$ & 0.10 / 0.90 & 81.00 ($+$44\%) & 9.00 ($-$52\%) & 9.00 ($-$52\%) & 1.00 ($-$84\%) \\
        $\mu_{0.85}$ & 0.15 / 0.85 & 72.25 ($+$28\%) & 12.75 ($-$32\%) & 12.75 ($-$32\%) & 2.25 ($-$64\%) \\
        $\mu_{0.80}$ & 0.20 / 0.80 & 64.00 ($+$14\%) & 16.00 ($-$15\%) & 16.00 ($-$15\%) & 4.00 ($-$36\%) \\
        \textbf{$\mu_{0.75}$} & \textbf{0.25 / 0.75} & \textbf{56.25 (0\%)} & \textbf{18.75 (0\%)} & \textbf{18.75 (0\%)} & \textbf{6.25 (0\%)} \\
        $\mu_{0.70}$ & 0.30 / 0.70 & 49.00 ($-$13\%) & 21.00 ($+$12\%) & 21.00 ($+$12\%) & 9.00 ($+$44\%) \\
        $\mu_{0.65}$ & 0.35 / 0.65 & 42.25 ($-$25\%) & 22.75 ($+$21\%) & 22.75 ($+$21\%) & 12.25 ($+$96\%) \\
        $\mu_{0.60}$ & 0.40 / 0.60 & 36.00 ($-$36\%) & 24.00 ($+$28\%) & 24.00 ($+$28\%) & 16.00 ($+$156\%) \\
        $\mu_{0.55}$ & 0.45 / 0.55 & 30.25 ($-$46\%) & 24.75 ($+$32\%) & 24.75 ($+$32\%) & 20.25 ($+$224\%) \\
        $\mu_{0.50}$ & 0.50 / 0.50 & 25.00 ($-$56\%) & 25.00 ($+$33\%) & 25.00 ($+$33\%) & 25.00 ($+$300\%) \\
        \bottomrule
    \end{tabularx}
    \begin{tablenotes}[flushleft]
        \footnotesize
        \item \textit{Note.} Both thresholds move together as a single symmetric percentile $P$ (quartile = $P=0.25$), decay rate held fixed at 0.75 throughout. Bracketed figures show the relative \% change from the $\mu_{0.75}$ baseline (bold row).
    \end{tablenotes}
    \end{threeparttable}
\end{table}

Table \ref{tab:ThresholdSensitivityShares} presents the proportion of FTE in each category for different threshold values. The category definitions stay the same throughout; only the percentile at which the cuts are drawn changes. Raising $P$ moves FTE out of \textit{Insulated} (81\% of FTE at $P=10\%$, falling to 25\% at $P=50\%$) and into the other three categories. \textit{Displacement} changes most in relative terms, rising from 1\% to 25\% of FTE against 6.25\% at the $P=25\%$ baseline, while \textit{Augmentation} and \textit{Concentrated} each rise from 9\% to 25\%. \clearpage
\section{Job redesign full methodology and pilot testing}
\label{sec:AppendixJobRedesign}

\subsection{Methodology}
\label{sec:AppendixMethodology}

The redesign has two stages. Stage 1 replaces tasks within each role, and Stage 2 uses the time this frees to generate additional tasks. Stage 2 runs twice on the same roles, once with no guidance (the main redesign) and once with an added instruction to propose only tasks an AI system could not already do (the variant).

\begin{figure}[htbp]
    \centering
    \resizebox{\textwidth}{!}{%
    \begin{tikzpicture}[
        node distance=0.8cm and 1.3cm,
        box/.style={draw, rectangle, rounded corners, align=center, minimum height=1cm, text width=2.6cm, font=\small},
        model/.style={draw, rectangle, fill=blue!8, align=center, minimum height=1cm, text width=2.4cm, font=\small\itshape},
        decision/.style={draw, diamond, align=center, minimum width=2cm, inner sep=1pt, font=\small},
        arr/.style={-Stealth, thick},
        stagelabel/.style={font=\bfseries\small, align=left, inner sep=0pt, text width=2.1cm}
    ]

    \node[box] (s1in) at (0,0) {Job description \\ + current tasks};
    \node[model, right=of s1in] (redesign) {Redesign \\ Sonnet 4.6 \\ (\S\ref{sec:AppendixRedesignPrompt})};
    \node[box, right=of redesign] (s1out) {Removed task \\ + new task text};
    \node[model, right=of s1out] (rescore1) {Rescoring \\ GPT-OSS-120b \\ (\S\ref{sec:AppendixRescoringPrompt})};
    \node[box, right=of rescore1] (s1scored) {New task + \\ automation potential};
    \node[stagelabel, anchor=south west] (s1label) at ($(s1in.north west)+(0,0.6)$) {\hyphenpenalty=10000\exhyphenpenalty=10000 Stage 1: task redesign};

    \draw[arr] (s1in) -- (redesign);
    \draw[arr] (redesign) -- (s1out);
    \draw[arr] (s1out) -- (rescore1);
    \draw[arr] (rescore1) -- (s1scored);

    \node[box, below=2.4cm of s1in] (s2in) {Retained + new tasks \\ (advert order)};
    \node[stagelabel, anchor=south west] (s2label) at ($(s2in.north west)+(0,0.6)$) {\hyphenpenalty=10000\exhyphenpenalty=10000 Stage 2: generating new tasks};
    \node[model, right=of s2in] (insert) {Task-insertion ranking \\ Sonnet 4.6 \\ (\S\ref{sec:AppendixTaskInsertionPrompt})};
    \node[box, right=of insert] (hours) {Hours freed \\ per role};
    \node[decision, right=of hours] (netpos) {Net \\ positive?};
    \node[model, below=of netpos] (cleanjd) {Clean job description \\ Sonnet 4.6 \\ (\S\ref{sec:AppendixCleanJDPrompt})};
    \node[model, right=of cleanjd] (backfill) {Stage 2 task generation \\ (main and variant) \\ Sonnet 4.6 \\ (\S\ref{sec:AppendixBackfillPrompt})};
    \node[box, right=1.6cm of netpos] (nobackfill) {No further \\ action};
    \node[model, below=of backfill] (rescore2) {Stage 2 rescoring \\ (main and variant) \\ GPT-OSS-120b \\ (\S\ref{sec:AppendixRescoringPrompt})};

    \draw[arr] (s2in) -- (insert);
    \draw[arr] (insert) -- (hours);
    \draw[arr] (hours) -- (netpos);
    \draw[arr] (netpos) -- node[midway, above, fill=white, inner sep=1pt, font=\small] {no} (nobackfill);
    \draw[arr] (netpos) -- node[midway, right, fill=white, inner sep=1pt, font=\small] {yes} (cleanjd);
    \draw[arr] (cleanjd) -- (backfill);
    \draw[arr] (backfill) -- (rescore2);

    \draw[arr, dashed] (s1scored.south) -- ++(0,-0.35) -| ($(s2in.north)+(0.6,0)$);

    \end{tikzpicture}%
    }
    \caption{The two-stage job redesign pipeline. Solid arrows show the flow of a single role through the pipeline, and the dashed arrow shows Stage 1's output feeding Stage 2's input. Stage 2 runs twice on the same roles, once with no guidance (main) and once with an added instruction to propose only tasks an AI system could not already do (variant). Each model box cites the appendix subsection containing the full prompt used at that step.}
    \label{fig:JobRedesignPipeline}
\end{figure}

\subsubsection{Sampling}
\label{sec:AppendixSamplingMethodology}

The 10,000 roles are drawn from the Augmentation quadrant so that the sample represents the quadrant as a whole. The quadrant contains roles with at least three tasks whose decay-weighted mean exposure is in the top 25\% of the distribution and whose variance is above the 25th percentile among those roles. It comprises 57,179 roles and 89,727 full-time equivalents (FTE). In priority order, the sample is selected so that (1) its average exposure and variance match the FTE-weighted values for the quadrant, (2) its grade shares match the quadrant's FTE shares, and (3) its department and profession shares match the quadrant's FTE shares as closely as possible.

Mean exposure and variance are each divided into six bins using FTE-weighted quantiles within the quadrant. Grade totals are set in proportion to the quadrant's FTE shares by grade, capped at the number of roles available, with any shortfall redistributed to other grades. Within each grade, the total is split across the mean-by-variance bins in proportion to FTE and capped in the same way, and roles are then drawn at random within each cell. A swap search exchanges roles within a cell to move department and profession shares toward their targets while holding the sample's mean and variance at theirs. The procedure is deterministic given a fixed seed, and rerunning it returns the same sample.

The sample matches the quadrant's FTE-weighted average exposure (0.473) and variance (0.0193) to the reported precision, and its grade shares to within 0.02 percentage points (Table~\ref{tab:GradeAllocation}). Department and profession shares deviate from the quadrant's by 0.6 and 0.4 percentage points in total, largely because a few departments and the ``other'' profession have few roles in the quadrant. Because the quadrant contains relatively few senior roles, its grade profile differs from that of the whole Civil Service, with AO making up 41\% of the quadrant against 25\% of the Civil Service and G7 5\% against 12\%.

\begin{table}[htbp]
    \centering
    \small
    \begin{threeparttable}
    \caption{Target versus achieved grade shares in the 10,000-role sample.}
    \label{tab:GradeAllocation}
\begin{tabular}{lrr}
        \toprule
        Grade & Quadrant FTE share & Sample share \\
        \midrule
        AA        & 1.90\%  & 1.89\% \\
        AO        & 41.13\% & 41.15\% \\
        EO        & 25.10\% & 25.11\% \\
        HEO       & 15.61\% & 15.61\% \\
        SEO       & 10.30\% & 10.29\% \\
        G7        & 5.01\%  & 5.01\% \\
        G6        & 0.83\%  & 0.83\% \\
        SCS level & 0.12\%  & 0.11\% \\
        \bottomrule
    \end{tabular}
    \begin{tablenotes}[flushleft]
        \footnotesize
        \item \textit{Note.} The target is the grade distribution of the Augmentation quadrant, weighted by FTE.
    \end{tablenotes}
    \end{threeparttable}
\end{table}

\subsubsection{Role-level exposure}
\label{sec:AppendixRoleLevelExposure}

Role-level exposure is the time-weighted mean of the GPT-OSS-120b scores of a role's tasks, and role-level variance is the corresponding time-weighted variance. Before redesign, each task is weighted by the decay rate (0.75 per position in the advert's task list), scaled so that the weights sum to a 37-hour week. After Stage 1, retained tasks keep their original weights and each replacement takes the weight of the position the task-insertion step ranks it at. Each Stage 2 task is weighted by its estimated hours per week. Because the replacements are ranked at different positions from the tasks they replace, the weights after Stage 1 sum to less than 37 hours in roles where time is freed, and this difference is the freed time that Stage 2 fills. Unweighted versions, which treat every task equally, give the same pattern.

\providecommand{\hd}[1]{\begin{tabular}[b]{@{}c@{}}#1\end{tabular}}

\subsection{Pilot testing}
\label{sec:AppendixPilotTesting}

To test two circularity risks, we compare three approaches to prompting the model on a 100-role sample before the full run. The first uses our prompt as is. The second withholds the exposure scores, to check whether showing the model the original scores biases which tasks it chooses to regenerate, even under the explicit instruction not to react to them. The third forcibly regenerates all tasks above an exposure score of 0.7, to identify whether only ever touching preselected high scorers inflates the measured effect by construction, independent of redesign quality.

\begin{table}[htbp]
    \centering
    \small
    \begin{threeparttable}
    \caption{Smoke-test comparison of the three circularity-risk conditions on the same 100-vacancy pilot sample.}
    \label{tab:SmokeTestComparison}
\begin{tabular}{lrrrrr}
        \toprule
        Condition & n vacancies & n=0 rate & n swaps & Mean delta & \% swaps lower \\
        \midrule
        Scored              & 100 & 33\% & 76  & $-$0.161 & 81.6\% \\
        No-scores           & 100 & 21\% & 98  & $-$0.088 & 56.1\% \\
        Forced ($\geq$0.7)  & 100 & 84\% & 18  & $-$0.306 & 94.4\% \\
        \bottomrule
    \end{tabular}
    \end{threeparttable}
\end{table}

The results indicate that, although the model is told not to react to the exposure score, showing it biases the model toward removing tasks that have higher scores. The measured reduction roughly halves once scores are withheld (from $-$0.161 to $-$0.088) and the share of swaps that score lower falls from 82\% to 56\%, which is barely above chance. Forcibly regenerating all tasks above an exposure score of 0.7 mechanically inflates the impact significantly ($-$0.306, 94\% lower) compared to the approaches where tasks are not preselected. Based on these results, we withhold exposure scores and use the default prompt for the job redesign exercise on the 10,000-role sample.

To assess sensitivity to the choice of generation model, we repeat the scored condition on the same 100-vacancy sample using GPT-OSS-120b in place of Sonnet 4.6.

\begin{table}[htbp]
    \centering
    \small
    \begin{threeparttable}
    \caption{Generation-model sensitivity check, with Sonnet 4.6 and GPT-OSS-120b as the redesign generator in the scored condition, on the same 100-vacancy sample.}
    \label{tab:GenerationModelSensitivity}
\begin{tabular}{lrrrr}
        \toprule
        Generation model & n swaps & Mean delta & \% swaps lower & \hd{\% needing\\JSON fallback} \\
        \midrule
        Sonnet 4.6 (scored)     & 76  & $-$0.161 & 81.6\% & 0\% \\
        GPT-OSS-120b (scored)   & 137 & $-$0.012 & 46.7\% & 24\% \\
        \bottomrule
    \end{tabular}
    \end{threeparttable}
\end{table}

The measured effect collapses (mean delta $-$0.012, with 47\% of swaps scoring lower), and 24\% of responses require text-based JSON recovery rather than a structured tool call, versus 0\% for Sonnet. We take this as evidence that the measured effect depends on genuine generation quality and is not an artifact of the pipeline structure, since a weaker model run through the identical non-circular architecture does not reproduce the finding.

These checks address two routes by which circularity could enter the design, but they do not make the generation and scoring models independent. Both are language models and may share a view of which tasks are automatable, so some link between the tasks that are removed and the scores assigned to them is to be expected.

\subsection{Additional results}
\label{sec:AppendixAdditionalResults}

\subsubsection{Stage 1 headline results}
\label{sec:AppendixStage1Results}

Table~\ref{tab:Stage1Headline} gives the headline figures for the task replacements described in Section~\ref{sec:ResultsRoleRedesign}.

\begin{table}[htbp]
    \centering
    \small
    \begin{threeparttable}
    \caption{Stage 1 headline results, 10,000-role sample.}
    \label{tab:Stage1Headline}
\begin{tabular}{lr}
        \toprule
        Metric & Value \\
        \midrule
        Vacancies & 10,000 \\
        Original tasks & 105,452 \\
        n=0 (no redesign) & 421 (4.2\%) \\
        Tasks removed and replaced & 12,725 \\
        New tasks scored & 12,725 \\
        Removed-task mean exposure & 0.599 \\
        New-task mean exposure & 0.517 \\
        Retained-task mean exposure & 0.411 \\
        Mean delta (new $-$ removed) & $-$0.082 \\
        Swaps scoring lower & 62.2\% \\
        GPT-OSS-120b vs 8-model ensemble correlation & 0.895 \\
        \bottomrule
    \end{tabular}
    \begin{tablenotes}[flushleft]
        \footnotesize
        \item \textit{Note.} Task-level means are unweighted averages of task scores.
    \end{tablenotes}
    \end{threeparttable}
\end{table}

\subsubsection{Stage 1 results by grade}
\label{sec:AppendixGradeResults}

Table~\ref{tab:Stage1ByGrade} gives the grade-level figures. The fall in exposure is smallest at AA and AO and grows with seniority. In practice this means the benefit of this kind of redesign reaches senior roles more than junior ones. The AO figures include 1,316 swaps in HM Courts and Tribunals Service, where the replacements are more exposed than the tasks removed, and the SCS figures rest on only 13 swaps.

\begin{table}[htbp]
    \centering
    \small
    \begin{threeparttable}
    \caption{Stage 1 results by grade, task-level exposure and change per swap.}
    \label{tab:Stage1ByGrade}
\begin{tabular}{lrrrrr}
            \toprule
            Grade & n & Removed mean & New mean & Mean delta & \% lower \\
            \midrule
            AA        & 220 & 0.591 & 0.531 & $-$0.059 & 53.6\% \\
            AO        & 5,428 & 0.564 & 0.526 & $-$0.038 & 56.9\% \\
            EO        & 3,045 & 0.633 & 0.521 & $-$0.111 & 65.5\% \\
            HEO       & 1,997 & 0.628 & 0.512 & $-$0.117 & 66.0\% \\
            SEO       & 1,308 & 0.617 & 0.497 & $-$0.120 & 67.9\% \\
            G7        & 612 & 0.609 & 0.482 & $-$0.127 & 70.8\% \\
            G6        & 102 & 0.607 & 0.478 & $-$0.128 & 64.7\% \\
            SCS level & 13 & 0.631 & 0.439 & $-$0.192 & 76.9\% \\
            \bottomrule
        \end{tabular}
    \begin{tablenotes}[flushleft]
        \footnotesize
        \item \textit{Note.} \textit{n} is the number of tasks replaced at that grade. \textit{Removed mean} and \textit{New mean} are the average exposure scores of the tasks the model removes and of the tasks it writes to replace them. \textit{Mean delta} is the average change per swap (new minus removed), so a negative value means the replacement is less exposed. \textit{\% lower} is the share of swaps where the replacement scores lower than the task it replaces. Scores run from 0 to 1, and higher means more automatable.
    \end{tablenotes}
    \end{threeparttable}
\end{table}

\subsubsection{Reading the review pattern}
\label{sec:AppendixReviewPattern}

Nearly two thirds of the Stage 1 replacements ask the postholder to review, validate or check AI output. A task phrased this way scores as less automatable, but the scorer sees only the wording. Some of the fall in exposure may therefore come from putting a human in the loop on paper and not from a real change in what the person does. This design cannot separate the two. The replacements are not interchangeable, since each is about the role's own domain, but most of them share the same shape.

\subsubsection{Which tasks the model removes}
\label{sec:AppendixRemovedTasks}

Because the exposure scores are withheld from the model, we check whether the tasks it removes are nonetheless the most exposed ones in the role. Among the 9,579 roles in which the model replaces some but not all tasks, the removed tasks score on average 0.20 higher than the tasks retained in the same role, and sit at the 77th percentile of the role's task scores. In 63\% of these roles at least one removed task is among the role's highest-scoring tasks, against 19\% if the same number of tasks were chosen at random. In the 6,793 roles with a single swap, the removed task is a highest-scoring task in 62\% of roles, against 17\% at random. Scoring tasks with the eight-model ensemble instead of GPT-OSS-120b gives the same result (63\% against 16\%). The model therefore removes the most exposed work most of the time, but not always, and the other 37\% of roles are cases where it removes a task that is exposed to a lesser degree.

This has two implications for the Stage 1 results. First, the fall in exposure at the task level is partly by construction, since the replaced tasks start from the top of each role's distribution and any replacement is likely to score lower. This is why the role-level measure in Section~\ref{sec:ResultsRoleRedesign}, which also includes the tasks that are not replaced, is the more informative one. Second, it supports the view that the model identifies the tasks AI could take over, even without seeing the scores.

\subsubsection{Stage 2 and overall results}
\label{sec:AppendixStage2Results}

Table~\ref{tab:Stage2Headline} gives the headline figures for Stage 2 and for the whole role. Of the 9,579 roles with at least one swap, 48\% are left with no net time freed, because the replacement is ranked as taking as much time as the original or more. Where time is freed it averages 2.72 hours a week, roughly 7\% of a 37-hour week. That is the room Stage 2 has to work with.

\begin{table}[htbp]
    \centering
    \small
    \begin{threeparttable}
    \caption{Stage 2 and overall headline results for the main redesign and the variant.}
    \label{tab:Stage2Headline}
\begin{tabular}{lrr}
        \toprule
        Metric & Main & Variant \\
        \midrule
        Roles with $\geq$1 swap & \multicolumn{2}{c}{9,579} \\
        Net-positive hours freed & \multicolumn{2}{c}{4,965 (51.8\%)} \\
        Mean hours freed & \multicolumn{2}{c}{2.72/week} \\
        Roles given no new tasks & 750 (15.1\%) & 758 (15.3\%) \\
        New tasks proposed & 7,894 & 7,486 \\
        New-task mean exposure & 0.550 & 0.468 \\
        Freed hours used, median (roles given tasks) & 92.2\% & 92.1\% \\
        Freed hours used, 75th percentile (roles given tasks) & 97.1\% & 97.1\% \\
        Largest ratio of proposed to freed hours & 1.15 & 1.09 \\
        Role-level exposure, before & \multicolumn{2}{c}{0.456} \\
        Role-level exposure, after Stage 1 only & \multicolumn{2}{c}{0.443} \\
        Role-level exposure, after Stages 1 and 2 & 0.447 & 0.444 \\
        Role-level mean delta (after Stages 1 and 2 $-$ before) & $-$0.009 & $-$0.012 \\
        Role-level variance, before & \multicolumn{2}{c}{0.0277} \\
        Role-level variance, after Stage 1 only & \multicolumn{2}{c}{0.0238} \\
        Role-level variance, after Stages 1 and 2 & 0.0242 & 0.0239 \\
        Roles with lower exposure after Stages 1 and 2 & 64.7\% & 66.9\% \\
        \bottomrule
    \end{tabular}
    \begin{tablenotes}[flushleft]
        \footnotesize
        \item \textit{Note.} Role-level exposure and variance are time-weighted across all 10,000 roles, with retained and Stage 1 tasks weighted by the decay rate and Stage 2 tasks by their estimated hours per week. New-task mean exposure is an unweighted average of task scores.
    \end{tablenotes}
    \end{threeparttable}
\end{table}

To see what each stage contributes, Table~\ref{tab:Round2Marginal} restricts the sample to the roles that are given new tasks in Stage 2 (4,215 in the main redesign and 4,207 in the variant), since these are the only roles in which Stage 2 can have any effect. Stage 1 takes these roles from 0.454 to 0.435. Adding the main-redesign tasks from Stage 2 pushes exposure back up to 0.444, giving back about half of Stage 1's gain, while adding the variant tasks gives 0.438, so the guidance removes most of what Stage 2 gives back but not all of it. Stage 1 lowers exposure in 76\% of these roles, but Stage 2 lowers it in only 14\% of roles in the main redesign and 40\% in the variant, so most roles go up in both cases and the average rises.

\begin{table}[htbp]
    \centering
    \small
    \begin{threeparttable}
    \caption{Role-level exposure for the roles that are given new tasks in Stage 2 (n=4,215 for the main redesign and n=4,207 for the variant).}
    \label{tab:Round2Marginal}
\begin{tabular}{lrrr}
        \toprule
        Stage & Mean exposure & Average variance & \% of roles lower \\
        \midrule
        Before redesign          & 0.454 & 0.0296 & --- \\
        After Stage 1 only       & 0.435 & 0.0243 & 76.1\% \\
        After Stages 1 and 2 (main)    & 0.444 & 0.0251 & 14.0\% \\
        After Stages 1 and 2 (variant)    & 0.438 & 0.0246 & 40.2\% \\
        \bottomrule
    \end{tabular}
    \begin{tablenotes}[flushleft]
        \footnotesize
        \item \textit{Note.} Tasks are weighted by the share of the working week they take. The final column shows the share of roles where exposure falls, relative to before redesign for Stage 1 and relative to after Stage 1 for the two Stage 2 runs.
    \end{tablenotes}
    \end{threeparttable}
\end{table}

\subsubsection{Quadrant assignment after redesign}
\label{sec:AppendixQuadrantCalibration}

The quadrants are defined using the mean score of eight models, whereas the redesigned tasks are scored with GPT-OSS-120b alone. Over the 2.98 million scored tasks (the full scored corpus, before de-duplication and the sample filters described in Appendix~\ref{sec:GRIDCleaning}), the two scores correlate at 0.89, but GPT-OSS-120b has a wider spread (standard deviation 0.187 against 0.158 for the ensemble) and the two have similar means (0.386 and 0.391). Assigning redesigned roles to quadrants directly from GPT-OSS-120b scores would therefore mix a change of scorer with the effect of redesign. It would also place only 74\% of the sampled roles in Augmentation before any redesign, even though all were selected on the ensemble. Selecting roles on one noisy measure and then measuring them with a different one pulls them toward the average, so this share cannot be 100\% on the second measure.

To avoid this, we map GPT-OSS-120b scores onto the ensemble scale with a quantile map fitted on all original tasks in the population, so that a score at a given percentile of the GPT-OSS-120b distribution is replaced by the ensemble score at the same percentile. For each role we compute the change in time-weighted mean and variance at each stage on this adjusted scale and add it to the role's ensemble mean and variance before redesign. Each role therefore starts in Augmentation, and any movement reflects the redesign. Table~\ref{tab:QuadrantShares} reports the result. A linear map gives similar figures, with 77\% and 76\% of roles remaining in Augmentation after the main redesign and the variant.

\begin{table}[htbp]
    \centering
    \small
    \begin{threeparttable}
    \caption{Share of the 10,000 roles in each quadrant before and after redesign, using the thresholds in Table~\ref{tab:CategoryConditions}.}
    \label{tab:QuadrantShares}
\begin{tabular}{lrrrr}
        \toprule
        Quadrant & Before & \hd{After\\Stage 1} & \hd{After Stages 1\\and 2 (main)} & \hd{After Stages 1\\and 2 (variant)} \\
        \midrule
        Augmentation & 100\% & 69.8\% & 73.4\% & 71.9\% \\
        Displacement & 0\% & 12.9\% & 12.6\% & 12.5\% \\
        Insulated & 0\% & 12.8\% & 10.1\% & 11.5\% \\
        Concentrated & 0\% & 4.5\% & 4.0\% & 4.1\% \\
        \midrule
        Mean exposure & 0.473 & 0.462 & 0.465 & 0.463 \\
        Mean variance & 0.0193 & 0.0165 & 0.0168 & 0.0166 \\
        \bottomrule
    \end{tabular}
    \begin{tablenotes}[flushleft]
        \footnotesize
        \item \textit{Note.} Exposure and variance are on the ensemble scale, calculated as each role's ensemble value before redesign plus the change in its adjusted GPT-OSS-120b value.
    \end{tablenotes}
    \end{threeparttable}
\end{table}

Most of the movement out of Augmentation is into Displacement and Insulated, which are the low-variance quadrants. Mean exposure falls by only 0.01, so most roles keep a mean above the threshold of 0.417, but the average variance falls by about 14\%. The roles that move to Displacement have a median variance of 0.014 before redesign, close to the Augmentation threshold of 0.010, and their variance falls by 0.008 on average. These roles are therefore close to the threshold, and the replacement and additional tasks, which cluster in a narrow band of exposure, lower their spread enough to cross it. This shows that exposure remains high in these roles but is less varied. It does not show that they become more automatable.

\subsubsection{Stage 2 task content clustering}
\label{sec:AppendixBackfillClustering}

The task categories described below sort every task into a fixed set of nine categories, so that the additional tasks can be compared with the original ones. They cannot show what kinds of task the model writes beyond those categories, or why the variant's tasks score lower. For that purpose, we group the additional tasks by their content without a predefined scheme. We pool the 15,380 task descriptions from the main redesign and the variant, and cluster them using term frequency-inverse document frequency (TF-IDF) vectorization, dimensionality reduction with truncated singular value decomposition (SVD), and k-means clustering into 12 groups. Fitting the two runs together makes the clusters comparable across them. Table~\ref{tab:Stage2Clusters} shows each cluster's share of each run, ordered from the largest fall to the largest rise, along with its mean exposure in each run and the share of its tasks carrying the AI-output marker, which flags tasks that explicitly mention AI-generated or AI-drafted output. The labels are short descriptions of each cluster's top terms and example tasks.

\begin{table}[htbp]
    \centering
    \footnotesize
    \begin{threeparttable}
    \caption{Content clusters of the pooled additional tasks from the main redesign and the variant (k=12).}
    \label{tab:Stage2Clusters}
\begin{tabular}{p{4.3cm}rrrrr}
        \toprule
        & \multicolumn{2}{c}{Share of tasks} & \multicolumn{2}{c}{Mean exposure} & \\
        \cmidrule(lr){2-3} \cmidrule(lr){4-5}
        Theme & Main & Variant & Main & Variant & AI marker \\
        \midrule
        Trackers and weekly updates & 12.7\% & 7.3\% & 0.62 & 0.57 & 0.4\% \\
        Error logs for AI tools & 15.2\% & 10.4\% & 0.62 & 0.59 & 32.1\% \\
        Guidance and procedures & 11.2\% & 6.4\% & 0.56 & 0.55 & 14.2\% \\
        Briefings and horizon scanning & 12.2\% & 8.7\% & 0.63 & 0.58 & 5.5\% \\
        Reviews and written findings & 16.3\% & 13.3\% & 0.56 & 0.51 & 30.2\% \\
        Coaching and training & 4.0\% & 4.0\% & 0.37 & 0.35 & 37.1\% \\
        Casework sampling & 9.5\% & 9.8\% & 0.54 & 0.49 & 39.5\% \\
        Peer support & 3.0\% & 4.2\% & 0.30 & 0.28 & 13.8\% \\
        Court-user support & 3.2\% & 4.9\% & 0.37 & 0.32 & 10.5\% \\
        Outbound calls & 7.0\% & 9.8\% & 0.50 & 0.38 & 6.9\% \\
        One-to-one meetings & 3.8\% & 9.9\% & 0.40 & 0.39 & 17.6\% \\
        Stakeholder sessions & 1.8\% & 11.3\% & 0.46 & 0.40 & 4.0\% \\
        \bottomrule
    \end{tabular}
    \begin{tablenotes}[flushleft]
        \footnotesize
        \item \textit{Note.} Shares are of each run's tasks. The AI marker column is the share of the cluster's tasks, across both runs, that explicitly mention AI-generated or AI-drafted output.
    \end{tablenotes}
    \end{threeparttable}
\end{table}

No single theme dominates. The largest cluster covers 14.9\% of tasks and the top five together cover 57.9\%. The written-output themes (trackers and weekly updates, error logs, guidance and procedures, and briefings) make up about half of the main-redesign tasks (51\%) and a third of the variant tasks (33\%). The contact-based themes (stakeholder sessions, one-to-one meetings, coaching, peer support, court-user support and outbound calls) grow from 23\% to 44\% of tasks, with the largest rise in stakeholder sessions, from 2\% to 11\%. The theme built around sampling AI-assisted casework is as common in the variant (9.8\%) as in the main redesign (9.5\%), so the instruction does not remove oversight work.

Exposure is lower in the variant in every theme. About 44\% of the fall in mean exposure comes from the shift between themes, and the rest from lower exposure within themes, which may reflect how tasks of the same type are written as well as what they involve. The largest themes in the main redesign are mostly recurring reporting and monitoring work, which fits the main-redesign tasks scoring higher on exposure than the Stage 1 replacements (0.550 against 0.517). A plausible reading is that when the model is asked to fill hours rather than replace a task it thinks AI handles, it reaches for routine administrative work.

\subsubsection{Task categories}
\label{sec:AppendixTaskCategories}

The clustering above is built from the additional tasks alone. To compare them with the work the roles contained to begin with, every task, original and new, is assigned to one of nine predefined categories of work, or to an ``other'' category, so that the mix of work can be tracked across all groups of tasks. The nine categories are drafted from a random sample of the original tasks, and are Administration, Enquiries, Casework, Stakeholder engagement, Analysis and reporting, Advice, Planning and delivery, People management and Assurance. GPT-OSS-120b assigns each task using the same prompt for all tasks (Appendix~\ref{sec:AppendixTaskCategoryPrompt}). The prompt contains only the category definitions and the task text, and does not say whether a task is an original or a new one. Tasks from all groups are shuffled together within each request. On a sample of 1,720 tasks that GPT-OSS-120b and Sonnet 4.6 both classify, the two agree on 78\% (Cohen's kappa 0.75), with most disagreements between neighboring categories such as Analysis and reporting and Assurance. Table~\ref{tab:TaskCategories} gives each category's share of each group of tasks, counting every task equally. Table~\ref{tab:TaskCategoryShares} in Section~\ref{sec:ResultsRoleRedesign} gives the corresponding shares of a role's working time.

\begin{table}[htbp]
    \centering
    \small
    \begin{threeparttable}
    \caption{Share of tasks in each category of work.}
    \label{tab:TaskCategories}
\begin{tabular}{lrrrrr}
        \toprule
        Category & Original & Removed & \hd{Stage 1\\new} & \hd{Stage 2\\new (main)} & \hd{Stage 2\\new (variant)} \\
        \midrule
        Administration & 28.8\% & 50.9\% & 11.5\% & 11.7\% & 4.9\% \\
        Enquiries & 7.2\% & 3.8\% & 2.3\% & 4.4\% & 5.3\% \\
        Casework & 7.2\% & 2.6\% & 5.4\% & 5.0\% & 8.5\% \\
        Stakeholder engagement & 11.9\% & 1.6\% & 1.7\% & 4.5\% & 15.1\% \\
        Analysis and reporting & 12.2\% & 29.3\% & 28.0\% & 36.3\% & 25.4\% \\
        Advice & 4.7\% & 1.2\% & 1.9\% & 6.8\% & 6.4\% \\
        Planning and delivery & 9.5\% & 3.8\% & 2.3\% & 3.8\% & 4.8\% \\
        People management & 3.8\% & 0.2\% & 0.2\% & 4.5\% & 7.5\% \\
        Assurance & 10.9\% & 4.9\% & 46.4\% & 22.5\% & 21.8\% \\
        Other & 3.6\% & 1.7\% & 0.1\% & 0.5\% & 0.3\% \\
        \midrule
        Tasks classified & 105,452 & 12,725 & 12,725 & 7,894 & 7,486 \\
        \bottomrule
    \end{tabular}
    \begin{tablenotes}[flushleft]
        \footnotesize
        \item \textit{Note.} The removed tasks are a subset of the original tasks. Each column sums to 100\%.
    \end{tablenotes}
    \end{threeparttable}
\end{table}

The tasks the model removes in Stage 1 are mostly administration (51\%) and analysis and reporting (29\%), which together are 80\% of what it removes against 41\% of the original tasks. The Stage 1 replacements are mostly assurance (46\%) and analysis and reporting (28\%). The main redesign is dominated by analysis and reporting (36\%) and assurance (22.5\%), with little stakeholder engagement (4.5\%). The variant moves toward stakeholder engagement (15\%), people management (7.5\%) and casework (8.5\%), and away from administration (4.9\%) and analysis and reporting (25\%), while assurance is unchanged (22\%).
\clearpage
\section{Team restructuring exercise}\label{sec:AppendixTeamRestructure}

We are interested in how AI automation affects whole teams, not just individuals or isolated workflows, because the effect of automating a task does not stay with the person who performs it: it depends on the colleagues, roles and hand-offs around it. We therefore exploit the team-level data in our dataset to explore to what extent AI can assist with team redesign. The aim is to illustrate what is possible, rather than to forecast an outcome that is proportionate to current exposure or likely to be realized soon.

\subsection{Methodology}
\label{sec:AppendixTeamMethodology}

\begin{figure}[htbp]
    \centering
    \resizebox{\textwidth}{!}{%
    \begin{tikzpicture}[
        node distance=0.8cm and 1.3cm,
        box/.style={draw, rectangle, rounded corners, align=center, minimum height=1cm, text width=2.6cm, font=\small},
        model/.style={draw, rectangle, fill=blue!8, align=center, minimum height=1cm, text width=2.4cm, font=\small\itshape},
        decision/.style={draw, diamond, align=center, minimum width=2cm, inner sep=1pt, font=\small},
        arr/.style={-Stealth, thick},
        stagelabel/.style={font=\bfseries\small, align=left, inner sep=0pt, text width=7cm}
    ]

    \node[box] (teamin) at (0,0) {Team: roles, posts, \\ job descriptions, \\ weighted tasks};
    \node[model, right=of teamin] (choose) {Step 1: role choice \\ Opus 5.5 \\ (\S\ref{sec:AppendixTeamStage1Prompt})};
    \node[box, right=of choose] (chosen) {Role losing one post \\ + protected tasks};
    \node[model, right=of chosen] (triage) {Step 2: automation triage \\ Opus 5.5 \\ (\S\ref{sec:AppendixTeamStage2Prompt})};
    \node[box, right=of triage] (remaining) {Automated tasks removed; \\ remaining workload};
    \node[stagelabel, anchor=south west] (s1label) at ($(teamin.north west)+(0,0.6)$) {\hyphenpenalty=10000\exhyphenpenalty=10000 Steps 1 and 2: removing a post and automating tasks};

    \draw[arr] (teamin) -- (choose);
    \draw[arr] (choose) -- (chosen);
    \draw[arr] (chosen) -- (triage);
    \draw[arr] (triage) -- (remaining);

    \node[box, below=2.4cm of teamin] (s3in) {Remaining + protected \\ tasks};
    \node[stagelabel, anchor=south west] (s3label) at ($(s3in.north west)+(0,0.6)$) {\hyphenpenalty=10000\exhyphenpenalty=10000 Steps 3 and 4: regrouping and backfilling};
    \node[model, right=of s3in] (regroup) {Step 3: regrouping \\ Opus 5.5 \\ (\S\ref{sec:AppendixTeamStage3Prompt})};
    \node[decision, right=of regroup] (fits) {Fits \\ capacity?};
    \node[model, right=1.6cm of fits] (backfill) {Step 4: human-only backfill \\ Opus 5.5 \\ (\S\ref{sec:AppendixTeamStage4Prompt})};
    \node[model, right=of backfill] (rescore) {Blind rescoring \\ GPT-OSS-120b \\ (\S\ref{sec:AppendixScorePrompt})};

    \node[box, below=1.4cm of fits] (retry) {Retry once, \\ with feedback};
    \node[box] (excluded) at (backfill |- retry) {Still over capacity: \\ team excluded};
    \node[box] (final) at (rescore |- retry) {Redesigned team: \\ new roles + tasks};

    \draw[arr] (s3in) -- (regroup);
    \draw[arr] (regroup) -- (fits);
    \draw[arr] (fits) -- node[midway, above, fill=white, inner sep=1pt, font=\small] {yes} (backfill);
    \draw[arr] (backfill) -- (rescore);
    \draw[arr] (rescore) -- (final);
    \draw[arr] (fits) -- node[midway, right, fill=white, inner sep=1pt, font=\small] {no} (retry);
    \draw[arr] (retry.west) -| (regroup.south);
    \draw[arr] (retry) -- node[midway, above, fill=white, inner sep=1pt, font=\small] {fails} (excluded);

    \draw[arr, dashed] (remaining.south) -- ++(0,-0.35) -| ($(s3in.north)+(0.6,0)$);

    \end{tikzpicture}%
    }
    \caption{The team restructuring pipeline. Solid arrows show the flow of a single team through the pipeline, and the dashed arrow shows the remaining workload from Stage 2 feeding Stage 3. Stage 3 may be retried once if a new role's workload exceeds the capacity of its posts. Teams whose remaining work already (almost) fills the reduced headcount skip Stages 3 and 4, and are not shown. Each model box cites the appendix subsection containing the full prompt used at that step.}
    \label{fig:TeamRestructurePipeline}
\end{figure}

The exercise uses 106 teams, built in two steps. First, we identified business areas with high AI exposure. For each area (a department and the detailed business area label from the advert, with at least 10 postings), we calculated the weighted mean of vacancy exposure and its weighted variance across the area's roles. We kept areas in the top quartile of mean exposure, drawing from both the high-dispersion quadrant (uneven exposure across roles) and the low-dispersion quadrant (even exposure across roles). Second, because a business area is too broad to be a team, we asked Claude Sonnet 4.6 to read every advert in these areas and name the specific team the post sits in. We required a verbatim quote from the advert as evidence and rejected any name that could not be found in the text. A second pass then merged variants of the same name and removed labels that were not teams, such as job-title abbreviations. We merged teams across business-area labels within a department and cleaned job titles of locations, organization and grade prefixes so that re-advertised posts counted once. We then retained teams with between 5 and 40 distinct roles and at least two grade-and-profession combinations. We excluded by hand networks of sites, whole organizations, generic names that pool unrelated units and sub-teams of larger shortlisted teams. This gives 106 teams (958 vacancies, 1,480 advertised posts, 94 business areas), of which 37 come from the high-dispersion quadrant, 59 from the low-dispersion quadrant and 10 from both. A team's roster is the union of its adverts over time rather than a snapshot of headcount, and teams are only found where an advert names them. The shortlist is therefore a sample and not the whole Civil Service. 

\begin{figure}[htbp]
  \centering
  \caption{Exposure landscape of business areas}
  \label{fig:BusinessAreaExposureLandscape}
  \begin{threeparttable}
    \includegraphics[width=0.8\textwidth]{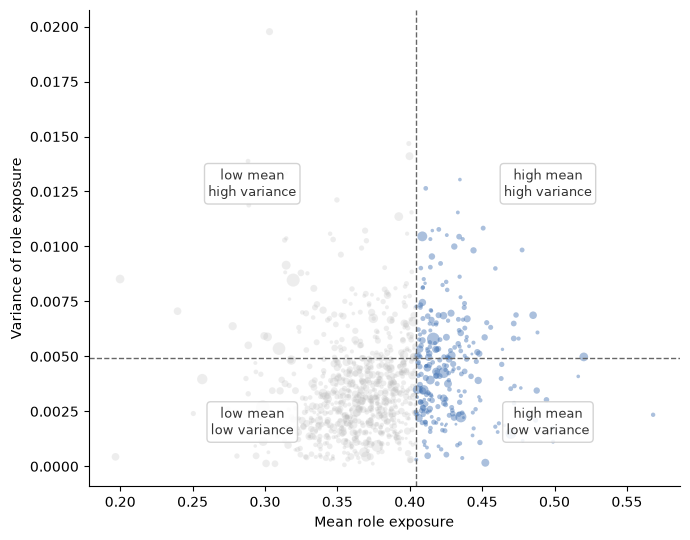}
    \begin{tablenotes}[flushleft]
      \footnotesize
        \item Each point is a business area (a department and business-area label from the job advert) with at least 10 postings; 1,000 areas are shown. Mean and variance are weighted by each vacancy's estimated FTE, and point size is proportional to the square root of the area's total FTE. Dashed lines mark the 75th percentile of each axis across areas. Areas to the left of the vertical line (gray) fall below the top quartile of mean exposure and are not used as a source of teams. Areas in the two blue, high-mean quadrants are the source of teams: 104 with high variance, where exposure is uneven across roles, and 146 with low variance, where it is more even.
        \end{tablenotes}
  \end{threeparttable}
\end{figure}

Each team is described by its roles, the number of posts (people) holding each role, and each role's list of tasks. We convert each task list into weekly hours by weighting tasks according to the order in which they appear in the job advert, so that the hours for one post-holder sum to a 37-hour working week, and multiply by the number of posts to obtain the workload of each role. We set a simple counterfactual: every team loses exactly one post, and AI absorbs enough work for the remaining staff to cover what is left. A large language model (Claude Opus 5.5) makes the judgments at each stage described below, with one draw per team. At no point does the model see our task-level exposure scores, so its choices are independent of the estimates reported in the main text.

\paragraph{Step 1:} The model is shown the whole team (every role with its grade, number of posts, job description and weighted tasks) and selects exactly one role to lose exactly one post. It is asked to prefer the role whose work looks most amenable to AI, but may also weigh duplication with other roles, fixed-term cover and the ease of consolidating work. For the chosen role, it also identifies any tasks that cannot be automated away, such as those that depend on a personal relationship or a physical presence. These tasks are protected and carried forward in full, while the rest of that role's work leaves with the departing post. The prompt is reproduced in Appendix~\ref{sec:AppendixTeamStage1Prompt}.

\paragraph{Step 2:} The model then reviews all remaining tasks in the team and marks each as either something AI can plausibly do away with or something that must remain. Tasks marked for automation are removed from the team's workload. The prompt is reproduced in Appendix~\ref{sec:AppendixTeamStage2Prompt}.

\paragraph{Step 3:} The surviving tasks, together with the protected tasks from Stage 1, are regrouped into a new set of roles with a new allocation of posts that sums to the original total minus one. The model may leave a role unchanged, merge roles that overlap, or move tasks between roles. To keep the result realistic, each new role's workload cannot exceed what its post-holders can deliver in a working week. If the first attempt breaks this limit, the model is given feedback and tries once more. The prompt, including the feedback given on the retry, is reproduced in Appendix~\ref{sec:AppendixTeamStage3Prompt}.

\paragraph{Step 4:} Removing tasks and a post leaves some roles with spare capacity. Rather than treat this slack as pure saving, we ask the model to fill it where there is credible work that still needs a person. This can be a new task or additional time on an existing one, and each addition is tagged with the reason a human is needed (for example, verifying or handling exceptions to AI output, maintaining relationships, taking legal accountability, or physical presence). Backfill is capped at each role's spare capacity, so the redesigned role is never overloaded. As a final check, a different model (GPT-OSS-120B) scores each task in the new roles for automation potential using the same rubric as our main task scores, again without sight of how the task arose. These scores are recorded for analysis but are not used to remove tasks. The backfill prompt is reproduced in Appendix~\ref{sec:AppendixTeamStage4Prompt}, and the scoring prompt in Appendix~\ref{sec:AppendixScorePrompt}.

A small tolerance is allowed around the capacity limit, since a team that is only marginally over capacity need not be treated as a failure. We therefore sort teams into three groups. \emph{Viable} teams complete all four stages with a regrouped structure that fits within capacity. In \emph{no-reorganization} teams the remaining work already (almost) fills the reduced headcount, so there is too little slack to regroup or backfill; the post is removed, the structure is otherwise left as it is, and Stages 3 and 4 are skipped. The remaining teams are \emph{excluded}, either because the remaining work cannot be fit into the reduced headcount even after the retry, or because the regrouped structure only fits within tolerance and so cannot be treated as a clean redesign. Our results focus on the viable teams.

\subsection{Results}\label{sec:AppendixTeamRestructureResults}

We focus on the 80 viable teams, which together hold 1,256 posts in 603 distinct roles (median team size 10.5 posts, ranging from 5 to 158). We first set out the headline effects on teams, then how their structure changes, and finally three anonymized illustrations. All figures describe a single model draw per team and should be read as indicative of what is feasible rather than as estimates.

Removing one post from each team takes 80 posts out of the 1,256 (6.4\%). The number of roles falls much further, from 603 to 496 (17.7\%), so the average role is held by 2.4 FTE rather than 2.1. In 66 of the 80 teams the model merges at least two roles. In the 14 teams where nothing is merged, the number of roles is unchanged, and in none of the 80 does it rise. Of the 496 new roles, 361 (73\%) are an old role carried over whole and 120 (24\%) merge two or more old roles, 81 at one grade and 39 across grades. The remaining 15 (3\%) take all their work from a single old role whose work is also split across other new roles, so they are neither carried over whole nor merged; they occur in 13 teams. The model chose the team's most exposed role in 36 of 80 viable teams, and in the top 3 most exposed roles for 70 teams. Therefore, despite masking the exposure scores, the model chooses often the most exposed, and nearly always a highly exposed role to remove.

Table~\ref{tab:RestructureHours} shows where the work goes. Of the 46,472 weekly hours in these teams, 5.1\% leaves with the removed posts and a further 9.2\% is judged automatable, so that the remaining workload is 85.7\% of the original, against 93.6\% for the capacity of the reduced headcount. The gap, 3,665 hours per week or 8.4\% of the remaining capacity, is the slack that the model's automation judgments create beyond the removal of the post. Backfilling with human-only work adds back 3,369 of these hours, leaving just 296 (0.7\% of capacity) unused, so that, once it is added, total work falls by 7.0\%, close to the 6.4\% fall in headcount. The slack is therefore large relative to the cut, but it is closed almost entirely by the model's own proposals for new human work, which are capped at the available slack. The results should thus be read as showing that plausible human-only work can be found to fill the capacity that automation frees (not that this work would exist, or be valued, in practice).

\begin{table}[h]
    \centering
    \small
    \begin{threeparttable}
    \caption{Weekly hours across the 80 viable teams, before and after the restructuring.}
    \label{tab:RestructureHours}
\begin{tabular}{lrr}
\toprule
 & Hours per week & \% of original \\
\midrule
Original workload (posts $\times$ 37 hours) & 46,472 & 100.0 \\
\quad Leaves with the removed posts (Step 1) & $-$2,347 & $-$5.1 \\
\quad Automated (Step 2) & $-$4,278 & $-$9.2 \\
Remaining workload & 39,847 & 85.7 \\
\quad Human-only work added (Step 4) & $+$3,369 & $+$7.2 \\
Redesigned workload & 43,216 & 93.0 \\
\midrule
Capacity of the reduced headcount & 43,512 & 93.6 \\
\bottomrule
\end{tabular}
    \begin{tablenotes}[flushleft]
        \footnotesize
        \item \textit{Note.} Capacity is the remaining posts multiplied by a 37-hour week.
    \end{tablenotes}
    \end{threeparttable}
\end{table}

The share of hours automated varies across teams, with a median of 8.2\% and a maximum of 38\%; two teams have no work automated at all. It is modestly higher in teams with greater average exposure (correlation of 0.29), and lower in the largest teams (a median of 4.8\% in teams with 21 or more posts, against 7 to 10\% in smaller teams).

The reduction in headcount is delivered mainly through consolidation, as discussed in Section \ref{sec:ResultsTeamRedesign}. Merged roles are as full as the roles carried over whole before any backfill (91\% against 95\% of capacity on average), so they are not merely the remnants of roles that automation emptied. The share of roles that merge falls with team size, from 39\% in teams of five or six posts to about 21\% in teams of eleven or more, so larger teams absorb the cut with fewer changes to the roles around it, although most of them still consolidate somewhere.

The role chosen to lose a post is at HEO in 28 teams, EO in 20, SEO in 19 and AO in 13, and never at G6 or G7 (Table~\ref{tab:RestructureGrades}). The cut post and automation remove 6,625 hours of work a week. Of this, 2,960 hours is the capacity the cut itself removes (80 posts at 37 hours), leaving 3,665 hours of spare capacity. The model uses almost all of it (92\%, 3,369 hours) to add human-only work, in 73 of the 80 teams and none in the remaining seven (in two of which there is no spare capacity to fill). The median team fills 94\% of its spare capacity. Because additions are capped at spare capacity, these hours reflect the room available rather than an independent estimate of the demand for human work. 

In total the model proposes 541 additions, of which 46\% are more time on an existing task and 54\% are new tasks. Almost four in ten additions (39\%) concern verifying or handling exceptions to automated output, followed by maintaining relationships (15\%), coordination and negotiation (13\%), and accountability or legal authority (12\%); the rest relate to tacit knowledge, ethics and duty of care, physical presence, judgment and creativity. In other words, much of the new work given to people is the supervision of the automation that replaced the rest.

Having examined the effects in aggregate, to show how these effects combine within a team, Table~\ref{tab:RestructureExamples} summarizes three viable teams of different sizes, and we describe each in turn. To avoid identifying any team or role, we report only their size, the structural change, and aggregate hours.

\providecommand{\hd}[1]{\begin{tabular}[b]{@{}c@{}}#1\end{tabular}}
\begin{table}[!htb]
    \centering
    \small
    \begin{threeparttable}
    \caption{Three anonymized viable teams.}
    \label{tab:RestructureExamples}
\begin{tabular}{@{}lccccccc@{}}
\toprule
 & Posts & Roles & \multicolumn{5}{c}{Hours per week} \\
\cmidrule(lr){4-8}
Team & before $\to$ after & before $\to$ after & Original & \hd{Leaves\\with post} & Automated & Added & Capacity \\
\midrule
A & 5 $\to$ 4 & 5 $\to$ 4 & 185 & 34 & 34 & 31 & 148 \\
B & 8 $\to$ 7 & 7 $\to$ 6 & 296 & 25 & 61 & 47 & 259 \\
C & 12 $\to$ 11 & 6 $\to$ 4 & 444 & 28 & 47 & 35 & 407 \\
\bottomrule
\end{tabular}
    \begin{tablenotes}[flushleft]
        \footnotesize
        \item \textit{Note.} ``Leaves with post'' is work that goes with the removed post, ``Added'' is human-only work added in Step 4, and capacity is the remaining posts multiplied by a 37-hour week.
    \end{tablenotes}
    \end{threeparttable}
\end{table}

\emph{Team A} is a small operational team of five single-post roles at junior grades. The model removes the post from the role with the most rules-based and administrative work, reasoning that it carries little decision-making authority and that its duties largely repeat those of two colleagues. Almost all of that role's work leaves with the post, and its one protected task, which requires someone to be physically present, joins the other junior administrative role, which absorbs it in a merged role. The three higher-grade casework and finance roles are unchanged. Just under a fifth of the team's hours are automated, mostly routine reconciliation, templated correspondence and reporting. The slack this creates is then largely filled with checking the automated output, handling cases it cannot resolve, and identifying vulnerable cases, so that every remaining role is close to fully loaded.

\emph{Team B} is a mid-sized support and assurance team in which one role, the only one with two post-holders, loses a post. Much of its work is data analysis, information management and reporting, which the model judges AI can largely absorb, and it overlaps with another support role at the same grade. The two support roles are merged into one two-post role, while the specialist roles keep their shape, and a handful of tasks move to a specialist role where related work already sits. This is the team with the highest automated share of the three, and it adds the most human-only work: challenging reported data, signing off AI-drafted reports, and site visits and relationship work that cannot be done remotely.

\emph{Team C} is a larger analytical team whose overall automated share is lower. Here the model removes a post from an analyst role judged to have the most automatable mix of work, but protects its tasks involving commissioning, external relationships and briefing senior stakeholders. The regrouping consolidates six roles into four, merging two near-identical senior leadership roles and two overlapping analyst roles into a single five-post role, leaves one senior role untouched, and keeps a single post in a reduced version of the chosen role. The work added is mainly checking AI-produced statistics, reviewing AI-drafted briefings, and negotiating access to data.

In all three teams the average automation-potential score of the work that remains, weighted by hours, is lower after the restructuring than before (from 0.48 to 0.36, 0.43 to 0.38, and 0.45 to 0.39 respectively). The work left to people is therefore less exposed than the work that was there originally, mainly because the most exposed tasks have been automated away. The human-only additions are not uniformly low-exposure, however: the blind scores for the verification tasks, the largest category, average around 0.5, which suggests that some of this work remains itself a candidate for later automation, for example through LLM-as-a-judge workflows.

\begin{figure}[htbp]
    \centering
    \begin{subfigure}[t]{0.49\textwidth}
        \centering
        \includegraphics[width=\textwidth]{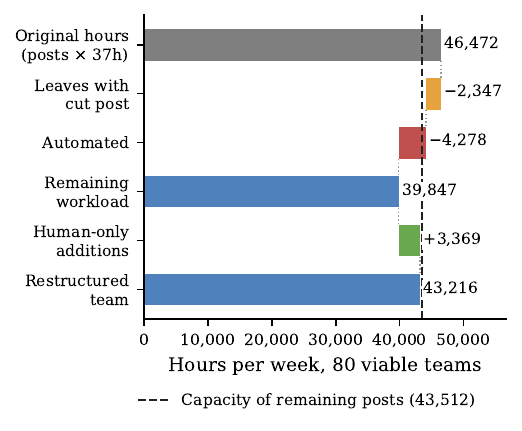}
        \caption{Where the hours go}
        \label{fig:RestructureHoursWaterfall}
    \end{subfigure}
    \hfill
    \begin{subfigure}[t]{0.49\textwidth}
        \centering
        \includegraphics[width=\textwidth]{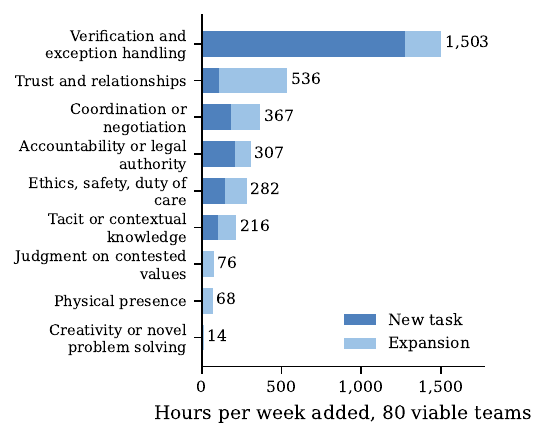}
        \caption{Why a human is still needed}
        \label{fig:RestructureBackfillReasons}
    \end{subfigure}
    \caption{Hours in the 80 viable teams. Panel (a) follows weekly hours from the original workload (posts multiplied by a 37-hour week) through the work that leaves with the removed posts and the work the model judges can be automated, to the remaining workload, then adds the human-only work the model proposes. The dashed line is the capacity of the remaining posts. Panel (b) shows the added hours by the reason the model gives for why a human is needed, split between new tasks and additional time on an existing task (``Expansion'').}
    \label{fig:RestructureHoursBackfill}
\end{figure}

Taken together, the exercise suggests that the effect of automation on a team is larger and more structural than a headcount calculation implies. Removing one post in every team (6.4\% of posts) is accompanied by AI absorbing a further 9.2\% of the team's hours, and the roles that remain are rebuilt rather than thinned: the number of roles falls by almost three times as much as the number of posts (17.7\%), and in most teams the model reaches the smaller headcount by merging overlapping roles rather than by trimming each one. Spare capacity is not simply banked. The model redeploys almost all of it (92\%) to human-only work, and that work is dominated by supervising the automation itself: checking AI output, handling its exceptions, and holding the accountability and relationships that cannot be delegated. The work that remains is, on average, less exposed than the work the team started with, but the added work is not exposure-free. The restructuring is therefore best read as an existence proof that team-level slack can plausibly be reorganized into coherent roles, and as a description of what an LLM judges that reorganization would look like, not as a forecast of what departments will do.

Several features of the design limit how far these results can be taken. First, the redesign is unvalidated. Every choice rests on one model's judgment, and we have not checked it against HR professionals, line managers or the institutional constraints of the Civil Service, the kind of external check that the study of algorithm-augmented work would motivate \citep{AllenChoudhury2022AlgorithmAugmentedWork}.
Second, the hours-based account treats work as transferable between people. Strategic human capital theory stresses that valuable capabilities are often specific to individuals, teams and contexts \citep{PloyhartMoliterno2011HumanCapitalResource, CampbellCoffKryscynski2012HumanCapitalAdvantage, WrightCoffMoliterno2014StrategicHumanCapital}. Merging roles, as the model does in most teams, assumes a post-holder can take on a colleague's tasks at the same productivity, which may not hold where the knowledge is tacit or tied to a relationship. The constraint we impose is on hours, not on the knowledge required to do the work.
Third, the exercise does not price coordination. Merged roles and moved tasks create epistemic interdependence that requires post-holders to understand each other's work \citep{PuranamRaveendranKnudsen2012}, and we choose automation (Step 2) and augmentation (Step 4) in separate steps although the two interact \citep{RaischKrakowski2021AutomationAugmentationParadox}.
Fourth, the added human work assumes that AI deepens rather than flattens expertise. Experimental evidence suggests AI can make outputs more homogeneous across specializations \citep{DellAcquaEtAl2025CyberneticTeammate, DoshiHauser2024GenerativeAICreativityDiversity}, in which case the specialized, human-centric roles we describe might not materialize. Fifth, the redesign is imposed top-down. It says nothing about how post-holders would craft their own roles \citep{WrzesniewskiDutton2001CraftingAJob}, nor about grading structures, unions or redundancy rules. Removing one post is a counterfactual device, not a recommendation to remove staff.
Sixth, the exercise holds the volume of work fixed. If AI raises productivity and the demand for output rises, or if exposure is concentrated enough to allow reallocation within roles, the net effect on labor demand could differ \citep{HampoleEtAl2025AILaborMarket, AghionEtAl2025DifferentUsesAIShapeLaborDemand}. Our backfill is a within-team analogue of that reallocation channel, not a test of it.
Finally, teams are constructed from advertised vacancies, so each team is only the part of a team that was recruiting. 
\clearpage
\section{Worked example}\label{sec:AppendixExampleWalkthrough}

This appendix follows a single job advert through every stage of the analysis. To avoid identifying the team, we describe the role generically: a Senior Executive Officer (SEO) post in the Human Resources profession, leading stakeholder engagement for a people-analytics team in a central government department, with a single post advertised.

\subsection{Entry in GRID}\label{sec:ExampleGRID}

The advert enters our data as one row of the raw GRID extract, with a title, department, grade, a job description, a job summary, and posting and closing dates, the profession, and business unit. To provide the models with the relevant context, we concatenate the job description (1,185 characters after cleaning) and job summary (1,905 characters) into a single advert text of 3,105 characters. This is the text the models read. The duties are described in a short paragraph: the team provides regular reporting and insight on workforce, diversity, pay, engagement, recruitment, retention and absence; the postholder engages with stakeholders, acts as the focal point for customers, communicates with senior leaders, and gathers requirements for, and builds, automated solutions. The remainder of the advert describes the wider shared-services organization and application guidance.

\subsection{Task Extraction and scoring by Sonnet 4.6}\label{sec:ExampleSonnet}

Claude Sonnet 4.6 receives the advert text in a single call (the prompt is reproduced in Appendix~\ref{sec:AppendixExtractAndScorePrompt}). It extracts the tasks the postholder would perform, in the order the advert gives them, and for each writes a one-sentence rationale before assigning an automation-potential score on an eleven-point scale from 0.0 to 1.0, defined as the share of the task's substantive content that current AI systems could perform. For this advert it extracts six tasks (Table~\ref{tab:ExampleTasks}). The first, regular reporting and insight on HR topics, scores 0.6: the model reasons that AI can aggregate data, generate standard reports and surface patterns, but that a person is needed to contextualize findings and turn them into recommendations. The last, designing and building automated solutions, scores 0.5. The three stakeholder-facing tasks score 0.2, because the model judges the substance to be relationship-building and judgment, and requirements gathering scores 0.3.

\subsection{Scoring by the other models}\label{sec:ExampleModels}

The extracted task list is then scored by seven other models using the same scoring rubric, so that no exposure estimate rests on one model's judgment (Appendix~\ref{sec:AppendixScorePrompt}). The models broadly agree on the ordering of task scores: reporting is the most automatable and the stakeholder tasks the least. They differ on levels. For the reporting task, scores range from 0.6 (Sonnet 4.6, GPT-OSS-120B, Nemotron) to 0.8 (Qwen), and one model scores the two technical tasks well above Sonnet. Disagreement, measured as twice the standard deviation of the eight scores, is highest on the task of coordinating with customers (0.25) and lowest on the first two tasks (0.14). The \emph{ensemble score} for each task is the mean across the eight models (Table~\ref{tab:ExampleTasks}).

\begin{table}[!htbp]
    \centering
    \small
    \begin{threeparttable}
    \caption{Tasks extracted from the example advert, with Sonnet 4.6 and eight-model ensemble automation-potential scores, and the hours implied by the order of the advert (decay parameter 0.75, 37-hour week).}
    \label{tab:ExampleTasks}
\begin{tabular}{@{}c>{\raggedright\arraybackslash}p{5.2cm}cccc@{}}
\toprule
 & & \multicolumn{2}{c}{Automation potential} & & \\
\cmidrule(lr){3-4}
Task & Summary & Sonnet 4.6 & Ensemble & Hours/week & Weight \\
\midrule
1 & Regular reporting and insight on HR issues & 0.6 & 0.68 & 11.25 & 0.304 \\
2 & Engaging with stakeholders & 0.2 & 0.28 & 8.44 & 0.228 \\
3 & Coordinating with customers as focal point & 0.2 & 0.31 & 6.33 & 0.171 \\
4 & Communicating with senior leaders & 0.2 & 0.26 & 4.75 & 0.128 \\
5 & Gathering requirements for automated solutions & 0.3 & 0.45 & 3.56 & 0.096 \\
6 & Designing and building automated solutions & 0.5 & 0.63 & 2.67 & 0.072 \\
\bottomrule
\end{tabular}
    \end{threeparttable}
\end{table}

\subsection{Weighting}\label{sec:ExampleWeighting}

Three weights turn the task scores into an estimate for the Civil Service as a whole.

\paragraph{Within the vacancy.} Adverts tend to list duties roughly in order of importance, so we weight task $i$ in proportion to $0.75^{\,i-1}$, normalized to sum to one, and multiply by a 37-hour week to obtain the hours per postholder in Table~\ref{tab:ExampleTasks}. The vacancy's exposure is the weighted mean of the ensemble task scores, 0.444, and its variance across tasks is 0.032 (standard deviation 0.180). Because the most automatable task is listed first, exposure rises as the decay becomes steeper: it is 0.433 with no decay, 0.434 at a decay of 0.9 and 0.493 at 0.5. Mean cross-model disagreement for the vacancy is 0.173.

\paragraph{Across vacancies.} Each vacancy represents a number of Civil Service full-time equivalents (FTE). Weights start at the number of posts advertised and are raked so that the adverts reproduce the published department, grade and profession totals, giving this advert a weight of 1.449 FTE for one advertised post. Under the alternative of matching published department-by-grade-by-profession cells, its weight is 1.029 FTE.

\paragraph{Salary and value.} The average salary for this department, grade and profession cell, \pounds 47,770, is used because that cell is large enough; the department-by-grade mean (\pounds 49,360) and the grade mean (\pounds 49,300) are the fall-backs. The paybill the vacancy represents is therefore $1.449 \times \pounds 47{,}770 \approx \pounds 69{,}215$, and the AI opportunity, paybill multiplied by mean exposure, is $\pounds 69{,}215 \times 0.444 \approx \pounds 30{,}701$.

\subsection{Assignment to a category}\label{sec:ExampleCategory}

Each vacancy is placed on two axes: its mean exposure and the variance of exposure across its tasks. Using the FTE-weighted distribution, the cut points are a mean exposure of 0.417 (75th percentile), a variance of 0.0101 (25th percentile among high-mean roles) and a variance of 0.0183 (75th percentile among low-mean roles). The example role has a mean of 0.444, above the first threshold, and a variance of 0.032, well above the second, so it is classified as \emph{Augmentation-type}: exposure is high on average but concentrated in a few tasks (here, the reporting and technical tasks), while the stakeholder work remains mostly human. It lies close to the vertical cut, so a slightly lower mean would place it among the lower-exposure categories. Augmentation-type roles account for 18.8\% of classified FTE. For roles of this type the counted saving is the opportunity (paybill multiplied by mean exposure), \pounds 30,701 here, rather than the full paybill.

\subsection{Individual role redesign}\label{sec:ExampleIndividual}

Roles in the Augmentation quadrant were passed through the individual redesign described in Section~\ref{sec:MethodologyRedesign}, which considers each role in isolation.

\paragraph{Stage 1: swap.} Given the advert and task list, but not the scores, Sonnet 4.6 proposes to replace the reporting task with a review-and-interpretation task: reviewing AI-generated HR reports, validating data quality, identifying significant trends or anomalies, and writing narratives that explain the implications and recommended actions. Its stated reasoning is that AI can now generate routine, recurring reports at scale, so the postholder's most valuable contribution shifts from producing reports to interpreting them. GPT-OSS-120B scores the new task at 0.6, and it is placed fifth of six in the role's time ranking (a rank fraction of 0.8). Under our reconstruction of the time conversion, the new task occupies about 3.6 hours per week, against 11.25 for the task it replaces, freeing about 7.7 hours.

\paragraph{Stage 2: filling the freed time.} The model is then asked to propose additional tasks, either without guidance (main) or with the instruction that each new task must be one an AI system could not already do well on its own (variant); these are alternative versions of the step, not cumulative. Both propose 7.5 hours of new work in four tasks. Without guidance (main), it proposes maintaining a people-analytics product roadmap (2.5 hours, score 0.5), a weekly data-quality audit (2.0 hours, 0.6), monthly user-research sessions with stakeholders (1.5 hours, 0.3) and prototyping a new analytical method (1.5 hours, 0.7). With the added instruction (variant), it proposes discovery sessions with HR business partners and senior leaders (2.5 hours, 0.2), data-literacy workshops and coaching (2.0 hours, 0.3), an ethics and data-quality assurance review of the automated pipeline (2.0 hours, 0.5), and attending cross-government communities of practice (1.0 hour, 0.1).

\paragraph{Outcome.} Scoring all versions with the same model (GPT-OSS-120B) and weighting by hours, the original role has an exposure of 0.35. After the Stage 1 swap alone it is 0.29, because the highest-scoring task is replaced by one that occupies less time. After filling the freed time it is 0.34 (main) or 0.29 (variant): unguided additions partly reintroduce automatable work, while the guided additions are mainly stakeholder, training and assurance tasks.

\subsection{Team redesign}\label{sec:ExampleTeam}

The same role belongs to a team of eight posts in six roles, comprising a head of the function (G6), a systems and data lead (G7), a systems and data manager (SEO), the example role (SEO) and two analyst roles at HEO grade with two posts each. The team restructuring exercise (Appendix~\ref{sec:AppendixTeamRestructure}) asks what the team does when it loses one post and AI absorbs part of the workload.

\paragraph{Stage 1: which role loses the post.} Claude Opus 5.5 chooses one of the two-post HEO data analyst roles. Its reasoning is that around 26 of the role's 37 weekly hours go on routine reporting, dashboards and data visualization, which current AI tools can largely do, and that another HEO analyst role does similar dashboard work and could absorb what remains. It notes that the leadership, solution-building and stakeholder roles each have only one post, so a cut there would remove a whole capability. Two of the chosen role's tasks (agreeing needs with senior leaders and drawing out requirements for strategic reports) are set aside as protected, because they rest on trust and on negotiation between competing priorities. Our worked-example role is therefore not cut: it keeps its post and all of its tasks proceed to Stage 2.

\paragraph{Stage 2: which tasks are automated.} The model marks nine tasks across the team as automatable, amounting to 84.9 of the team's 296 weekly hours. The average GPT-OSS-120B score of these tasks is 0.57, against 0.32 for the rest, even though the model never saw scores. One of the nine is the example role's first task, regular HR reporting, which accounts for 11.25 hours per week: the model's stated reason is that routine reporting, including recruitment metrics, can be produced substantially by AI from HR data.

\paragraph{Stage 3: regrouping.} The remaining work is about 182 hours per week against a capacity of 259 hours for seven posts, so no role is overloaded. The model keeps all six roles, and takes the one post from the analyst tier, noting that the second post goes to the other analyst role because its remaining 35.4 hours would almost fill a single post. The example role is kept unchanged, with its five remaining tasks totaling 25.75 hours against a capacity of 37, leaving 11.25 hours of slack.

\paragraph{Stage 4: backfill.} The model proposes 11.0 hours of human-only work for the role, in five additions that account for almost all of the slack (Table~\ref{tab:ExampleBackfill}). Three of them are about controlling what automation produces: checking AI-generated insight packs for plausible but wrong conclusions (for example, an attrition spike caused by a data-load error), acting as the named sign-off for disclosure control on diversity, pay and engagement outputs, and explaining anomalies that the data cannot, such as restructures or transfers between departments. The other two are about relationships: coaching colleagues in how to read and question the dashboards, and extra time presenting insights to senior leaders in person. Scored blind with GPT-OSS-120B, the role's hours-weighted exposure is 0.35 originally, 0.25 after automation and before backfill, and 0.28 once the backfill is added.

\begin{table}[!htbp]
    \centering
    \small
    \begin{threeparttable}
    \caption{Work added to the example role in Stage 4 of the team redesign, with the reason given for why a human is needed and the blind GPT-OSS-120B automation-potential score.}
    \label{tab:ExampleBackfill}
\begin{tabular}{@{}>{\raggedright\arraybackslash}p{7.2cm}>{\raggedright\arraybackslash}p{4.4cm}cc@{}}
\toprule
Added work & Reason a human is needed & Hours & Score \\
\midrule
Check AI-generated insight packs and commentary before release & Verification and exception handling & 3.0 & 0.6 \\
Named sign-off for disclosure control on sensitive outputs & Accountability or legal authority & 2.0 & 0.4 \\
Explain anomalies flagged by automated reporting & Tacit or contextual knowledge & 2.0 & 0.3 \\
Coach colleagues on reading and questioning dashboards & Trust and relationships & 2.0 & 0.2 \\
More time presenting insights to senior leaders (expansion of existing task) & Trust and relationships & 2.0 & 0.2 \\
\bottomrule
\end{tabular}
    \end{threeparttable}
\end{table}

\subsection{Comparing the two redesigns}\label{sec:ExampleComparison}

Both approaches arrive at a similar place (Table~\ref{tab:ExampleComparison}): around 11 hours of new work built around a stable core of stakeholder tasks, and hours-weighted exposure of 0.28--0.29, down from 0.35, provided the individual redesign is steered towards work that AI cannot already do. They differ in how the reporting task is handled. The individual redesign keeps a reduced version of it as a review task, whereas the team redesign removes it altogether because the reports can be produced by AI. They also differ in the character of the new work. The individual redesign, working from the advert alone, leans towards discovery, training and assurance. The team redesign is driven by what the automation leaves exposed, and so concentrates on verification, accountability and trust. Unguided individual redesign (main) barely reduces exposure (0.34), which reflects the point that new work proposed without a constraint is itself often automatable.

\begin{table}[!htbp]
    \centering
    \small
    \begin{threeparttable}
    \caption{Hours-weighted exposure of the example role under each redesign, scored by GPT-OSS-120B.}
    \label{tab:ExampleComparison}
\begin{tabular}{lccc}
\toprule
 & Tasks & New-work hours & Exposure \\
\midrule
Original advert & 6 & 0 & 0.35 \\
Individual redesign (main) & 10 & 11.1 & 0.34 \\
Individual redesign (variant) & 10 & 11.1 & 0.29 \\
Team redesign & 10 & 11.0 & 0.28 \\
\bottomrule
\end{tabular}
    \begin{tablenotes}[flushleft]
        \footnotesize
        \item \textit{Note.} New-work hours include the replacement task in the individual redesign (3.6 hours) and the expanded hours in the team redesign.
    \end{tablenotes}
    \end{threeparttable}
\end{table}
\clearpage

\end{document}